\newif\iffigs\figstrue
\documentclass[a4paper,11pt]{article}
\usepackage{latexsym,amssymb,lscape,graphics}
\usepackage{graphicx}        
\usepackage{longtable}
\usepackage{multirow}
\usepackage{color}
\usepackage{slashed,epsfig}
\usepackage{amsfonts}

\newcommand{\e}{\textrm{e}}

\newtheorem{definizione}{Definition}[section]

\newtheorem{statement}{Statement}[section]

\newcommand{\bd}{\begin{definizione}}
\newcommand{\ed}{\end{definizione}}

\def\IC{\relax\,\hbox{$\inbar\kern-.3em{\rm C}$}}
\def\IG{\relax\,\hbox{$\inbar\kern-.3em{\rm G}$}}
\def\IB{\relax{\rm I\kern-.18em B}}
\def\ID{\relax{\rm I\kern-.18em D}}
\def\IL{\relax{\rm I\kern-.18em L}}
\def\IF{\relax{\rm I\kern-.18em F}}
\def\IH{\relax{\rm I\kern-.18em H}}
\def\II{\relax{\rm I\kern-.17em I}}
\def\IN{\relax{\rm I\kern-.18em N}}
\def\IP{\relax{\rm I\kern-.18em P}}
\def\IQ{\relax\,\hbox{$\inbar\kern-.3em{\rm Q}$}}
\def\bfzero{\relax\,\hbox{$\inbar\kern-.3em{\rm 0}$}}
\def\IK{\relax{\rm I\kern-.18em K}}
\def\IG{\relax\,\hbox{$\inbar\kern-.3em{\rm G}$}}
 \font\cmss=cmss10 \font\cmsss=cmss10 at 7pt
\def\IR{\relax{\rm I\kern-.18em R}}
\def\ZZ{\relax\ifmmode\mathchoice
{\hbox{\cmss Z\kern-.4em Z}}{\hbox{\cmss Z\kern-.4em Z}}
{\lower.9pt\hbox{\cmsss Z\kern-.4em Z}} {\lower1.2pt\hbox{\cmsss
Z\kern-.4em Z}}\else{\cmss Z\kern-.4em Z}\fi}
\def\bfone{\relax{\rm 1\kern-.35em 1}}

\def\inbar{\vrule height1.5ex width.4pt depth0pt}
\def\bfzero{\relax{\rm I\kern-.18em 0}}
\def\bfone{\relax{\rm 1\kern-.35em 1}}

\DeclareFontFamily{U}{rsf}{} \DeclareFontShape{U}{rsf}{m}{n}{
  <5> <6> rsfs5 <7> <8> <9> rsfs7 <10-> rsfs10}{}
\DeclareMathAlphabet\Scr{U}{rsf}{m}{n}

\def\e{\epsilon}

\newcommand{\SO}{\mathop{\rm SO}}

\newcommand{\U}{\mathop{\rm {}U}}

\newcommand{\ft}[2]{{\textstyle\frac{#1}{#2}}}
\def\tilde{\widetilde}

\def\1bar{1\hskip -.275cm -}
\def\2bar{2\hskip -.275cm -}
\def\3bar{3\hskip -.275cm -}

\newsavebox{\uuunit}
\sbox{\uuunit}
                 {\setlength{\unitlength}{0.825em}
                      \begin{picture}(0.6,0.7)
                                      \thinlines
                                      \put(0,0){\line(1,0){0.5}}
                                      \put(0.15,0){\line(0,1){0.7}}
                                      \put(0.35,0){\line(0,1){0.8}}
                                     \multiput(0.3,0.8)(-0.04,-0.02){10}{\rule{0.5pt}{0.5pt}}
                      \end {picture}}

\makeatletter \@addtoreset{equation}{section} \makeatother

\def\bfone{\relax{\rm 1\kern-.35em 1}}

\def\bfone{\relax{\rm 1\kern-.35em 1}}
\font\cmss=cmss10 \font\cmsss=cmss10 at 7pt

\newcommand{\so}{\mathfrak{so}}
\newcommand{\su}{\mathfrak{su}}

\newcommand{\slal}{\mathfrak{sl}}

\begin{document}
\begin{titlepage}
\vskip 0.2cm
\begin{center}
{\Large {\bf Black Hole Nilpotent Orbits \\
and Tits Satake Universality Classes}}\\[1cm]
{\large Pietro Fr\'e$^{a}$\footnote{Prof. Fr\'e is presently fulfilling the duties of Scientific Counsellor of the Italian Embassy in the Russian Federation, Denezhnij pereulok, 5, 121002 Moscow, Russia.}, Alexander S. Sorin$^{b}$ and Mario Trigiante$^{c}$}
{}~\\
\quad \\
{{\em $^{a}$  Dipartimento di Fisica Teorica, Universit\'a di Torino,}}
\\
{{\em $\&$ INFN - Sezione di Torino}}\\
{\em via P. Giuria 1, I-10125 Torino, Italy}~\quad\\
{\tt fre@to.infn.it}
{}~\\
\quad \\
{{\em $^{b}$ Bogoliubov Laboratory of Theoretical Physics,}}\\
{{\em Joint Institute for Nuclear Research,}}\\
{\em 141980 Dubna, Moscow Region, Russia}~\quad\\
{\tt sorin@theor.jinr.ru}{}~\\
\quad \\
{{\em $^e$  Dipartimento di Fisica Politecnico di Torino,}}\\
{\em C.so Duca degli Abruzzi, 24, I-10129 Torino, Italy}~\quad\\
{\tt mario.trigiante@gmail.com}
\quad \\
\end{center}
~{}
\begin{abstract}
In this paper we consider the problem of classification of nilpotent orbits for the pseudo-quaternionic coset manifolds $\mathrm{U/H}^\star$ obtained in the time-like dimensional reduction of $\mathcal{N} = 2$ supergravity models based on homogeneous symmetric special geometries. Within the $D=3$ approach this classification amounts to a classification of regular and singular extremal black hole solutions of supergravity. We show that the pattern of such orbits is a universal property depending only on the Tits-Satake universality class of the considered model, the number of such classes being  five.  We present a new algorithm for the classification and construction of the nilpotent orbits for each universality class which is based on an essential use of the Weyl group $\mathcal{W}$ of the Tits Satake subalgebra $\mathbb{U}_{\mathrm{TS}} \subset \mathbb{U}$ and on a certain subgroup thereof  $\mathcal{W}_H \subset \mathcal{W}$. The splitting of orbits of the full group $\mathrm{U}$ into suborbits with respect to the stability subgroup $\mathrm{H}^\star \subset \mathrm{U}$ is shown to be governed by the structure of the discrete coset $\mathcal{W} / \mathcal{W}_H$. For the case of the universality class $\mathrm{SO(4,5) /SO(2,3) \times SO(2,2)}$  we derive the complete list of nilpotent orbits which happens to contain $37$ elements. We also show how the universal orbits are regularly embedded in  all the members of the class that are  infinite in number. As a matter of check we apply our new algorithm also to the Tits Satake class $\mathrm{G_{(2,2)}/SL(2) \times SL(2)}$ confirming the previously obtained result encompassing  $7$ nilpotent orbits. Perspectives for future developments based on the obtained results are outlined.
\end{abstract}
\end{titlepage}
\tableofcontents
\newpage
\section{Introduction}
The topic of spherically symmetric, asymptotically flat, extremal, black hole solutions of
supergravity has already a  long history. In the mid
nineties a broad interest was raised by the two almost parallel discoveries of the  attractor mechanism \cite{ferrarakallosh,Gibbons:1996af} and of
the first statistical interpretation of black-hole entropy \cite{stromingerBH}. These two discoveries have a strong conceptual link pivoted around the interpretation of the entropy as the square root of the quartic symplectic invariant $\mathfrak{I}_4(p,q)$ of the unified duality group $\mathrm{U_{D=4}}$ acting on the quantized charges of the black hole $(p,q)$. Indeed the quantized charges provide the clue to construct $D$-brane configurations yielding the considered  black-hole solution and on its turn these $D$-brane constructions provide the means to single out the underlying string microstates. This is a particular instance of the general deep relation between the continuous $\mathrm{U}$-duality symmetries of supergravity and the exact discrete dualities mapping different string theories and different string vacua into each other. Indeed the group of string dualities was conjectured to be the restriction to integers $\mathrm{U(\mathbb{Z})}$ of the supergravity duality group \cite{dualitiessg}.
In view of these perspectives, the search and analysis of supergravity BPS black hole solutions was extensively pursued
in the  nineties in all versions of extended supergravity \cite{olderBHliterature}. The basic tool in these analyses was the use of the first order  \textit{Killing spinor equations} obtained by imposing that a certain fraction of the original supersymmetry should be preserved by the classical solution \cite{firstorderSUSY,Behrndt:1997ny,iomatteo,ortino}. Allied tool in this  was the use of the harmonic function construction of $p$-brane solutions of higher dimensional supergravities (see for instance \cite{harmfunpbrane} and references therein). In parallel to this study of classical supergravity solutions an extended investigation of the black-hole microstates within string theory \cite{microstateliterature} was pursued.
\par
The  bridge between the two aspects of the problem, namely the
macroscopic and the microscopic one, was constantly provided by the geometric
and algebraic structure of supergravity theories dictating the
properties of the $\mathrm{U}$-duality group and of the
supersymmetry field-dependent central charges $Z^A$. In this
context the richest and most interesting case of study is that of
$\mathcal{N}=2$ supergravity where the geometric structure of the
scalar sector, i.e. \textit{Special K\"ahler Geometry} \cite{SKG,Andrianopoli:1996cm,specHomgeo},
on one side provides  a challenging mathematical framework to
formulate and investigate all the fundamental questions about
black-hole construction and properties, on the other side it
directly relates these latter to string-compactifications on
three-folds of vanishing first Chern class, i.e. Calabi-Yau
threefolds \cite{CYandBH} or their singular orbifold limits
\cite{orbifoldlimits}.
\par
Renewed interest in the topics of spherically symmetric supergravity
black-holes and a new wave of extended research activities developed
in the last decade as soon as it was realized that the attractor
mechanism is not limited to the BPS black-holes but occurs also for
the non BPS ones \cite{criticalrefs}. In this context there emerged
the concept of \textit{fake-superpotential} \cite{fakeprepo,fakeprepoHJ,Lopes Cardoso:2007ky,fakeprepo2}. The first order
differential equations obtained by imposing the existence of Killing
spinors are just particular instance of a more general class of ``gradient-flow''
equations which are reminiscent  of the Hamilton-Jacobi formulation of classical mechanics.
\par
An answer to the issue of whether  black-hole equations might be put into the form of a dynamical system came with the development of the $D=3$ approach to black-hole solutions \cite{Breitenlohner:1987dg,pioline,Gaiotto:2007ag,Bergshoeff:2008be,Bossard:2009at}.
\par
The fundamental algebraic root of this development is located in the so named $c$-map \cite{cmappa} from Special K\"ahler Manifolds of complex dimension $n$ to quaternion manifolds of real dimension $4n+4$:
\begin{equation}\label{cmappus}
    \mbox{c-map} \quad : \quad \mathcal{SK}_{n} \, \rightarrow \,  \mathcal{QM}_{(4n+4)}
\end{equation}
This latter follows from the systematic procedure of dimensional reduction from a $D=4,\mathcal{N}=2$ supergravity theory to a $D=3$ $\sigma$-model endowed with $\mathcal{N}=4$ three-dimensional supersymmetry. Naming $z^i$ the scalar fields that fill the special K\"ahler manifold $\mathcal{SK}_{n}$ and $g_{i{\bar \jmath}}$ its metric, the $D=3$ $\sigma$-model which encodes all the supergravity field equations after dimensional reduction on a space-like direction admits, as target manifold, a quaternionic manifold whose $4n+4$ coordinates we name as follows:
\begin{equation}\label{finnico}
    \underbrace{\{U,a\}}_{2}\, \bigcup \,\underbrace{\{ z^i\}}_{2n} \, \bigcup\, \underbrace{\mathbf{Z} \, = \, \{ Z^\Lambda \, , \, Z_\Sigma \}}_{2n+2}
\end{equation}
and whose quaternionic metric has a  general form that we will shortly present.
\par
The brilliant discovery related with the $D=3$ approach to
supergravity black-holes consists in the following. The radial
dependence of all the relevant functions parameterizing the
supergravity solution can be viewed as the field equations of
another one-dimensional $\sigma$-model where the evolution
parameter $\tau$ is actually a monotonic function of the radial
variable $r$ and where the target manifold is a
\textit{pseudo-quaternionic manifold}
$\mathcal{Q}^\star_{(4n+4)}$ related to the quaternionic manifold
$\mathcal{Q}_{(4n+4)}$ in the following way. The coordinates of
$\mathcal{Q}^\star_{(4n+4)}$ are the same as those of $\mathcal{Q}_{(4n+4)}$,
while the two metrics
differ   only by a change of sign. Indeed we have
\begin{eqnarray}
    ds^2_{\mathcal{Q}} & = &\frac{1}{4} \, \left [ d{U}^2+2\,g_{i{\bar \jmath}}\,d{z}^i\,d{{\bar z}}^{\bar \jmath}
+ \e^{-2\,U}\,(d{a}+{\bf Z}^T\mathbb{C}d{{\bf
Z}})^2\,-\,2 \, e^{-U}\,d{{\bf
Z}}^T\,\mathcal{M}_4(z,{\bar z})\,d{{\bf Z}}\right ]\label{quatermetric}\\
& \Downarrow & \mbox{Wick rot.} \label{viccus}\\
 ds^2_{\mathcal{Q}^\star} & = & \frac{1}{4} \, \left [ d{U}^2+2\,g_{i{\bar \jmath}}\,d{z}^i\,d{{\bar z}}^{\bar \jmath}
+ \e^{-2\,U}\,(d{a}+{\bf Z}^T\mathbb{C}d{{\bf
Z}})^2\,+\,2 \, e^{-U}\,d{{\bf
Z}}^T\,\mathcal{M}_4(z,{\bar z})\,d{{\bf Z}}\right ]\label{pseudoquatermetric}
\end{eqnarray}
In eq.s (\ref{quatermetric},\ref{pseudoquatermetric}), $\mathbb{C}$ denotes the $(2n+2)\times(2n+2)$  antisymmetric matrix defined over the fibers of the symplectic bundle characterizing special geometry, while the \textit{negative definite}, $(2n+2)\times(2n+2)$ matrix $\mathcal{M}_4(z,{\bar z})$ is an object uniquely defined by the geometric set up of special geometry (see ref.\cite{noig22} for a review of the construction of $\mathcal{M}_4$ tailored to our purposes).
The pseudo-quaternionic  metric is non-Euclidean and it has the following signature:
\begin{equation}\label{segnaturaqstarro}
    \mbox{sign}\,\left(ds^2_{\mathcal{Q}^\star}\right) \, = \, \left(\underbrace{+\, , \,  \dots \, , \, + \,}_{2n+2} , \underbrace{\, - \, , \, \dots \, , \, - }_{2n+2}\right )
\end{equation}
The indefinite signature (\ref{segnaturaqstarro}) introduces a clear-cut distinction between non-extremal and extremal black-holes. As solutions of the $\sigma$-model defined by the metric (\ref{pseudoquatermetric}), all
spherically symemtric black-holes correspond to geodesics: the non-extremal ones   to time-like geodesics, while the extremal black-holes are associated with light-like ones. Space-like geodesics produce supergravity solutions with naked singularities \cite{Breitenlohner:1987dg}.
\par
In those cases where the Special Manifold $\mathcal{SK}_n$ is a symmetric space $\mathrm{U_{D=4}/H_{D=4}}$ also the quaternionic manifold defined by the metric
      (\ref{quatermetric}) is a symmetric coset manifold:
      \begin{equation}\label{cosettusD3}
        \frac{\mathrm{U_{D=3}}}{\mathrm{H_{D=3}}}
      \end{equation}
      where $\mathrm{H_{D=3}}\subset \mathrm{U_{D=3}}$ is the \textit{maximal compact subgroup} of the $\mathrm{U}$-duality group, in three dimensions $\mathrm{U_{D=3}}$. The change of sign in the metric (\ref{segnaturaqstarro}) simply turns the coset (\ref{cosettusD3}) into a new one:
      \begin{equation}\label{cosettusD3bis}
        \frac{\mathrm{U_{D=3}}}{\mathrm{H_{D=3}^\star}}
      \end{equation}
      where $\mathrm{H_{D=3}}^\star \subset \mathrm{U_{D=3}}$ is another \textit{non-compact maximal subgroup} of the $\mathrm{U}$-duality group whose Lie algebra $\mathbb{H}^\star$ happens to be a different real form of the complexification of the Lie algebra $\mathbb{H}$ of $\mathrm{H_{D=3}}$. That such a different real form always exists within $\mathrm{U_{D=3}}$ is one of the group theoretical miracles of supergravity.
\subsection{The Lax pair description}
Once the problem of black-holes is reformulated in terms of geodesics within the coset manifold (\ref{cosettusD3bis}) a rich spectrum of additional mathematical techniques becomes available for its study and solution.
\par
The most relevant of these techniques is the Lax pair representation of the supergravity field equations. According to a formalism that we reviewed  in papers \cite{Chemissany:2010zp,noig22}, the fundamental evolution equation takes the following form:
\begin{equation}\label{primolasso}
    \frac{d}{d\tau} \, L(\tau) \, + \, \left[ W(\tau) \, , \, L(\tau)\right] \, = \, 0
\end{equation}
where the so named Lax operator $L(\tau)$ and the connection $W(\tau)$ are  Lie algebra elements of $\mathbb{U}$ respectively lying
in the orthogonal subspace $\mathbb{K}$ and in the subalgebra $\mathbb{H}^\star$ in relation with the decomposition:
\begin{equation}\label{finofresco}
    \mathbb{U} \, = \, \mathbb{H}^\star \oplus \mathbb{K}
\end{equation}
As it was proven by us in \cite{sahaedio, Fre:2009dg,Fre':2007hd, noiultimo,
marioetal} and \cite{Chemissany:2010zp}, both for the case of the
coset (\ref{cosettusD3}) and the coset (\ref{cosettusD3bis}), the Lax
pair representation (\ref{primolasso}) allows the construction
of an explicit integration algorithm which provides the finite
form of any supergravity solution in terms of two initial
conditions, the Lax $L_0 =L(0)$  and the solvable coset
representative $\mathbb{L}_0 \, = \, \mathbb{L}(0)$ at  radial infinity $\tau=0$. \par
The action of the global symmetry group $\U_{D=3}$ on a geodesic can be described as follows:
By means of  a transformation $\U_{D=3}/\mathrm{H}^\star$ we can move the ``initial point'' at $\tau=0$ (described by $\mathbb{L}_0$) anywhere on the manifold, while for a fixed initial point  we can act by means of $\mathrm{H}^\star$ on the ``initial velocity vector'', namely on $L_0$. Since the action of  $\U_{D=3}/\mathrm{H}^\star$  is transitive on the manifold, we can always bring the initial point to coincide with the origin (where all the scalar fields vanish)
and classify the geodesics according to the $\mathrm{H}^\star$-orbit of the Lax matrix at radial infinity $L_0$.
Since the evolution of the Lax operator occurs via a
similarity transformation of $L_0$ by means of a time evolving
element of the subgroup $\mathrm{H}^\star$, it will unfold within a same $\mathrm{H}^\star$-orbit.
Our main purpose is then to classify  all possible solutions by means of $\mathbb{H}^\star$-orbits within $\mathbb{K}$
which, in every $\mathcal{N}=2$ supergravity based on homogeneous
symmetric special geometries, is a well defined irreducible
representation of $\mathbb{H}^\star$.
\subsection{Nilpotent Orbits and Tits Satake Universality Classes}
As it was discussed in \cite{noig22} and in previous literature, regular extremal black-holes are associated with Lax operators $L(\tau)$ that are nilpotent at all times of their evolution. Hence the classification of extremal black-holes requires a classification of the orbits of nilpotent elements of the $\mathbb{K}$ space with respect to the stability subgroup $\mathbb{H}^\star \subset \mathrm{U}_{\mathrm{D=3}}$. This is a well posed, but difficult, mathematical problem. In \cite{noig22} it was solved for the case of the special K\"aher manifold $\frac{\mathrm{SU(1,1)}}{\mathrm{U(1)}}$ which, upon time-like dimensional reduction to $D=3$, yields the pseudo quaternionic manifold $\frac{\mathrm{G_{(2,2)}}}{\mathrm{SU(1,1)} \times \mathrm{SU(1,1)}}$. It would  be desirable to extend the classification of such nilpotent orbits to supergravity models based on all the other special symmetric manifolds. Although these latter fall into a finite set of series, some of them are infinite and it might seem that we need to examine an infinite number of cases. This is not so because of a very important property of special geometries and of their quaternionic descendants.
\par
This relates to the Tits-Satake (TS) projection of  \textit{special homogeneous (SH) manifolds}:
\begin{equation}
  \mathcal{SH} \, \stackrel{\mbox{Tits-Satake}}{\Longrightarrow} \,
  \mathcal{SH}_{\rm TS}
\label{titssatake}
\end{equation}
which was analysed in detail in \cite{titsusataku}, together with  the allied concept of \textit{Paint Group} that had been   introduced previously
in \cite{Fre':2005sr}. What it is meant by this wording is the following.
It turns out that one can define an algorithm, the Tits-Satake projection
$\pi_{\rm TS}$, which works on the space of homogeneous manifolds with a
solvable transitive group of motions $\mathcal{G}_M$, and with
any such manifold associates another one of the same type. This map has a series of
very strong distinctive features:
\begin{enumerate}
  \item $\pi_{\rm TS}$ is a projection operator, so that several different
  manifolds $\mathcal{SH}_i$ ($i=1,\dots ,r$) have the same image $\pi_{\rm TS}\left(\mathcal{SH}_i \right)
  $.
  \item $\pi_{\rm TS}$ preserves the rank of $\mathcal{G}_M$ namely the
  dimension of the maximal Abelian semisimple subalgebra (Cartan
  subalgebra) of $\mathcal{G}_M$.
  \item $\pi_{\rm TS}$ maps special homogeneous into special homogeneous
  manifolds. Not only. It preserves the two classes of
  manifolds discussed above, namely maps  \textit{special K{\"a}hler} into \textit{special
  K{\"a}hler} and maps \textit{Quaternionic} into \textit{Quaternionic}
  \item{ $\pi_{\rm TS}$ commutes with  $c$--map, so that we obtain the
  following commutative diagram:}
  \begin{equation}
  \begin{array}{ccc}
   \mbox{Special K{\"a}hler} &
    \stackrel{\mbox{$c$-map}}{\Longrightarrow} & \mbox{Quaternionic-K{\"a}hler} \\
    \begin{array}{cc}
      \pi_{\rm TS} & \Downarrow \\
    \end{array} & \null & \begin{array}{cc}
      \pi_{\rm TS} & \Downarrow \\
    \end{array} \\
    \left( \mbox{Special K{\"a}hler}\right) _{\rm TS} &
    \stackrel{\mbox{$c$-map}}{\Longrightarrow} & \left( \mbox{Quaternionic-K{\"a}hler}\right) _{\rm TS} \
  \end{array}
\label{diagrammo}
\end{equation}
\end{enumerate}
The main consequence of the above features is that the whole set of
special homogeneous manifolds and hence of associated supergravity models
is distributed into a set of \textit{universality classes} which turns
out to be composed of extremely few elements.
\par
If we confine ourselves to homogenous symmetric special geometries, which are those for which we can implement the integration algorithm based on the Lax pair representation, then the list of special symmetric manifolds contains only eight items among which two infinite series. They are displayed in the first column of table \ref{skTS}.  The $c$-map produces just as many quaternionic (K\"ahler) manifolds, that  are displayed in the second column of the same table.
\begin{table}[h!]
\begin{center}
$$
\begin{array}{||c|c|c||}
\hline
\hline
\null & \null & \null \\
  \mbox{Special K\"ahler}& \mbox{Quaternionic} & \mbox{Tits Satake projection of Quater.} \\
  \mathcal{SK}_n & \mathcal{QM}_{4n+4} & \mathcal{QM}_{\mathrm{TS}}\\
  \null & \null & \null \\
  \hline
  \null & \null & \null \\
  \frac{\mathrm{U(s+1,1)}}{\mathrm{U(s+1)}\times\mathrm{ U(1)}} & \frac{\mathrm{U(s+2,2)}}{\mathrm{U(s+2)}\times\mathrm{ U(2)}}  & \frac{\mathrm{U(3,2)}}{\mathrm{U(3)}\times\mathrm{ U(2)}} \\
  \null & \null & \null \\
  \hline
  \null & \null & \null \\
  \frac{\mathrm{SU(1,1)}}{\mathrm{U(1)}}&\frac{\mathrm{G_{(2,2)}}}{\mathrm{SU(2)} \times \mathrm{SU(2)}}& \frac{\mathrm{G_{(2,2)}}}{\mathrm{SU(2)} \times \mathrm{SU(2)}} \\
  \null & \null & \null \\
  \hline
  \null & \null & \null \\
  \frac{\mathrm{SU(1,1)}}{\mathrm{U(1)}}\times \frac{\mathrm{SU(1,1)}}{\mathrm{U(1)}}&\frac{\mathrm{SO(3,4)}}{\mathrm{SO(3)}\times \mathrm{SO(4)}} & \frac{\mathrm{SO(3,4)}}{\mathrm{SO(3)}\times \mathrm{SO(4)}} \\
  \null & \null & \null \\
  \hline
  \null & \null & \null \\
  \frac{\mathrm{SU(1,1)}}{\mathrm{U(1)}}\times \frac{\mathrm{SO(p+2,2)}} {\mathrm{SO(p+2)}\times \mathrm{SO(2)}}& \frac{\mathrm{SO(p+4,4)}}{\mathrm{SO(p+4)}\times \mathrm{SO(4)}} & \frac{\mathrm{SO(5,4)}}{\mathrm{SO(5)}\times \mathrm{SO(4)}} \\
  \null & \null & \null \\
  \hline
 \begin{array}{c}
 \null\\
     \frac{\mathrm{Sp(6)}}{\mathrm{U(3)}}\\
     \null \\
     \frac{\mathrm{SU(3,3)}}{\mathrm{SU(3)\times SU(3) \times U(1)}} \\
     \null\\
     \frac{\mathrm{SO^\star(12)}}{\mathrm{SU(6)\times  U(1)}} \\
     \null \\
   \frac{\mathrm{E_{(7,-25)}}}{\mathrm{E_{(6,-78)}\times SU(2)}}
   \end{array}
 & \begin{array}{c}
 \null\\
     \frac{\mathrm{F_{(4,4)}}}{\mathrm{Usp(6)\times SU(2)}}\\
     \null\\
     \frac{\mathrm{E_{(6,-2)}}}{\mathrm{SU(6)\times SU(2)}}\\
     \null\\
     \frac{\mathrm{E_{(7,-5)}}}{\mathrm{SO(12)\times SU(2)}}\\
     \null \\
      \frac{\mathrm{E_{(8,-24)}}}{\mathrm{E_{(7,-133)}\times SU(2)}}
   \end{array} & \frac{\mathrm{F_{(4,4)}}}{\mathrm{Usp(6)\times SU(2)}} \\
 \null & \null & \null \\
\hline
\end{array}
$$
\caption{The eight series of homogenous symmetric special K\"ahler manifolds (infinite and finite), their quaternionic counterparts and the grouping of the latter into five Tits Satake universality classes.\label{skTS}}
\end{center}
\end{table}
Upon the Tits-Satake projection, this infinite set of models is organized into just five universality classes that are displayed on the third column of table
\ref{skTS}.
The key-feature of the projection, relevant to our purposes is that all of its properties extend also to the \textit{pseudo-quaternionic} manifolds produced by a time-like dimensional reduction. We can say that there exists a $c^\star$-map defined by this type of reduction, which associates a pseudo-quaternionic manifold with each special K\"ahler manifold. The Tits-Satake projection commutes also with the $c^\star$-map and we have another commutative diagram:
 \begin{equation}
  \begin{array}{ccc}
   \mbox{Special K{\"a}hler} &
    \stackrel{\mbox{$c^\star$-map}}{\Longrightarrow} & \mbox{Pseudo-Quaternionic-K{\"a}hler} \\
    \begin{array}{cc}
      \pi_{\rm TS} & \Downarrow \\
    \end{array} & \null & \begin{array}{cc}
      \pi_{\rm TS} & \Downarrow \\
    \end{array} \\
    \left( \mbox{Special K{\"a}hler}\right) _{\rm TS} &
    \stackrel{\mbox{$c^\star$-map}}{\Longrightarrow} & \left( \mbox{Pseudo-Quaternionic-K{\"a}hler}\right) _{\rm TS} \
  \end{array}
\label{diagrammo2}
\end{equation}
By means of this token, we obtain table \ref{skTSstarro}, perfectly analogous to table \ref{skTS} where the Pseudo-Quaternionic manifolds associated which each symmetric special geometry are organized into five distinct Tits Satake universality classes.
\begin{table}[h!]
\begin{center}
$$
\begin{array}{||c|c|c||}
\hline
\hline
\null & \null & \null \\
  \mbox{Special K\"ahler}& \mbox{Pseudo-Quaternionic} & \mbox{Tits Satake proj. of Pseudo Quater.} \\
  \mathcal{SK}_n & \mathcal{QM}^\star_{4n+4} & \mathcal{QM}^\star_{\mathrm{TS}}\\
  \null & \null & \null \\
  \hline
  \null & \null & \null \\
  \frac{\mathrm{U(s+1,1)}}{\mathrm{U(s+1)}\times\mathrm{ U(1)}} & \frac{\mathrm{U(s+2,2)}}{\mathrm{U(s+1,1)}\times\mathrm{ U(1,1)}}  & \frac{\mathrm{U(3,2)}}{\mathrm{U(2,1)}\times\mathrm{ U(1,1)}} \\
  \null & \null & \null \\
  \hline
  \null & \null & \null \\
  \frac{\mathrm{SU(1,1)}}{\mathrm{U(1)}}&\frac{\mathrm{G_{(2,2)}}}{\mathrm{SU(1,1)} \times \mathrm{SU(1,1)}}& \frac{\mathrm{G_{(2,2)}}}{\mathrm{SU(1,1)} \times \mathrm{SU(1,1)}} \\
  \null & \null & \null \\
  \hline
  \null & \null & \null \\
  \frac{\mathrm{SU(1,1)}}{\mathrm{U(1)}}\times \frac{\mathrm{SU(1,1)}}{\mathrm{U(1)}}&\frac{\mathrm{SO(3,4)}}{\mathrm{SO(2,1)}\times \mathrm{SO(2,2)}} & \frac{\mathrm{SO(3,4)}}{\mathrm{SO(1,2)}\times \mathrm{SO(2,2)}} \\
  \null & \null & \null \\
  \hline
  \null & \null & \null \\
  \frac{\mathrm{SU(1,1)}}{\mathrm{U(1)}}\times \frac{\mathrm{SO(p+2,2)}} {\mathrm{SO(p+2)}\times \mathrm{SO(2)}}& \frac{\mathrm{SO(p+4,4)}}{\mathrm{SO(p+2,2)}\times \mathrm{SO(2,2)}} & \frac{\mathrm{SO(5,4)}}{\mathrm{SO(3,2)}\times \mathrm{SO(2,2)}} \\
  \null & \null & \null \\
  \hline
 \begin{array}{c}
 \null\\
     \frac{\mathrm{Sp(6)}}{\mathrm{U(3)}}\\
     \null \\
     \frac{\mathrm{SU(3,3)}}{\mathrm{SU(3)\times SU(3) \times U(1)}} \\
     \null\\
     \frac{\mathrm{SO^\star(12)}}{\mathrm{SU(6)\times  U(1)}} \\
     \null \\
   \frac{\mathrm{E_{(7,-25)}}}{\mathrm{E_{(6,-78)}\times SU(2)}}
   \end{array}
 & \begin{array}{c}
 \null\\
     \frac{\mathrm{F_{(4,4)}}}{\mathrm{Sp(6)\times SU(1,1)}}\\
     \null\\
     \frac{\mathrm{E_{(6,-2)}}}{\mathrm{SU(3,3)\times SU(1,1)}}\\
     \null\\
     \frac{\mathrm{E_{(7,-5)}}}{\mathrm{SO^\star(12)\times SU(1,1)}}\\
     \null \\
      \frac{\mathrm{E_{(8,-24)}}}{\mathrm{E_{(7,-25)}\times SU(1,1)}}
   \end{array} & \frac{\mathrm{F_{(4,4)}}}{\mathrm{Sp(6)\times SU(1,1)}} \\
 \null & \null & \null \\
\hline
\end{array}
$$
\caption{The eight series of homogenous symmetric special K\"ahler manifolds (infinite e finite),  their Pseudo-Quaternionic counterparts and the grouping of the latter into five Tits Satake universality classes. \label{skTSstarro}}
\end{center}
\end{table}
\par
The main result of the present paper is contained in the following :
\begin{statement}
\label{miprude}
The number, structure and properties of  $\mathrm{H}^\star$ orbits of $\mathbb{K}$ nilpotent elements depend only on the Tits Satake universality class and it is an intrinsic property of the class.
\end{statement}
So it suffices to determine the classification of nilpotent orbits for the five manifolds appearing in the third column of table \ref{skTSstarro}.
\par
We will provide evidence for statement \ref{miprude} by working out in full detail the classification of nilpotent orbits in one of the five cases of table
\ref{skTSstarro}, namely that of the special geometry series:
\begin{equation}\label{skgseries}
    \mathcal{SKO}_{2s+2} \, \equiv \, \frac{\mathrm{SU(1,1)}}{\mathrm{U(1)}} \, \times \, \frac{\mathrm{SO(2,2+2s)}}{\mathrm{SO(2) } \times \mathrm{SO(2+2s)}}
\end{equation}
that describes one of the possible couplings of $2+2s$ vector multiplets.
\par
Upon space-like dimensional reduction to $D=3$ and dualization of all the vector fields, a supergravity model of this type becomes a $\sigma$-model with the following quaternionic manifold as target space:
\begin{equation}
  \mathcal{QM}_{(4,4+2s)}\, \equiv \, \frac{\mathrm{U_{D=3}}}{\mathrm{H}} \, = \, \frac{\mathrm{SO(4,4+2s)}}{\mathrm{SO(4)} \times \mathrm{SO(4+2s)}}~.
\label{p2qman}
\end{equation}
as mentioned in table \ref{skTS}.
If we perform instead a time-like dimensional reduction, as it is relevant for the construction of black-hole solutions, we obtain an Euclidean $\sigma$-model where, as mentioned in table \ref{skTSstarro} the target space is the following Pseudo-Quaternionic  manifold:
\begin{equation}
  \mathcal{QM}^\star_{(4,4+2s)}\, \equiv \, \frac{\mathrm{U_{D=3}}}{\mathrm{H}^\star} \, = \, \frac{\mathrm{SO(4,4+2s)}}{\mathrm{SO(2,2)} \times \mathrm{SO(2,2+2s)}}~.
\label{p2qmanstarro}
\end{equation}
The Tits Satake projection of all such manifolds is:
\begin{equation}
  \mathcal{QM}^\star_{\mathrm{TS}}\, = \, \frac{\mathrm{U_{D=3}^{TS}}}{\mathrm{H}^\star_{TS}} \, = \, \frac{\mathrm{SO(4,5)}}{\mathrm{SO(2,3)} \times \mathrm{SO(2,2)}}~.
\label{Titussatakus}
\end{equation}
\subsection{Scope of the paper}
In order to obtain the desired classification of nilpotent orbits we have devised a new algorithm which combines the method of standard triples with new techniques based on the Weyl group. Our main result is a list of $37$ nilpotent orbits for the considered model which we claim to be exhaustive.
\par
Equally important is the mechanism of Tits Satake universality which we clearly see at work within our framework.
\par
As a calibration of our new algorithm we reconsidered the nilpotent orbits for the $\mathfrak{g}_{(2,2)}$ case, reobtaining the same classification presented in \cite{noig22}.
\par
We also considered the extension of the method of tensor classifiers introduced in \cite{noig22} and we came to the conclusion that, although useful, they are not able to separate all the distinct orbits in a complete way as it happens in the  $\mathfrak{g}_{(2,2)}$ case.
\par
The perspectives opened by our result, together with the plan of further investigations that it suggests are discussed in the conclusive section \ref{concludo}.
\section{A practitioner approach to the standard triple method for the classification of nilpotent orbits}

The construction and classification of nilpotent orbits in semi-simple Lie algebras is a relatively new field of mathematics which has already generated a vast literature. Notwithstanding this, a well established set of results ready to use by physicists is not yet available mainly because existing classifications are concerned with orbits with respect to the full complex group $\mathrm{G}_\mathbb{C}$ or of one of its real forms $\mathrm{G}_{\mathbb{R}}$ \cite{nilorbits}, which is not exactly what  the problem of supergravity black-holes requires (i.e. the classification of the nilpotent $\mathrm{H}^\star$-orbits in $\mathbb{K}$).  Furthermore the complexity of the existing mathematical papers and books is rather formidable and their reading not too easy. Yet the main mathematical idea underlying all classification schemes is very simple and intuitive and can be rephrased in a language very familiar to physicists, namely that of angular momentum. Such rephrasing allows for what we named a \textit{practitioner's approach} to the method of triples. In other words after decoding this method in terms of angular momentum we can derive case by case the needed results by using a relatively elementary algorithm supplemented with some hints borrowed from mathematical books.
\subsection{Presentation of the method}
\label{praticone}
In this section we shall denote  the isometry group $\mathrm{U}_{D=3}$  by $\mathrm{G}_{\mathbb{R}}$ to emphasize that it is a real form of some complex semisimple Lie group. \par We will present the practitioner's argument in the form of an ordered list.
\begin{enumerate}
\item
The basic theorem  proved by mathematicians (the Jacobson-Morozov theorem \cite{nilorbits}) is that any nilpotent element  of a Lie algebra $X \in \mathfrak{g}$ can be regarded as belonging to a triple of elements $\left \{ x,y,h \right \}$ satisfying the standard commutation relations of the $\slal(2)$ Lie algebra, namely:
\begin{equation}\label{basictriple}
   \left [ h \, , \, x\right ] \, = \, x \quad ; \quad \left [ h \, , \, y\right ] \, = -\, y
   \quad ; \quad \left [ x \, , \, y \right ] \, = \, 2 \, h
\end{equation}
Hence the classification of nilpotent orbits is just the classification of embeddings of an $\slal(2)$ Lie algebra in the ambient one, modulo conjugation by the full group $\mathrm{G}_\mathbb{R}$ or by one of its subgroups. In our case the relevant subgroup is $\mathrm{H}^\star \subset \mathrm{G}_\mathbb{R}$.
\item
The second relevant point in our decoding is that embeddings of subalgebras $\mathfrak{h} \subset \mathfrak{g}$ are characterized by the branching law of any representation of $\mathfrak{g}$ into irreducible representations of $\mathfrak{h}$. Clearly two embeddings might be conjugate only if their branching laws are identical. Embeddings with different branching laws necessarily belong to different orbits. In the case of the $\slal(2) \sim \so(1,2)$ Lie algebra, irreducible representations are uniquely identified by their spin $j$, so that the branching law is expressed by listing the angular momenta $\left\{ j_1 , j_2 , \dots j_n\right \}$ of the irreducible blocks into which any representation of the original algebra, for instance the fundamental, decomposes with respect to the embedded subalgebra. The dimensions of each irreducible module is $2j+1$ so that an a priori constraint on the labels $\left\{ j_1 , j_2 , \dots j_n\right \}$ characterizing an orbit is the summation rule:
\begin{equation}\label{summarulla}
    \sum_{i=1}^{n} (2 j_i +1) \, = \, N\, = \, \mbox{dimension of the fundamental representation}
\end{equation}
Taking into account that $j_i$ are integer or half integer numbers, the sum rule (\ref{summarulla}) is actually a partition of $N$ into integers and this explains why mathematicians classify nilpotent orbits starting from partitions of $N$ and use Young tableaux in the process.
\item
The next observation  is that the central element $h$ of any triple is by definition a diagonalizable (semisimple) non-compact element of the Lie algebra and as such it can always be rotated into the Cartan subalgebra by means of a $\mathrm{G}_\mathbb{R}$ transformation. In the case of interest to us, the Cartan subalgebra $\mathcal{C}$ can be chosen, as we will do, inside the subalgebra $\mathbb{H}^\star$ and consequently we can argue that for any standard triple $\left \{ x,y,h \right \}$ the central element is inside that subalgebra:
\begin{equation}\label{hinsideH}
    h \, \in \, \mathbb{H}^\star
\end{equation}
Since we shall work with  real representations of $\mathrm{G}_{\mathbb{R}}$, we choose a basis in which $h$ is a symmetric matrix.
Indeed there are two possibilities: either $x\in \mathbb{H}^\star$ or $x\in \mathbb{K}$. In the first case we have $y \in \mathbb{H}^\star$, while in the second we have $y \in \mathbb{K}$. This follows from matrix transposition. Given $x$, the element $y$ is just its transposed $y=x^T$ and transposition maps $\mathbb{H}^\star$ into $\mathbb{H}^\star$ and $\mathbb{K}$ into $\mathbb{K}$.  Since it is already in $\mathbb{H}^\star$, in order to rotate the central element $h$  into the Cartan subalgebra it suffices an $\mathrm{H}^\star$ transformation. Therefore to classify $\mathrm{H}^\star$ orbits of nilpotent $\mathbb{K}$ elements we can start by considering central elements $h$ belonging to the Cartan subalgebra $\mathcal{C}$ chosen inside $\mathbb{H}^\star$.
\item The central element $h$ of the standard triple, chosen inside the Cartan subalgebra, is identified by its eigenvalues and by their ordering with respect to a standard basis. Since $h$ is the third component of the angular momentum, \textit{i.e.} the operator $J_3$, its eigenvalues in a representation of spin $j$ are $-j,-j+1,\dots,j-1,j$. Hence if we choose a branching law $\left\{ j_1 , j_2 , \dots j_n\right \}$, we also decide the eigenvalues of $h$ and consequently its components along a standard basis of simple roots. The only indeterminacy which remains to be resolved is the order of the available eigenvalues.
\item The question which remains to be answered is how much we can order the eigenvalues of Cartan elements by means of $\mathrm{H}^\star$ group rotations. The answer is given in terms of the generalized Weyl group $\mathcal{GW}$ and the Weyl group $\mathcal{W}$.
 \item The generalized Weyl group (see \cite{Fre':2007hd}) is the discrete group generated by all matrices of the form:
     \begin{equation}\label{Oalpa}
        \mathcal{O}_{\alpha} \, = \,\exp \left [ \theta_\alpha \, \left ( E^\alpha - E^{-\alpha}\right ) \right ]
     \end{equation}
 where $E^{\pm \alpha}$ are the step operators associated witht the roots $\pm \alpha$ and the angle $\theta_\alpha$ is chosen in such a way that it realizes the $\alpha$-reflection on a Cartan subalgebra element
 $\beta \cdot \mathcal{H}$ associated with a vector $\beta$ :
 \begin{eqnarray}
   O_\alpha \, \beta\cdot \mathcal{H} \,  O_\alpha ^{-1}&=& \sigma_\alpha(\beta) \cdot \mathcal{H}\nonumber\\
   \sigma_\alpha (\beta)  &\equiv & \beta - 2 \, \frac {(\alpha \, , \, \beta)}{(\alpha \, , \, \alpha)} \, \alpha
 \end{eqnarray}
 The generalized Weyl group has the property that for each of its elements $\gamma \, \in \, \mathcal{GW}$ and for each element $h \, \in \, \mathcal{C}$ of the Cartan subalgebra $\mathcal{C}$, we have:
 \begin{equation}\label{furbone}
    \gamma \, h \, \gamma^{-1} \, = \, h^\prime \, \in \, \mathcal{C}
 \end{equation}
\item The generalized Weyl group contains a normal subgroup $\mathcal{HW} \subset \mathcal{GW}$, named the Weyl stability group and defined by the property that for each element $\xi \,\in \, \mathcal{HW}$ and for each Cartan subalgebra element $h \, \in \, {\mathcal{HW}}$ we have:
    \begin{equation}\label{furbissimo}
    \gamma \, h \, \gamma^{-1} \, = \, h
 \end{equation}
\item The proper Weyl group is defined as the quotient of the  generalized Weyl group with respect to the Weyl stability subgroup:
\begin{equation}\label{ronaldino}
    \mathcal{W} \, \equiv \, \frac {\mathcal{GW}}{\mathcal{HW}}
\end{equation}
\item The above definition of the Weyl group shows that we can distinguish among its elements those that can be realized by ${\mathrm{H}}^\star$ transformations, namely those whose corresponding generalized Weyl group elements satisfy the condition $O^T\eta O \, = \, \eta$ and those that are outside of ${\mathrm{H}}^\star$.
\item If we were to consider nilpotent orbits with respect to the whole group $\mathrm{G}$  we would just have to mod out all Weyl transformations. In the case of $\mathrm{H}^\star$ orbits this is too much since the entire Weyl group is not contained in $\mathrm{H}^\star$ as we just said.   The rotations that have to be modded out are those of the intersection of  the generalized Weyl group $\mathcal{GW}_H $ with $\mathrm{H}^\star$, namely:
    \begin{equation}\label{intersezione}
        \mathcal{GW}_H \, \equiv \, \mathcal{GW} \bigcap \mathrm{H}^\star
    \end{equation}
It should be noted that the Weyl stability subgroup is always contained in $\mathrm{H}^\star$ so that, by definition, it is also a subgroup of $\mathcal{GW}_H$:
\begin{equation}\label{fratolino}
    \mathcal{HW} \, \subset \, \mathcal{GW}_H
\end{equation}
which happens to be normal. Hence we can define the ratio
\begin{equation}\label{finocchius}
    \mathcal{W}_H \, \equiv \, \frac{\mathcal{GW}_H }{\mathcal{HW} }
\end{equation}
which is a subgroup of the Weyl group.
\item There is a simple method to find directly $\mathcal{W}_H$.  The Weyl group is the symmetry group of the root system $ \Delta$. When we choose the Cartan subalgebra inside $\mathrm{H}^\star$ the root system splits into two disjoint subsets:
    \begin{equation}\label{subsetti}
        \Delta \, = \, \Delta_H \bigoplus \Delta_K
    \end{equation}
  respectively containing the roots represented in $\mathbb{H}^\star$ and  those represented in $\mathbb{K}$. Clearly the looked for subgroup $\mathcal{W}_H \subset \mathcal{W}$ is composed by those Weyl elements which do not mix $\Delta_H$ with $\Delta_K$ and thus respect the splitting (\ref{subsetti}). According to this viewpoint, given a Cartan element $h$ corresponding to a partition
  $\left\{ j_1 , j_2 , \dots j_n\right \}$, we consider its Weyl orbit and we split this Weyl orbit into $m$ suborbits corresponding to the $m$ cosets:
  \begin{equation}\label{cosettiweilici}
    \frac{\mathcal{W}}{\mathcal{W}_H} \quad ; \quad m\, \equiv \, \frac{|\mathcal{W}|}{|\mathcal{W}_H|}
  \end{equation}
  Each Weyl suborbit corresponds to an $\mathrm{H}^\star$-orbit of the neutral elements $h$ in the standard triples. We just have to separate those triples whose $x$ and $y$ elements lie in $\mathbb{K}$ from those whose $x$ and $y$ elements lie in $\mathbb{H}^\star$. By construction if the $x$ and $y$ elements of one triple lie in $\mathbb{K}$, the same is true for all the other triples in the same $\mathcal{W}_H$ orbit. Weyl transformations outside $\mathcal{W}_H$ mix instead $\mathbb{K}$-triples with $\mathbb{H}^\star$ ones.
  \item The construction described in the above points fixes completely the choice of the central element $h$ in a standard triple providing a standard representative of an $\mathrm{H}^\star$ orbit. The work would be finished if the choice of $h$ uniquely fixed also $x$ and $y=x^T$ that are our main target. This is not so. Given $h$ one can impose the commutation relations:
      \begin{eqnarray}
        \left[ h \, , \, x \right] &=& x \label{pagnato1}\\
        \left[ x \, , \, x^T\right] &=& 2 \, h \label{pagnato2}
      \end{eqnarray}
 as a set of algebraic equations for $x$. Typically these equations admit more than one solution\footnote{Such solutions actually correspond to different $\mathrm{G}_{\mathbb{R}}$-orbits \cite{nilorbits}.}. The next task is that of arranging such solutions in orbits with respect to the stability subgroup  $\mathcal{S}_h \subset \mathrm{H}^\star$  of the central element. Typically such a group is the product, direct or semidirect, of the discrete group
 $\mathcal{HW}$, which stabilizes any Cartan Lie algebra element, with a continuous subgroup of $\mathrm{H}^\star$ which stabilizes only the considered central element $h$. The presence of such a  continuous part of the stabilizer $\mathcal{S}_h$ manifests itself in the presence of continuous parameters in the solution of the second equation (\ref{pagnato2}) at fixed $h$.
 \item When there are no continuous parameters in the solution of eq.(\ref{pagnato2}) what we have to do is quite simple.
 We just need to verify which solutions are related to which by means of $\mathcal{HW}$ transformations and we immediately construct the $\mathcal{HW}$-orbits. Each $\mathcal{HW}$ orbit of $x$  solutions  corresponds to an independent $\mathrm{H}^\star$ orbit of nilpotent operators.
 \item When continuous parameters are left over in the solutions space, signaling the existence of a continuous part in  the $\mathcal{S}_h$ stabilizer, the direct construction of $\mathcal{S}_h$ orbits is more involved and time consuming. An alternative method, however, is available to distribute the obtained solutions into distinct orbits which is based on invariants. Let us define the non-compact operator:
     \begin{equation}\label{betlab1}
       X_c \equiv {\rm i} \left(x \, -\, x^T\right)
     \end{equation}
 and  consider its adjoint action on the maximal compact subalgebra $\mathbb{H} \subset \mathbb{U}$ which, by construction, has the same dimension as $\mathbb{H}^\star$. We name $\beta$-labels the spectrum of eigenvalues of that adjoint matrix\footnote{In the literature, see \cite{nilorbits}, $\beta$-labels are defined as the value of the simple roots $\beta^i$ of the complexification $\mathbb{H}_\mathbb{C}$ of $\mathbb{H}^\star$ on the non-compact element $X_c$, viewed as a Cartan element of $\mathbb{H}_\mathbb{C}$ in the Weyl chamber of $(\beta^i)$. We find it more  practical to work with the equivalent characterization (\ref{betlab2}).}:
 \begin{equation}\label{betlab2}
   \beta-\mbox{label} \, = \, \mbox{Spectrum} \left[ \mathrm{adj}_{\mathbb{H}}\left(X_c\right) \right]
 \end{equation}
 Since the spectrum is an invariant property with respect to conjugation, $x$-solutions that have different $\beta$-labels belong to different $\mathrm{H}^\star$ orbits necessarily. Actually they even belong to different orbits with respect to the full group $\mathrm{U}$. In fact there exists a one-to-one correspondence between nilpotent $\mathrm{U}$ orbits in $\mathbb{U}$ and  $\beta$-labels, which directly follows from the celebrated Kostant-Sekiguchi theorem \cite{nilorbits}. So we arrange the different solutions of eq.(\ref{pagnato2}) into orbits by grouping them according to their $\beta$-labels.
 \item The set of possible $\beta$-labels at fixed choice of the partition $\left\{ j_1 , j_2 , \dots j_n\right \}$ is predetermined since it corresponds to the set of $\gamma$-labels \cite{bruxelles}. Let us define these latter. Given the central element $h$ of the triple, we consider its adjoint action on the subalgebra $\mathbb{H}^\star$ and we set:
     \begin{equation}\label{gamlab}
   \gamma-\mbox{label} \, = \, \mbox{Spectrum} \left[ \mathrm{adj}_{\mathbb{H}^\star}\left(h\right) \right]
 \end{equation}
 Obviously all $h$-operators in the same $\mathcal{W}_H$-orbit have the same $\gamma$-label.  Hence the set of possible
 $\gamma$-labels corresponding to the same partition $\left\{ j_1 , j_2 , \dots j_n\right \}$ contains at most as many elements as the order of lateral classes
 $\frac{\mathcal{W}}{\mathcal{W}_H}$. The actual number can be less when some $\mathcal{W}_H$-orbits of $h$-elements coincide\footnote{Note that the action of certain Weyl group elements $g \in \mathcal{W}$ on specific $h$.s can be the identity: $g\cdot h = h$. When such stabilizing group elements $g$ are inside $\mathcal{W}_H$ the number of different $h$.s inside each lateral classes is accordingly reduced. If there are stabilizing elements $ g$  that are not inside $\mathcal{W}_H$ than  two or more $\mathcal{W}_H$ orbits  coincide.}. Given the set of $\gamma$-labels pertaining to one
 $\left\{ j_1 , j_2 , \dots j_n\right \}$-partition the set of possible $\beta$-labels pertaining to the same partition is the same. We know a priori that the solutions to eq.(\ref{pagnato2}) will distribute in groups corresponding to the available $\beta$-labels. Typically all available $\beta$-labels will be populated, yet for some partition $\left\{ j_1 , j_2 , \dots j_n\right \}$ and for some chosen $\gamma$-label one or more $\beta$-labels might be empty.
 \item The above discussion shows that by naming $\alpha$-label the partition $\left\{ j_1 , j_2 , \dots j_n\right \}$ (branching rule of the fundamental representation of $\mathbb{U}$ with respect to the embedded $\slal(2)$) the orbits can be classified and named with a triple of indices:
     \begin{equation}\label{abclab}
        \mathcal{O}^\alpha_{\gamma\beta}
     \end{equation}
 the set of $\gamma\beta$-labels available for each $\alpha$-label being determined by means of the action of the Weyl group as we have thoroughly explained.
\end{enumerate}
What we have described in the above list is a concrete algorithm to single out standard triple representatives of nilpotent $\mathrm{H}^\star$ orbits of $\mathbb{K}$ operators. In the next section we apply it to the  known example of the $\mathfrak{g}_{(2,2)}$ model in order to show how it works.
\section{The nilpotent orbits of the $\mathfrak{g}_{(2,2)}$ model revisited}
In a recent paper \cite{noig22} we thoroughly discussed the static spherical symmetric Black-Hole solutions of the simplest $\mathcal{N}=2$ supergravity model with one vector vector multiplet coupling, often named the $S^3$-model in the current literature. In that case the relevant $D=3$ group is $\mathrm{G_{(2,2)}}$ and its $\mathrm{H}^\star$-subgroup is $\su(1,1) \times \su(1,1)$. In \cite{noig22} we
showed that a complete classification of the nilpotent $\mathrm{H}^\star$-orbits of $\mathbb{K}$-operators can be effected using the signatures of a certain set of Tensor Classifiers introduced there. Our results were consistent with previous ones in \cite{bruxelles}. In the present section we revisit the classification of nilpotent $\mathrm{H}^\star$-orbits in $\mathfrak{g}_{(2,2)}$ by using the algorithm described in the previous section. The outcome confirms the results of our previous paper, with which it fully agrees.
\subsection{The Weyl  and the generalized Weyl groups for $\mathfrak{g}_{(2,2)}$}
According to our general discussion the most important tools for the orbit classification are the generalized Weyl groups and its subgroups.
\par
We begin with the  structure of the Weyl group for the $\mathfrak{g}_{(2,2)}$ root system $\Delta_{g2}$. By definition this is the group of  rotations in a two-dimensional plane generated by the reflections along all the roots contained in $\Delta_{g2}$. Abstractly the structure of the group is given by the semidirect product of the permutation group of three object $\mathrm{S_3}$ with a $\mathbb{Z}_2$ factor:
\begin{equation}\label{weylg22}
    \mathcal{W} \, = \, \mathrm{S_3} \ltimes \mathbb{Z}_2
\end{equation}
 Correspondingly the order of the group is:
\begin{equation}\label{orderg}
    |\mathcal{W}| \, = \, 12
\end{equation}
An explicit realization by means of $2\times 2 $ orthogonal matrices is the following one:
\begin{equation}\label{weylelementi}
    \begin{array}{ccccccccccc}
                          \mathrm{Id} & = & \left(
\begin{array}{ll}
 1 & 0 \\
 0 & 1
\end{array}
\right) & ; & \alpha_1 & = & \left(
\begin{array}{ll}
 -1 & 0 \\
 0 & 1
\end{array}
\right)& ; & \alpha_2 & = & \left(
\begin{array}{ll}
 -\frac{1}{2} & \frac{\sqrt{3}}{2} \\
 \frac{\sqrt{3}}{2} & \frac{1}{2}
\end{array}
\right) \\
                          \alpha_3 & = & \left(
\begin{array}{ll}
 \frac{1}{2} & \frac{\sqrt{3}}{2} \\
 \frac{\sqrt{3}}{2} & -\frac{1}{2}
\end{array}
\right)& ; & \alpha_4 & = & \left(
\begin{array}{ll}
 \frac{1}{2} & -\frac{\sqrt{3}}{2} \\
 -\frac{\sqrt{3}}{2} & -\frac{1}{2}
\end{array}
\right)& ; & \alpha_5 & = & \left(
\begin{array}{ll}
 -\frac{1}{2} & -\frac{\sqrt{3}}{2} \\
 -\frac{\sqrt{3}}{2} & \frac{1}{2}
\end{array}
\right)\\
                          \alpha_6& = & \left(
\begin{array}{ll}
 1 & 0 \\
 0 & -1
\end{array}
\right) & ; & \xi_1 & = & \left(
\begin{array}{ll}
 -1 & 0 \\
 0 & -1
\end{array}
\right) & ; & \xi_2 & = & \left(
\begin{array}{ll}
 -\frac{1}{2} & -\frac{\sqrt{3}}{2} \\
 \frac{\sqrt{3}}{2} & -\frac{1}{2}
\end{array}
\right)\\
                          \xi_3 & = & \left(
\begin{array}{ll}
 -\frac{1}{2} & \frac{\sqrt{3}}{2} \\
 -\frac{\sqrt{3}}{2} & -\frac{1}{2}
\end{array}
\right) & ; & \xi_4 & = &\left(
\begin{array}{ll}
 \frac{1}{2} & -\frac{\sqrt{3}}{2} \\
 \frac{\sqrt{3}}{2} & \frac{1}{2}
\end{array}
\right)& ;  & \xi_5 & = & \left(
\begin{array}{ll}
 \frac{1}{2} & \frac{\sqrt{3}}{2} \\
 -\frac{\sqrt{3}}{2} & \frac{1}{2}
\end{array}
\right)\\
                        \end{array}
\end{equation}
where $Id$ is the identity element, $\alpha_i$ ($i=1,\dots,6$) denote the reflections along the corresponding roots and $\xi_i$ ($i=1,\dots,5$) are the additional elements created by products of reflections.
The multiplication table of this group is displayed below:
\begin{equation}\label{multatabla}
\begin{array}{|l|llllllllllll|}
\hline
 0 & \mathrm{Id} & \alpha _1 & \alpha _2 & \alpha _3 & \alpha
   _4 & \alpha _5 & \alpha _6 & \xi _1 & \xi _2 & \xi _3 &
   \xi _4 & \xi _5 \\
   \hline
 \mathrm{Id} & \mathrm{Id} & \alpha _1 & \alpha _2 & \alpha _3 &
   \alpha _4 & \alpha _5 & \alpha _6 & \xi _1 & \xi _2 & \xi
   _3 & \xi _4 & \xi _5 \\
 \alpha _1 & \alpha _1 & \mathrm{Id} & \xi _4 & \xi _2 & \xi
   _3 & \xi _5 & \xi _1 & \alpha _6 & \alpha _3 & \alpha _4
   & \alpha _2 & \alpha _5 \\
 \alpha _2 & \alpha _2 & \xi _5 & \mathrm{Id} & \xi _4 & \xi
   _1 & \xi _3 & \xi _2 & \alpha _4 & \alpha _6 & \alpha _5
   & \alpha _3 & \alpha _1 \\
 \alpha _3 & \alpha _3 & \xi _3 & \xi _5 & \mathrm{Id} & \xi
   _2 & \xi _1 & \xi _4 & \alpha _5 & \alpha _4 & \alpha _1
   & \alpha _6 & \alpha _2 \\
 \alpha _4 & \alpha _4 & \xi _2 & \xi _1 & \xi _3 &
   \mathrm{Id} & \xi _4 & \xi _5 & \alpha _2 & \alpha _1 &
   \alpha _3 & \alpha _5 & \alpha _6 \\
 \alpha _5 & \alpha _5 & \xi _4 & \xi _2 & \xi _1 & \xi _5 &
   \mathrm{Id} & \xi _3 & \alpha _3 & \alpha _2 & \alpha _6 &
   \alpha _1 & \alpha _4 \\
 \alpha _6 & \alpha _6 & \xi _1 & \xi _3 & \xi _5 & \xi _4 &
   \xi _2 & \mathrm{Id} & \alpha _1 & \alpha _5 & \alpha _2 &
   \alpha _4 & \alpha _3 \\
 \xi _1 & \xi _1 & \alpha _6 & \alpha _4 & \alpha _5 &
   \alpha _2 & \alpha _3 & \alpha _1 & \mathrm{Id} & \xi _5 &
   \xi _4 & \xi _3 & \xi _2 \\
 \xi _2 & \xi _2 & \alpha _4 & \alpha _5 & \alpha _1 &
   \alpha _3 & \alpha _6 & \alpha _2 & \xi _5 & \xi _3 &
   \mathrm{Id} & \xi _1 & \xi _4 \\
 \xi _3 & \xi _3 & \alpha _3 & \alpha _6 & \alpha _4 &
   \alpha _1 & \alpha _2 & \alpha _5 & \xi _4 & \mathrm{Id} &
   \xi _2 & \xi _5 & \xi _1 \\
 \xi _4 & \xi _4 & \alpha _5 & \alpha _1 & \alpha _2 &
   \alpha _6 & \alpha _4 & \alpha _3 & \xi _3 & \xi _1 & \xi
   _5 & \xi _2 & \mathrm{Id} \\
 \xi _5 & \xi _5 & \alpha _2 & \alpha _3 & \alpha _6 &
   \alpha _5 & \alpha _1 & \alpha _4 & \xi _2 & \xi _4 & \xi
   _1 & \mathrm{Id} & \xi _3\\
   \hline
\end{array}
\end{equation}
Next let us discuss the structure of the generalized Weyl group. In this case $\mathcal{GW}$ is composed by $48$ elements and its stability subgroup $\mathcal{HW} \, \sim \, \mathbb{Z}_2 \times \, \mathbb{Z}_2$ is made by the following four $7 \times 7$ matrices belonging to the $\mathrm{G_{(2,2)}}$ group:
{\scriptsize
\begin{equation}\label{HWg2}
    \begin{array}{ccccccc}
       hw_1 & = & \left(
\begin{array}{lllllll}
 -1 & 0 & 0 & 0 & 0 & 0 & 0 \\
 0 & 1 & 0 & 0 & 0 & 0 & 0 \\
 0 & 0 & -1 & 0 & 0 & 0 & 0 \\
 0 & 0 & 0 & 1 & 0 & 0 & 0 \\
 0 & 0 & 0 & 0 & -1 & 0 & 0 \\
 0 & 0 & 0 & 0 & 0 & 1 & 0 \\
 0 & 0 & 0 & 0 & 0 & 0 & -1
\end{array}
\right)& ; & hw_2 & = & \left(
\begin{array}{lllllll}
 0 & 0 & 0 & 0 & 0 & 0 & -1 \\
 0 & 0 & 0 & 0 & 0 & -1 & 0 \\
 0 & 0 & 0 & 0 & -1 & 0 & 0 \\
 0 & 0 & 0 & -1 & 0 & 0 & 0 \\
 0 & 0 & -1 & 0 & 0 & 0 & 0 \\
 0 & -1 & 0 & 0 & 0 & 0 & 0 \\
 -1 & 0 & 0 & 0 & 0 & 0 & 0
\end{array}
\right) \\
       hw_3 & = &\left(
\begin{array}{lllllll}
 0 & 0 & 0 & 0 & 0 & 0 & 1 \\
 0 & 0 & 0 & 0 & 0 & -1 & 0 \\
 0 & 0 & 0 & 0 & 1 & 0 & 0 \\
 0 & 0 & 0 & -1 & 0 & 0 & 0 \\
 0 & 0 & 1 & 0 & 0 & 0 & 0 \\
 0 & -1 & 0 & 0 & 0 & 0 & 0 \\
 1 & 0 & 0 & 0 & 0 & 0 & 0
\end{array}
\right) & ; & \mathrm{Id} & = &\left(
\begin{array}{lllllll}
 1 & 0 & 0 & 0 & 0 & 0 & 0 \\
 0 & 1 & 0 & 0 & 0 & 0 & 0 \\
 0 & 0 & 1 & 0 & 0 & 0 & 0 \\
 0 & 0 & 0 & 1 & 0 & 0 & 0 \\
 0 & 0 & 0 & 0 & 1 & 0 & 0 \\
 0 & 0 & 0 & 0 & 0 & 1 & 0 \\
 0 & 0 & 0 & 0 & 0 & 0 & 1
\end{array}
\right)
     \end{array}
\end{equation}}
In order to complete the description of the generalized Weyl group it is now sufficient to write one representative for each equivalence class of the quotient:
\begin{equation}\label{fronda}
    \frac{\mathcal{GW}}{\mathcal{HW}} \, \simeq \, \mathcal{W}
\end{equation}
We have:
{\tiny
\begin{equation}\label{prime2}
    \begin{array}{rclcrcl}
       \alpha_1& \sim & \left(
\begin{array}{lllllll}
 0 & 0 & -1 & 0 & 0 & 0 & 0 \\
 0 & 0 & 0 & 0 & 0 & 1 & 0 \\
 1 & 0 & 0 & 0 & 0 & 0 & 0 \\
 0 & 0 & 0 & -1 & 0 & 0 & 0 \\
 0 & 0 & 0 & 0 & 0 & 0 & -1 \\
 0 & 1 & 0 & 0 & 0 & 0 & 0 \\
 0 & 0 & 0 & 0 & 1 & 0 & 0
\end{array}
\right) & ; & \alpha_2 & \sim &
\left(
\begin{array}{lllllll}
 -\frac{1}{2} & 0 & 0 & -\frac{1}{\sqrt{2}} & 0 & 0 &
   \frac{1}{2} \\
 0 & \frac{1}{2} & -\frac{1}{2} & 0 & -\frac{1}{2} &
   -\frac{1}{2} & 0 \\
 0 & -\frac{1}{2} & -\frac{1}{2} & 0 & \frac{1}{2} &
   -\frac{1}{2} & 0 \\
 -\frac{1}{\sqrt{2}} & 0 & 0 & 0 & 0 & 0 &
   -\frac{1}{\sqrt{2}} \\
 0 & -\frac{1}{2} & \frac{1}{2} & 0 & -\frac{1}{2} &
   -\frac{1}{2} & 0 \\
 0 & -\frac{1}{2} & -\frac{1}{2} & 0 & -\frac{1}{2} &
   \frac{1}{2} & 0 \\
 \frac{1}{2} & 0 & 0 & -\frac{1}{\sqrt{2}} & 0 & 0 &
   -\frac{1}{2}
\end{array}
\right) \\
       \alpha_3 & \sim & \left(
\begin{array}{lllllll}
 -\frac{1}{2} & -\frac{1}{2} & 0 & 0 & 0 & -\frac{1}{2} &
   -\frac{1}{2} \\
 \frac{1}{2} & -\frac{1}{2} & 0 & 0 & 0 & \frac{1}{2} &
   -\frac{1}{2} \\
 0 & 0 & \frac{1}{2} & \frac{1}{\sqrt{2}} & \frac{1}{2} & 0
   & 0 \\
 0 & 0 & -\frac{1}{\sqrt{2}} & 0 & \frac{1}{\sqrt{2}} & 0 &
   0 \\
 0 & 0 & \frac{1}{2} & -\frac{1}{\sqrt{2}} & \frac{1}{2} & 0
   & 0 \\
 \frac{1}{2} & \frac{1}{2} & 0 & 0 & 0 & -\frac{1}{2} &
   -\frac{1}{2} \\
 -\frac{1}{2} & \frac{1}{2} & 0 & 0 & 0 & \frac{1}{2} &
   -\frac{1}{2}
\end{array}
\right)& ; & \alpha_4& \sim & \left(
\begin{array}{lllllll}
 -\frac{1}{2} & 0 & 0 & -\frac{1}{\sqrt{2}} & 0 & 0 &
   \frac{1}{2} \\
 0 & -\frac{1}{2} & \frac{1}{2} & 0 & \frac{1}{2} &
   \frac{1}{2} & 0 \\
 0 & \frac{1}{2} & \frac{1}{2} & 0 & -\frac{1}{2} &
   \frac{1}{2} & 0 \\
 -\frac{1}{\sqrt{2}} & 0 & 0 & 0 & 0 & 0 &
   -\frac{1}{\sqrt{2}} \\
 0 & \frac{1}{2} & -\frac{1}{2} & 0 & \frac{1}{2} &
   \frac{1}{2} & 0 \\
 0 & \frac{1}{2} & \frac{1}{2} & 0 & \frac{1}{2} &
   -\frac{1}{2} & 0 \\
 \frac{1}{2} & 0 & 0 & -\frac{1}{\sqrt{2}} & 0 & 0 &
   -\frac{1}{2}
\end{array}
\right) \\
       \alpha_5 & \sim & \left(
\begin{array}{lllllll}
 -\frac{1}{2} & -\frac{1}{2} & 0 & 0 & 0 & -\frac{1}{2} &
   -\frac{1}{2} \\
 -\frac{1}{2} & \frac{1}{2} & 0 & 0 & 0 & -\frac{1}{2} &
   \frac{1}{2} \\
 0 & 0 & -\frac{1}{2} & -\frac{1}{\sqrt{2}} & -\frac{1}{2} &
   0 & 0 \\
 0 & 0 & -\frac{1}{\sqrt{2}} & 0 & \frac{1}{\sqrt{2}} & 0 &
   0 \\
 0 & 0 & -\frac{1}{2} & \frac{1}{\sqrt{2}} & -\frac{1}{2} &
   0 & 0 \\
 -\frac{1}{2} & -\frac{1}{2} & 0 & 0 & 0 & \frac{1}{2} &
   \frac{1}{2} \\
 -\frac{1}{2} & \frac{1}{2} & 0 & 0 & 0 & \frac{1}{2} &
   -\frac{1}{2}
\end{array}
\right) & ; & \alpha_6 & = & \left(
\begin{array}{lllllll}
 0 & 0 & -1 & 0 & 0 & 0 & 0 \\
 0 & 0 & 0 & 0 & 0 & -1 & 0 \\
 -1 & 0 & 0 & 0 & 0 & 0 & 0 \\
 0 & 0 & 0 & -1 & 0 & 0 & 0 \\
 0 & 0 & 0 & 0 & 0 & 0 & 1 \\
 0 & -1 & 0 & 0 & 0 & 0 & 0 \\
 0 & 0 & 0 & 0 & 1 & 0 & 0
\end{array}
\right)\\
     \end{array}
\end{equation}
}
and
{\tiny
\begin{equation}\label{xitipo}
    \begin{array}{ccccccc}
       \xi_1 & \sim & \left(
\begin{array}{lllllll}
 -1 & 0 & 0 & 0 & 0 & 0 & 0 \\
 0 & -1 & 0 & 0 & 0 & 0 & 0 \\
 0 & 0 & 1 & 0 & 0 & 0 & 0 \\
 0 & 0 & 0 & 1 & 0 & 0 & 0 \\
 0 & 0 & 0 & 0 & 1 & 0 & 0 \\
 0 & 0 & 0 & 0 & 0 & -1 & 0 \\
 0 & 0 & 0 & 0 & 0 & 0 & -1
\end{array}
\right)& ; & \xi_2 & \sim & \left(
\begin{array}{lllllll}
 0 & 0 & -\frac{1}{2} & -\frac{1}{\sqrt{2}} & -\frac{1}{2} &
   0 & 0 \\
 \frac{1}{2} & \frac{1}{2} & 0 & 0 & 0 & -\frac{1}{2} &
   -\frac{1}{2} \\
 -\frac{1}{2} & -\frac{1}{2} & 0 & 0 & 0 & -\frac{1}{2} &
   -\frac{1}{2} \\
 0 & 0 & \frac{1}{\sqrt{2}} & 0 & -\frac{1}{\sqrt{2}} & 0 &
   0 \\
 \frac{1}{2} & -\frac{1}{2} & 0 & 0 & 0 & -\frac{1}{2} &
   \frac{1}{2} \\
 \frac{1}{2} & -\frac{1}{2} & 0 & 0 & 0 & \frac{1}{2} &
   -\frac{1}{2} \\
 0 & 0 & \frac{1}{2} & -\frac{1}{\sqrt{2}} & \frac{1}{2} & 0
   & 0
\end{array}
\right)\\
       \xi_3 & \sim & \left(
\begin{array}{lllllll}
 0 & -\frac{1}{2} & -\frac{1}{2} & 0 & \frac{1}{2} &
   -\frac{1}{2} & 0 \\
 0 & \frac{1}{2} & \frac{1}{2} & 0 & \frac{1}{2} &
   -\frac{1}{2} & 0 \\
 -\frac{1}{2} & 0 & 0 & -\frac{1}{\sqrt{2}} & 0 & 0 &
   \frac{1}{2} \\
 \frac{1}{\sqrt{2}} & 0 & 0 & 0 & 0 & 0 & \frac{1}{\sqrt{2}}
   \\
 -\frac{1}{2} & 0 & 0 & \frac{1}{\sqrt{2}} & 0 & 0 &
   \frac{1}{2} \\
 0 & -\frac{1}{2} & \frac{1}{2} & 0 & \frac{1}{2} &
   \frac{1}{2} & 0 \\
 0 & \frac{1}{2} & -\frac{1}{2} & 0 & \frac{1}{2} &
   \frac{1}{2} & 0
\end{array}
\right)& ; & \xi_4 & \sim & \left(
\begin{array}{lllllll}
 0 & -\frac{1}{2} & -\frac{1}{2} & 0 & \frac{1}{2} &
   -\frac{1}{2} & 0 \\
 0 & -\frac{1}{2} & -\frac{1}{2} & 0 & -\frac{1}{2} &
   \frac{1}{2} & 0 \\
 \frac{1}{2} & 0 & 0 & \frac{1}{\sqrt{2}} & 0 & 0 &
   -\frac{1}{2} \\
 \frac{1}{\sqrt{2}} & 0 & 0 & 0 & 0 & 0 & \frac{1}{\sqrt{2}}
   \\
 \frac{1}{2} & 0 & 0 & -\frac{1}{\sqrt{2}} & 0 & 0 &
   -\frac{1}{2} \\
 0 & \frac{1}{2} & -\frac{1}{2} & 0 & -\frac{1}{2} &
   -\frac{1}{2} & 0 \\
 0 & \frac{1}{2} & -\frac{1}{2} & 0 & \frac{1}{2} &
   \frac{1}{2} & 0
\end{array}
\right) \\
       \xi_5 & \sim & \left(
\begin{array}{lllllll}
 0 & 0 & -\frac{1}{2} & -\frac{1}{\sqrt{2}} & -\frac{1}{2} &
   0 & 0 \\
 -\frac{1}{2} & -\frac{1}{2} & 0 & 0 & 0 & \frac{1}{2} &
   \frac{1}{2} \\
 \frac{1}{2} & \frac{1}{2} & 0 & 0 & 0 & \frac{1}{2} &
   \frac{1}{2} \\
 0 & 0 & \frac{1}{\sqrt{2}} & 0 & -\frac{1}{\sqrt{2}} & 0 &
   0 \\
 -\frac{1}{2} & \frac{1}{2} & 0 & 0 & 0 & \frac{1}{2} &
   -\frac{1}{2} \\
 -\frac{1}{2} & \frac{1}{2} & 0 & 0 & 0 & -\frac{1}{2} &
   \frac{1}{2} \\
 0 & 0 & \frac{1}{2} & -\frac{1}{\sqrt{2}} & \frac{1}{2} & 0
   & 0
\end{array}
\right) & ; & \null & \null & \null
     \end{array}
\end{equation}
}
We can explicitly verify that all the elements of the $\mathcal{HW}$ subgroup are in $\mathrm{H}^\star = \su(1,1) \times \su(1,1)$ since they satisfy the condition:
\begin{equation}\label{condoeta}
    hw_i ^T\, \eta \, hw_i \, = \, \eta
\end{equation}
where
\begin{equation}\label{etatensore}
  \eta \, = \, \left(
\begin{array}{lllllll}
 -1 & 0 & 0 & 0 & 0 & 0 & 0 \\
 0 & -1 & 0 & 0 & 0 & 0 & 0 \\
 0 & 0 & 1 & 0 & 0 & 0 & 0 \\
 0 & 0 & 0 & 1 & 0 & 0 & 0 \\
 0 & 0 & 0 & 0 & 1 & 0 & 0 \\
 0 & 0 & 0 & 0 & 0 & -1 & 0 \\
 0 & 0 & 0 & 0 & 0 & 0 & -1
\end{array}
\right)
\end{equation}
is the invariant metric which defines the $\mathrm{H}^\star$ subgroup. Note that  here we use all the conventions and the definitions introduced in our previous paper \cite{noig22}.
\par
\begin{figure}[!hbt]
\begin{center}
\iffigs
 \includegraphics[height=60mm]{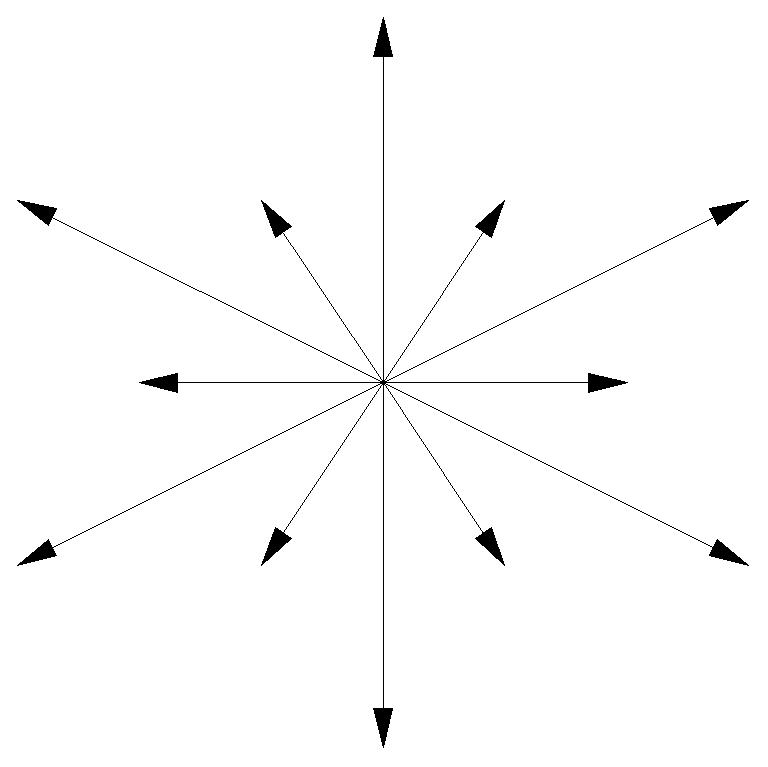}
\else
\end{center}
 \fi
\caption{\it
The $\mathfrak{g}_2$ root system $\Delta_{g2}$  is made of six positive roots and of their negatives}
\label{g2rutsys}
 \iffigs
 \hskip 1cm \unitlength=1.1mm
 \end{center}
  \fi
\end{figure}

The next required ingredient of our construction is the subgroup $\mathcal{W}_H$. As we showed in paper \cite{noig22},
when we diagonalize the adjoint action of a Cartan Subalgebra contained in the $\mathbb{H}^\star$ subalgebra, the root system of the $\mathfrak{g}_2$ Lie algebra (see fig.\ref{g2rutsys}), decomposes in two subsystems $\Delta_H$ and $\Delta_K$ such that the step operators corresponding to roots in $\Delta_H$ belong to $\mathbb{H}^\star$ while the step operators corresponding to roots in $\Delta_K$ belong to $\mathbb{K}$.
\begin{figure}[!hbt]
\begin{center}
\iffigs
 \includegraphics[height=60mm]{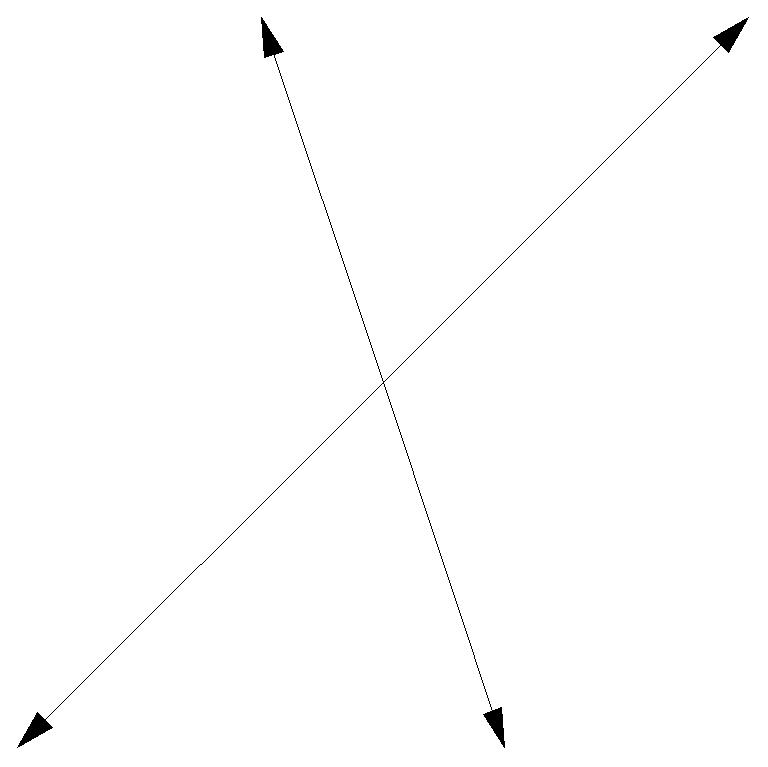}
 \includegraphics[height=60mm]{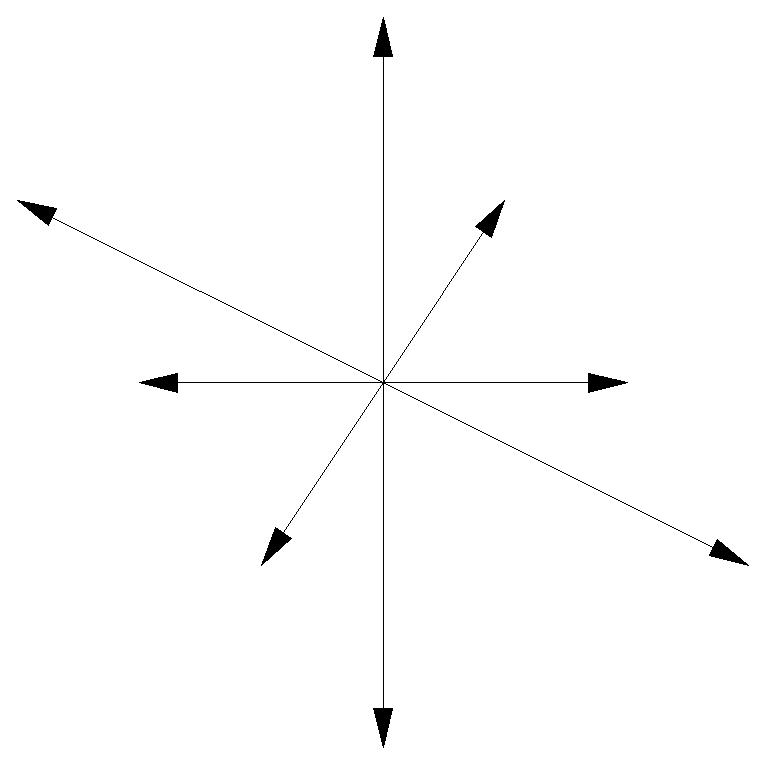}
\else
\end{center}
 \fi
\caption{\it
The root system $\Delta_{g2}$ splits in two subsystems, the system $\Delta_H$ on the left, the system $\Delta_K$ on the right}
\label{g2rutsys2}
 \iffigs
 \hskip 1cm \unitlength=1.1mm
 \end{center}
  \fi
\end{figure}
The subsystem $\Delta_H$ is composed by the roots $\pm \alpha_3 , \pm \alpha_5$, while $\Delta_K$ is made by the remaining ones. The subgroup $\mathcal{W}_H \subset \mathcal{W}$ can be easily derived. It is made by all those elements of the Weyl group which map $\Delta_H$ into itself and $\Delta_K$ into itself, as well.
Referring to the previously introduced notation, we easily see that:
\begin{equation}\label{whsubbo}
    \mathcal{W}_H \, = \, \left\{ \mbox{Id},\alpha_3 , \alpha_5 , \xi_1 \right \}
\end{equation}
Abstractly the structure of $\mathcal{W}_H$ is the following:
\begin{equation}\label{abstrawh}
    \mathcal{W}_H \, \sim \, \mathbb{Z}_2 \, \times \, \mathbb{Z}_2
\end{equation}
since all of its elements square to the identity.
\par
There are three lateral classes in $\mathcal{W}/\mathcal{W}_H$, respectively associated with the identity element and with the reflection along the two simple roots.
\begin{eqnarray}
  \left[\mbox{Id}\right ]&=&  \left \{\mbox{Id} , \alpha _3 , \alpha _5 , \xi _1 \right\} \label{latclas1}\\
 \left[\alpha_1\right ]&=& \left \{\alpha _1 , \alpha _6 , \xi _3 , \xi _4 \right \} \label{latclas2}\\
\left[\alpha_2\right ]&=&  \left \{\alpha _2 , \alpha _4 , \xi _2 , \xi _5 \right\}\label{latclas3}
\end{eqnarray}
It follows that for each partition $\left\{ j_1 , j_2 , \dots j_n\right \}$  ($\alpha$-label) there are three possible $\gamma$-labels and three possible $\beta$-labels. It remains to be seen for which combinations of these $\gamma$ and $\beta$-labels there exist an $x$-operator purely contained in $\mathbb{K}$ which completes the standard triple.
\subsection{The table of $\frac{\mathrm{G_{(2,2)}}}{\mathrm{SU(1,1)}\times \mathrm{SU(1,1)}}$ nilpotent orbits}
In order to derive the desired table of nilpotent orbits we begin from the first step namely from partitions or, said differently, from $\alpha$-labels.
\subsubsection{$\alpha$-labels}
Taking into account the restriction (see \cite{nilorbits}) that every half-integer spin $j$ should appear an even number of times  we easily conclude that the possible branching laws of the $7$-dimensional fundamental representation of $g_{(2,2)}$ into irreducible representations of $\slal(2)$ are the following ones:
\begin{eqnarray}
  \alpha_1-\mbox{label} &=& \mbox{[j=3]} \label{alab1}\\
  \alpha_2-\mbox{label} &=& \mbox{[j=1]$\times $2[j=1/2]} \label{alab2}\\
  \alpha_3-\mbox{label} &=& \mbox{2[j=1]$\times $[j=0]} \label{alab3}\\
  \alpha_4-\mbox{label} &=& \mbox{2[j=1/2]$\times $3 [j=0]} \label{alab4}
\end{eqnarray}
\subsubsection{$\gamma$-labels}
Analysing the two equations (\ref{pagnato1},\ref{pagnato2}) for the $x$-triple element at fixed $h$ we find the following result:
\begin{description}
  \item[$\alpha_1$] In this sector there are $x$ operators in $\mathbb{K}$ only for the second lateral class (\ref{latclas2}). This means that there is only one $\gamma$-label which has the following form:
      \begin{equation}\label{ga1}
        \gamma_1 \, = \, \left\{\pm 8, \pm 4, 0 , 0\right \} \, \equiv  \left\{8_1,4_1,0_1\right\}
      \end{equation}
    The notation introduced in equation (\ref{ga1}) is based on the following observation. The dimension of $\mathbb{H}$ or $\mathbb{H}^\star$ is six and every eigenvalue appears together with its negative. Hence it suffices to mention the non-negative eigenvalues (including the zero) with their multiplicity (all zeros appear in pairs as well). It follows that  the $\beta$-label is also unique so that in this sector there is only one nilpotent orbit.
  \item[$\alpha_2$] For this partition the $\mathcal{W}_H$ orbits (\ref{latclas1}) and (\ref{latclas2}) coincide: within them we find $x$ operators in $\mathbb{K}$. In the third $\mathcal{W}_H$ orbit there are no solutions for $x$ in $\mathbb{K}$. So we have only one $\gamma$-label:
       \begin{equation}\label{ga1p}
        \gamma_1 \, = \, \left\{3_1,1_1,0_1\right\}
      \end{equation}
      and consequently only one nilpotent orbit.
  \item[$\alpha_3$] For this partition the $\mathcal{W}_H$ orbits (\ref{latclas2}) and (\ref{latclas3}) coincide while the first is distinct. We find solutions for $x$ in $\mathbb{K}$ both for the first $\mathcal{W}_H$-orbit  (\ref{latclas1}) and for the coinciding subsequent two. That means that we have two $\gamma$-labels
      \begin{eqnarray}
        \gamma_1 & = &   \left\{4_1,0_2\right\} \label{g1lab}\\
        \gamma_2 & = &  \left\{2_2,0_1\right\} \label{g2lab}\\
      \end{eqnarray}
      Considering the solutions for $x$ both in the case of $\gamma_1$ and $\gamma_2$ they group in two non empty classes corresponding to $\beta$-labels $\beta_1$ and $\beta_2$. This means that we have a total of $4$ nilpotent orbits from this sector.
  \item[$\alpha_4$] For this partition the situation is similar to that of partition one and two. There are no $\mathbb{K}$ solutions for $x$ in the first $\mathcal{W}_H$ orbit while there are such solutions in the second and third $\mathcal{W}_H$-orbits, which coincide. Hence there is only one $\gamma$-label:
      \begin{equation}\label{ga1pU}
        \gamma_1 \, = \,    \left\{1_2,0_1\right\}
      \end{equation}
      and one nilpotent orbit.
\end{description}
In table \ref{tablettina} the results we have described are summarized.
\begin{table}
\begin{center}
{\scriptsize
$$
\begin{array}{||l||c|l|c|c|l||}
\hline
\hline
N & d_n &\alpha -\mbox{label} & \gamma\beta -\mbox{labels} & \mbox{Orbits} & \mathcal{W}_H -\mbox{classes}\\
\hline
\hline
1 & 7 &\mbox{[j=3]} &\gamma\beta_1 = \left\{8_1 4_1 0_1\right\} & \mathcal{O}_1^1& \left(\times,\gamma_1,\times\right)\\
 \hline
2 & 3 &\mbox{[j=1]$\times $2[j=1/2]} & \gamma\beta_1 = \left\{3_1 1_1 0_1 \right\}&\mathcal{O}_1^2 & \left(\gamma_1,\gamma_1,\times\right)\\
   \hline
7 & 3 &\mbox{2[j=1]$\times $[j=0]} & \begin{array}{lll}
                                    \gamma\beta_1& = & \left\{4_1 0_2 \right\} \\
                                    \gamma\beta_2 & =&\left\{ 2_2 0_1\right\} \\
                                  \end{array}&\begin{array}{l|cc}
                                                 & \beta_1 & \beta_2 \\
                                                \hline
                                                \gamma_1 &\mathcal{ O}^3_{1,1}& \mathcal{ O}^3_{1,2}  \\
                                                \gamma_2 & \mathcal{ O}^3_{2,1}& \mathcal{ O}^3_{2,2} \\
                                              \end{array}
                                  & \left(\gamma_1,\gamma_2,\gamma_2\right)\\
   \hline
4 &2 &\mbox{2[j=1/2]$\times $3 [j=0]} & \gamma\beta_1=\left\{1_2 0_1\right\}
  &\mathcal{O}_{1}^4 & \left(0,\gamma_1,\gamma_1\right)\\
  \hline
   \hline
\end{array}
$$
}
\caption{Classification of nilpotent orbits of $\frac{\mathrm{G_{(2,2)}}}{\mathrm{SU(1,1)}\times \mathrm{SU(1,1)}}$. \label{tablettina}}
\end{center}
\end{table}
\subsection{Comparison with the Tensor Classifiers}
In order to make contact with our previous results \cite{noig22} we considered the Tensor Classifiers  introduced in that paper and we calculated them on the representatives  found by means of the the Weyl group method. The result is displayed in table \ref{tablinag2} and shows that with the new method we exactly reproduce the same classification obtained there. In particular the splitting of the BPS and non BPS regular orbits in two sub-orbits according with the sign of the non-vanishing eigenvalues of the tensor classifiers is  justified in terms of  $\beta$-$\gamma$  labels.
\begin{table}
\begin{center}
{\small
\begin{tabular}{||c||c|c|c|c|c|c||}
\hline\hline
Orbit & Order  & Stab. & Sign.& Sign.&Sign.& Bivect\\
\null & Nilp. & subg. &$\mathcal{T}^{xy}$ & ${\mathfrak{T}^{xy}}$& ${\mathbb{T}^{ab}}$&$W^{a|x}$ \\
\hline
\hline
$\mathcal{O}^4_1$&$2$&$\mathrm{O(1,1)}\ltimes \mathbb{R}^2 $&$\{0,0,0\}$&$\{0,0,0\}$&$\{0,0,0\}$&$0$\\
\hline
$\mathcal{O}^2_1$ &$3$&$\mathrm{O(1,1)}\ltimes \mathbb{R} $&$\{0,0,0\}$&$\{0,0,0\}$&$\{0,0,0\}$&$\ne 0$\\
\hline
$\mathcal{O}^3_{2,2}$&$3$&$\mathbb{R} $&$\{0,0,+\}$&$\{0,0,0\}$&$\{0,0,0\}$&$\ne 0$\\
\hline
$\mathcal{O}^3_{2,1}$&$3$&$\mathbb{R} $&$\{0,0,-\}$&$\{0,0,0\}$&$\{0,0,0\}$&$\ne 0$\\
\hline
$\mathcal{O}^3_{1,1}$ &$3$&$\mathbb{R}$&$\{0,0,0\}$&$\{0,0,0\}$&$\{0,0,+\}$&$\ne 0$\\
\hline
$\mathcal{O}^3_{1,2}$ &$3$&$\mathbb{R}$&$\{0,0,0\}$&$\{0,0,0\}$&$\{0,0,-\}$&$\ne 0$\\
\hline
$\mathcal{O}^1_1$&$7$&$0$&$\{0,+,-\}$&$\{0,0,-\}$&$\{0,+,-\}$&$\ne 0$\\
\hline
\end{tabular}
}
\end{center}
\caption{Evaluation of the Tensor Classifiers on the nilpotent orbit representatives of $\frac{\mathrm{G_{(2,2)}}}{\mathrm{SU(1,1)}\times \mathrm{SU(1,1)}}$.\label{tablinag2}}
\end{table}
An important observation emerging from this exercise concerns the degree of nilpotency. It appears that:
\begin{equation}\label{banfo}
    d_n \, = \, 2 \, j_{max} \, + \, 1
\end{equation}
where $j_{max}$ is the highest spin appearing in the branching rule. Hence the regular and small black-hole solutions which require a degree of nilpotency equal to $3$ are associated with partitions where $j_{max} \, = \, 3$.
\par
Another observation which will be confirmed by our analysis of the $\so(4,4+2s)$ case is that both the BPS and non-BPS regular solutions arise from the same partition, namely from:
\begin{equation}\label{specialsolut}
   2 \mbox{[j=1]$\times (N-6)$ [j=1/2]}
\end{equation}
The BPS solutions correspond to one type of $\gamma$-labels while the non-BPS ones correspond to a second type.
\par
This fact is clearly inspiring  and might provide some new insight in the problem of fake superpotentials.
\section{Algebraic structure of the $\mathcal{SKO}_{2s+2}\Rightarrow \mathcal{QM}^\star_{(4,4+2s)}$ models}
Next we proceed to classify  extremal spherical black hole solutions in those supergravity models that are based on the  special geometry series (\ref{skgseries}).
According to the $D=3$ scheme, this problem is turned into that of classifying the $\mathrm{H}^\star$-orbits of nilpotent Lax operators for the coset manifolds (\ref{p2qmanstarro}). This requires an in depth analysis of  the $\so(4,4+2s)$ algebra and of its subalgebras.
\subsection{The  $\so(4,4+2s)$ algebra and its $\mathbb{H}$-subalgebra}
\label{sorr2salgebra}
The complex Lie algebra of which $\so(4,4+2s)$ is a non-compact real
section is just $D_\ell$ where
\begin{equation}
  \ell = 4 + s~.
\label{elldefi}
\end{equation}
The corresponding Dynkin
diagram is displayed in fig.\ref{Dell} and the associated root system
 is realized by the following set of vectors in
$\mathbb{R}^\ell$:
\begin{equation}
  \Delta \equiv \left\{ \pm \,\epsilon ^A \, \pm \epsilon ^B \right\} \quad ;
  \quad \mbox{card} \, \Delta \, = \, 2\left( \ell^2 \, - \, \ell \right)
\label{roots}
\end{equation}
where $\epsilon ^A$ denotes an orthonormal basis of unit
vectors. The set of positive roots is then easily defined as follows:
\begin{equation}
  \hat{\alpha} \, > \, 0 \quad \Rightarrow \, \hat{\alpha} \, \in \, \Delta_+ \,
  \equiv \, \left\{ \epsilon ^A \, \pm \epsilon ^B \right\} \quad ( A < B )~.
\label{Deltapiu}
\end{equation}
A standard basis of simple roots
representing the Dynkin diagram \ref{Dell} is given by
\begin{eqnarray}
  \hat{\alpha} _1 & = & \epsilon _1 \, - \, \epsilon _2~, \nonumber\\
  \hat{\alpha} _2 & = & \epsilon _2 \, - \, \epsilon _3~, \nonumber\\
  \dots & \dots & \dots~, \nonumber\\
   \hat{\alpha} _{\ell-1} & = & \epsilon _{\ell -1} \, - \, \epsilon _\ell~, \nonumber\\
   \hat{\alpha} _{\ell} & = & \epsilon _{\ell -1} \, + \, \epsilon _\ell~.
\label{simplerute}
\end{eqnarray}
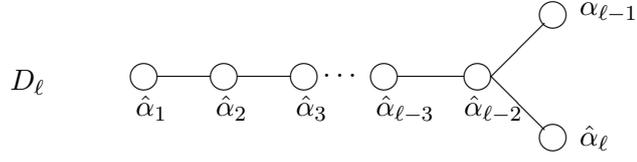
\begin{figure}
\caption{\it The Dynkin diagram of the  $D_\ell$ Lie algebra.
\label{Dell}}
\centering
\begin{picture}(60,130)
\put (-70,65){$D_\ell$}
\put (-20,70){\circle {10}}
\put (-23,55){$\hat{\alpha}_1$}
\put (-15,70){\line (1,0){20}}
\put (10,70){\circle {10}}
\put (7,55){$\hat{\alpha}_2$}
\put (15,70){\line (1,0){20}}
\put (40,70){\circle {10}}
\put (37,55){$\hat{\alpha}_3$}
\put (47,70){$\dots$}
\put (70,70){\circle {10}}
\put (67,55){$\hat{\alpha}_{\ell-3}$}
\put (75,70){\line (1,0){25}}
\put (105,70){\circle {10}}
\put (100,55){$\hat{\alpha}_{\ell-2}$}
\put (110,70){\line (1,1){20}}
\put (110,70){\line (1,-1){20}}
\put (133.2,93.2){\circle {10}}
\put (133.2,46.8){\circle {10}}
\put (143.2,93.2){$\hat{\alpha}_{\ell-1}$}
\put (143.2,43.8){$\hat{\alpha}_\ell$}
\end{picture}
\end{figure}
The maximally split real form of the $D_\ell$ Lie algebra is
$\so(\ell,\ell)$ and it is explicitly realized by the following
$2\ell \times 2\ell$ matrices. Let $e_{A,B}$ denote the $2\ell \times 2\ell$
matrix whose entries are all zero except the entry $A,B$ which is
equal to one. Then the Cartan generators $\mathcal{H}_A$ and the
positive root step operators $E^\alpha$ are represented as follows:
\begin{eqnarray}
\mathcal{H}_A & = & e_{A,A} \, - \, e_{A+\ell,A+\ell}~, \nonumber\\
E^{\epsilon _A \, - \, \epsilon _B} & = & e_{B,A} \, - \,
e_{A+\ell,B+\ell}~, \nonumber\\
E^{\epsilon _A \, + \, \epsilon _B} & = & e_{A+\ell,B} \, - \,
e_{B+\ell,A}~.
\label{trogut}
\end{eqnarray}
The solvable algebra of the maximally split coset
\begin{equation}
  \mathcal{M}_{(\ell,0)}\, = \, \frac{\mathrm{SO(\ell,\ell)}}{\mathrm{SO(\ell)} \times \mathrm{SO(\ell)}}
\label{ell2man}
\end{equation}
has therefore a very simple form in terms of matrices. Following
general constructive principles $Solv_{(\ell,\ell)}$ is just the algebraic
span of all the matrices (\ref{trogut}) so that
\begin{equation}
  Solv_{(\ell,\ell)} \, \ni \, M \, \Leftrightarrow \, M \, = \, \left(
  \begin{array}{c|c}
    T & B \\
    \hline
    0 & - T^T \
  \end{array} \right) \quad ; \quad \left \{\begin{array}{rcl}
  T & = & \mbox{upper triangular}~, \\
  B & = & - \, B^T \quad \mbox{antisymmetric.}
\end{array} \right.
\label{solvellell}
\end{equation}
The matrices of the form (\ref{solvellell}) clearly close a subalgebra
of the  $\so(\ell,\ell)$ algebra which, in this representation, is
defined as the set of matrices $\Lambda$ fulfilling the following
condition:
\begin{equation}
  \Lambda^T \, \left(
  \begin{array}{c|c}
    0 & \mathbf{1}_{l} \\
    \hline
    \mathbf{1}_{l} & 0 \
  \end{array} \right) \, + \, \left(
  \begin{array}{c|c}
    0 & \mathbf{1}_{l} \\
    \hline
    \mathbf{1}_{l} & 0 \
  \end{array} \right)\Lambda \, = \, 0~.
\label{fertilina}
\end{equation}
\subsection{The real form $\so(4,4+2s)$ of the $D_{4+s}$ Lie algebra and the $\mathbb{H}$ subalgebra}
The main point in order to apply the general Lax approach to the coset manifolds (\ref{p2qman}) or (\ref{p2qmanstarro} ) consists of introducing a
convenient basis of generators of the Lie algebra $\so(4,4+2s)$
where, in the fundamental representation, all elements of the
solvable Lie algebra associated with the coset under study turn out
to be given by upper triangular matrices. With some ingenuity such a
basis can be found by defining the $\so(4,4+2s)$ Lie algebra as the set
of matrices $\Lambda_t$ satisfying the following constraint:
\begin{equation}
  \Lambda_t^T \, \eta_t \, + \, \eta_t \, \Lambda_t \, = \, 0
\label{Lambdatdefi}
\end{equation}
where the symmetric invariant metric $\eta_t$ with $4+2s$ positive
eigenvalues $(+1)$
and $4$ negative ones $(-1)$ is given by the following matrix.
\begin{equation}
  \eta_t \, = \, \left(\begin{array}{c|c|c}
    0 & 0 & \varpi_4\\
    \hline
    0 & \mathbf{1}_{2s} & 0 \\
    \hline
    \varpi_4 & 0 & 0 \
  \end{array} \right)~.
\label{etaT}
\end{equation}
In the above equation the symbol $\varpi_4$ denotes the completely
anti-diagonal $4 \times 4$ matrix which follows:
\begin{equation}
  \varpi_4 \, = \, \left(
  \begin{array}{cccc}
    0 & 0 &  0 & 1 \\
    0 & 0 &  1 & 0 \\
    0 & 1 & 0 &  0 \\
    1 & 0 & 0 &0 \
  \end{array} \right)
\label{varpimatra}
\end{equation}
Obviously there is a simple orthogonal transformation which maps the
metric $\eta_t$ into the standard block diagonal metric $\eta_b$
written below
\begin{equation}
  \eta_b \, = \, \left( \begin{array}{c|c|c}
    \mathbf{1}_4 & 0 & 0 \\
    \hline
    0 & \mathbf{1}_{2s} & 0 \\
    \hline
    0 & 0 & -\mathbf{1}_4 \
  \end{array}\right)~.
\label{etab}
\end{equation}
Indeed we can write
\begin{equation}
  \Omega^T \, \eta_b \, \Omega \, = \, \eta_t
\label{omegatransf}
\end{equation}
where the explicit form of the matrix $\Omega$ is the following:
\begin{equation}
  \Omega \, = \, \left(\begin{array}{c|c|c}
    0 & \mathbf{1}_{2s} & 0 \\
    \hline
    \ft {1}{\sqrt{2}} \,\mathbf{1}_{4}  & 0 & \ft {1}{\sqrt{2}} \varpi_4 \\
    \hline
        \ft {1}{\sqrt{2}} \,\mathbf{1}_{4} & 0 & - \ft {1}{\sqrt{2}} \varpi_4 \
  \end{array} \right)~.
\label{Omega}
\end{equation}
Correspondingly the orthogonal transformation $\Omega$ maps the Lie
algebra and group elements of $\so(4,4+2s)$ from the standard basis
where the invariant metric is $\eta_b$ to the basis where it is
$\eta_t$
\begin{equation}
  \Lambda_t \, = \,  \Omega^T \, \Lambda_b \, \Omega \,.
\label{Lambdatb}
\end{equation}
In the $t$-basis the general form of an element of the solvable Lie
algebra which generates the coset manifold (\ref{p2qman}) has the
following appearance:
\begin{eqnarray}
  Solv\left(\frac{\mathrm{SO(4,4+2s)}}{\mathrm{SO(4)}\times \mathrm{SO(4+2s)}}\right)  \, \ni
  \, \Lambda_t & = & \left(\begin{array}{c|c|c}
    T & X & B \\
    \hline
    0 & 0 & X^T \, \varpi_4 \\
    \hline
        0 & 0 & -  \varpi_4 \, T^T \, \varpi_4 \
  \end{array} \right)
\label{Solmatra}
\end{eqnarray}
where
\begin{eqnarray}
T & = &  \left(\begin{array}{cccc}
    T_{1,1} & T_{1,2} & T_{1,3} & T_{1,4} \\
    0 & T_{2,2} &  T_{2,3} & T_{2,4} \\
    0 & 0 & T_{3,3} & T_{3,4} \\
    0 & 0 & 0  & T_{4,4} \
  \end{array} \right)  \, \quad \,\mbox{upper triangular $4 \times 4$~,} \nonumber\\
  B&=&-B^T \, \quad \, \mbox{antisymmetric $4\times 4$~,}\nonumber\\
  X & = & \mbox{arbitrary $4 \times 2s$}
\label{condizie}
\end{eqnarray}
while an element of the maximal compact subalgebra has instead the
following appearance:
\begin{eqnarray}
  \so(4) \, \oplus \, \so(4+2s) \, \ni
  \, \Lambda_t & = & \left(\begin{array}{c|c|c}
    Z & Y & C \, \varpi_4 \\
    \hline
    -Y^T & Q & - \, Y^T \, \varpi_4 \\
    \hline
        \varpi_4 \, C & \varpi_4 Y & -  \varpi_4\, Z^T \, \varpi_4 \
  \end{array} \right)
\label{sosomatra}
\end{eqnarray}
where
\begin{eqnarray}
Z & = & - \, Z^T \,  \mbox{antisymmetric $4 \times 4$}~, \nonumber\\
C & = & - \, C^T \,  \mbox{antisymmetric $4 \times 4$}~, \nonumber\\
Q & = & - \, Q^T \,  \mbox{antisymmetric $2s \times 2s$}~. \nonumber\\
Y & = & \mbox{arbitrary $4 \times 2s$}
\label{condesoso}
\end{eqnarray}
\subsection{The Tits Satake projection}
\label{tspara}
\begin{figure}
\centering
\begin{picture}(190,100)
\put (40,60){\circle {10}}
\put (37,45){$\alpha_1$}
\put (45,60){\line (1,0){19}}
\put (70,60){\circle {10}}
\put (67,45){$\alpha_{2}$}
\put (75,60){\line (1,0){20}}
\put (100,60){\circle {10}}
\put (97,45){$\alpha_{3}$}
\put (105,63){\line (1,0){20}}
\put (105,58){\line (1,0){20}}
\put (108,57){{\LARGE$>$}}
\put (130,60){\circle {10}}
\put (127,45){$\alpha_4$}
\end{picture}
\caption{\it The Dynkin diagram of the  $B_4$ Lie algebra.
\label{Biquattro}}
\end{figure}
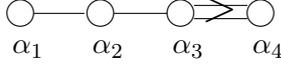

The above described form of the $\so(4,4+2s)$ Lie algebra matrices is well adapted to its Tits-Satake projection   which is as follows:
\begin{equation}\label{tistsat}
    \Pi_{TS} \left[ \so(4,4+2s) \right ] \, = \, \so(4,5)
\end{equation}
In terms of root systems the projection yields the $B_4$ system described by the Dynkin diagram of fig.\ref{Biquattro}
The projection is explicitly performed by dividing the range of the index $A=1,2,3,4,\dots , 4+s$ that labels components of the roots  in two subsets:
\begin{equation}\label{dividingrange}
    A=\left\{\underbrace{i}_{1,2,3,4}, \underbrace{p}_{5,\dots,4+s}\right\}
\end{equation}
The index $i$ enumerates the non-compact Cartan generators, while the index $p$ enumerates compact ones. For any root $\hat{\alpha}^A$ of the $D_{4+s}$ root system the corresponding Tits-Satake projection is obtained by suppressing the components $\alpha^p$ and keeping only the $\alpha^i$ ones.
\par
In this way we get all the roots of the $B_4$ system composed by $32$ four-component vectors:
\begin{equation}\label{fronte}
   \underbrace{\pm  \epsilon^i \pm \epsilon^j }_{\mbox{long roots}} \quad \quad \underbrace{\pm \epsilon^i}_{\mbox{short roots}}
\end{equation}
Within this projected system a basis of simple roots is provided by:
\begin{equation}\label{b4simple}
    \alpha_1 \, = \, \epsilon_1 - \epsilon_2 \quad ; \quad  \alpha_2 \, = \, \epsilon_2 - \epsilon_3 \quad ; \quad
    \alpha_3 \, = \, \epsilon_3 - \epsilon_4 \quad ; \quad\alpha_4 \, = \, \epsilon_4
\end{equation}
and a complete set of 16 positive roots can be presented as follows:
\begin{equation}\label{positivrut}
    \left(
\begin{array}{l}
 \alpha _1 \\
 \alpha _2 \\
 \alpha _3 \\
 \alpha _4 \\
 \alpha _1+\alpha _2 \\
 \alpha _2+\alpha _3 \\
 \alpha _3+\alpha _4 \\
 \alpha _1+\alpha _2+\alpha _3 \\
 \alpha _2+\alpha _3+\alpha _4 \\
 \alpha _3+2 \alpha _4 \\
 \alpha _1+\alpha _2+\alpha _3+\alpha _4 \\
 \alpha _2+\alpha _3+2 \alpha _4 \\
 \alpha _1+\alpha _2+\alpha _3+2 \alpha _4 \\
 \alpha _2+2 \alpha _3+2 \alpha _4 \\
 \alpha _1+\alpha _2+2 \alpha _3+2 \alpha _4 \\
 \alpha _1+2 \alpha _2+2 \alpha _3+2 \alpha _4
\end{array}
\right) \, = \, \left(
\begin{array}{l}
 \epsilon _1-\epsilon _2 \\
 \epsilon _2-\epsilon _3 \\
 \epsilon _3-\epsilon _4 \\
 \epsilon _4 \\
 \epsilon _1-\epsilon _3 \\
 \epsilon _2-\epsilon _4 \\
 \epsilon _3 \\
 \epsilon _1-\epsilon _4 \\
 \epsilon _2 \\
 \epsilon _3+\epsilon _4 \\
 \epsilon _1 \\
 \epsilon _2+\epsilon _4 \\
 \epsilon _1+\epsilon _4 \\
 \epsilon _2+\epsilon _3 \\
 \epsilon _1+\epsilon _3 \\
 \epsilon _1+\epsilon _2
\end{array}
\right)
\end{equation}
Having clarified the form of the Tits Satake projection and  the structure of the matrices representing the Lie
algebra elements in a basis well adapted to such a
projection, we can now discuss a convenient basis of well adapted generators.
\par
To this effect, let us denote by $\mathcal{I}_{ij}$ the
$4 \times 4$ matrices whose only non vanishing entry is
the $ij$-th one which is equal to $1$
\begin{equation}
  \mathcal{I}_{ij} \, = \, \left(\begin{array}{ccccl}
    0 &   \dots & \dots & 0 &\null\\
    0 &  \dots & 1 & 0 &\} \, \mbox{i-th row}\\
     0 &  \dots &  \dots & 0 \\
    0 & \dots & \underbrace{\dots}_{\mbox{j-th column}}  & 0 &\null \
  \end{array} \right)~.
\label{Imatri}
\end{equation}
Using this notation the $4$ non-compact Cartan generators are given by
\begin{equation}
  \mathcal{H}_i \, = \, \left(\begin{array}{c|c|c}
    \mathcal{I}_{ii} & 0 & 0 \\
    \hline
    0 & 0 & 0\\
    \hline
        0 & 0 & -  \varpi_4 \, \mathcal{I}_{ii} \, \varpi_4\
  \end{array} \right) \,
  \quad ; \quad \, (i=1,\dots , 4)~.
\label{cartani1}
\end{equation}
Next we introduce the coset generators associated with the long roots
of type: $\alpha = \epsilon^i - \epsilon^j$.
\begin{eqnarray}
\begin{array}{c}
   \mbox{\scriptsize{$\alpha = \epsilon^i - \epsilon^j$ }} \\
 \mbox{\scriptsize{ $i<j =1,\dots , 4$}}
\end{array}
  & \Rightarrow & K^{ij}_-
  =  \ft {1}{\sqrt{2}} \,\left( E^\alpha + E^{-\alpha} \right)  =
  \ft {1}{\sqrt{2}} \left(\begin{array}{c|c|c}
    \mathcal{I}_{ij}  +  \mathcal{I}_{ji} & 0 & 0 \\
    \hline
    0 & 0 & 0\\
    \hline
        0 & 0 & -  \varpi_4 \, \left( \mathcal{I}_{ij}  +  \mathcal{I}_{ji}\right)  \, \varpi_4 \
  \end{array} \right)\nonumber\\
\label{Kijm}
\end{eqnarray}
and the coset generators associated with the long roots of type $\alpha = \epsilon^i +
\epsilon^j$:
\begin{eqnarray}
\begin{array}{c}
   \mbox{\scriptsize{$\alpha = \epsilon^i + \epsilon^j$ }} \\
 \mbox{\scriptsize{ $i<j =1,\dots , 4$}}
\end{array}
  & \Rightarrow & K^{ij}_+
  =  \ft {1}{\sqrt{2}} \,\left( E^\alpha + E^{-\alpha} \right)  =
  \ft {1}{\sqrt{2}} \left(\begin{array}{c|c|c}
    0 & 0 & \left( \mathcal{I}_{ij}  -  \mathcal{I}_{ji}\right) \varpi_4 \\
    \hline
    0 & 0 & 0\\
    \hline
       \varpi_r \left( \mathcal{I}_{ji}  -  \mathcal{I}_{ij}\right) & 0 & 0 \
  \end{array} \right)~.\nonumber\\
\label{Kijp}
\end{eqnarray}
The short roots, after the Tits-Satake projection, are just $4$,
namely $\epsilon^i$. Each of them, however, appears with multiplicity
$2s$, due to the paint group. We introduce a $2s$-tuple of coset
generators associated to each of the short roots in such a way that
such $2s$-tuple transforms in the fundamental representation of
$\mathbf{G}_{paint} \, = \, \so(2s)$. To this effect let us define
the rectangular $4\times 2s$ matrices$ \mathcal{J}_{im}$ analogous to the
square matrices $\mathcal{I}_{ij}$, namely
\begin{equation}
  \mathcal{J}_{im} \, = \, \left(\begin{array}{ccccccccl}
    0 & 0 & \dots & \dots & \dots &  \dots & \dots & 0 &\null\\
    \dots & \dots & \dots & \dots & \dots & \dots \dots & \dots & \null\\
    0 & 0 & \dots & 1 & \dots &\dots & \dots & 0 &\} \, \mbox{i-th row}\\
    0 & 0 & \dots & \underbrace{\dots}_{\mbox{$m$-th column}} & \dots & \dots & \dots & 0 &\null \
  \end{array} \right)~.
\label{Jmatri}
\end{equation}
Then we introduce the following coset generators:
\begin{eqnarray}
\begin{array}{c}
   \mbox{\scriptsize{$\alpha = \epsilon^i $ }} \\
 \mbox{\scriptsize{ $i=1,\dots , 4$}}\\
 \mbox{\scriptsize{ $m=1,\dots , 2s$}}\\
\end{array}
  & \Rightarrow & K^{i}_m
  =
  \ft {1}{\sqrt{2}} \left(\begin{array}{c|c|c}
    0 & \mathcal{J}_{im} & 0 \\
    \hline
    \mathcal{J}_{im}^T & 0 & - \mathcal{J}_{im}^T \, \varpi_r\\
    \hline
        0 & - \varpi_r \,\mathcal{J}_{im}  & 0 \
  \end{array} \right)~.\nonumber\\
\label{Kim}
\end{eqnarray}
The remaining generators of the $\so(4,4+2s)$ algebra are all compact
and span the subalgebra $\so(4)\oplus \so(4+2s)\subset \so(4,4+2s)$.
According to the nomenclature of eq.(\ref{sosomatra}) we introduce
four sets of generators. The first set is associated with the long
roots of type $\alpha = \epsilon ^i -\epsilon ^j$ and is defined as
follows:
\begin{equation}
  Z^{ij} \, = \, \ft {1}{\sqrt{2}} \,\left( E^\alpha - E^{-\alpha} \right)  =
  \ft {1}{\sqrt{2}} \left(\begin{array}{c|c|c}
    \mathcal{I}_{ij}  -  \mathcal{I}_{ji} & 0 & 0 \\
    \hline
    0 & 0 & 0\\
    \hline
        0 & 0 & -  \varpi_4 \, \left( \mathcal{I}_{ij}  -  \mathcal{I}_{ji}\right)  \, \varpi_4 \
  \end{array} \right)~.
\label{Zij}
\end{equation}
The second set is associated with the long roots of type $\alpha = \epsilon ^i +\epsilon
^j$ and is defined as follows:
\begin{equation}
  C^{ij}
  =  \ft {1}{\sqrt{2}} \,\left( E^\alpha - E^{-\alpha} \right)  =
  \ft {1}{\sqrt{2}} \left(\begin{array}{c|c|c}
    0 & 0 & \left( \mathcal{I}_{ij}  -  \mathcal{I}_{ji}\right) \varpi_4 \\
    \hline
    0 & 0 & 0\\
    \hline
      - \varpi_r \left( \mathcal{I}_{ji}  -  \mathcal{I}_{ij}\right) & 0 & 0 \
  \end{array} \right)~.
\label{Cij}
\end{equation}
The above formulae can now be inverted in order to obtain the explicit form of the step-operators associated with long roots.
For the roots of type: $\widehat{\alpha}\,=\,\epsilon^i \, - \, \epsilon^j$ we have:
\begin{equation}\label{alfameno}
    E^{\pm \alpha} \, =\, \frac{1}{\sqrt{2}} \left( K^{ij}_- \pm Z^{ij}\right )
\end{equation}
while for the roots of type $\widehat{\alpha}\,=\,\epsilon^i \, + \, \epsilon^j$ we have:
\begin{equation}\label{alfapiu}
    E^{\pm \alpha} \, =\, \frac{1}{\sqrt{2}} \left( K^{ij}_+ \pm C^{ij}\right )
\end{equation}
The third group of compact generators spans the compact coset
\begin{equation}
  \frac{\mathrm{SO(4+2s)}}{\mathrm{SO(4)} \times \mathrm{SO(2s)}}
\label{Kcoset}
\end{equation}
and it is given by
\begin{equation}
  Y^{i}_{m}
  =
  \ft {1}{\sqrt{2}} \left(\begin{array}{c|c|c}
    0 & \mathcal{J}_{im} & 0 \\
    \hline
    -\mathcal{J}_{im}^T & 0 & - \mathcal{J}_{im}^T \, \varpi_r\\
    \hline
        0 &  \varpi_r \,\mathcal{J}_{im}  & 0 \
  \end{array} \right)~.
\label{Yim}
\end{equation}
\par
In this way we can define the set of step operators associated with the short roots of the Tits-Satake projection each of which has a multiplicity $2s$ and forms a vector under the action of the paint group $\mathrm{SO}(2s)$.
Hence for the short roots with multiplicity $\alpha_i(m)$ $(m=1,\dots,2s)$  we set:
\begin{equation}\label{frangione}
    E^{\pm \alpha}_m \, = \, \frac{1}{\sqrt{2}} \left ( K^{i}_m \pm  Y^{i}_{m} \right)
\end{equation}
The fourth set of compact generators spans the paint group Lie algebra
$\so(2s)$ and is given by
\begin{equation}
  Q_{mn} \, = \, \left(\begin{array}{c|c|c}
    0 & 0 & 0 \\
    \hline
    0 & \mathcal{Q}_{mn} -\mathcal{Q}_{nm} & 0\\
    \hline
     0 & 0 & 0 \
  \end{array} \right)
\label{Qmn}
\end{equation}
where $\mathcal{Q}_{mn}$ denotes the analogue of the
$\mathcal{I}_{ij}$ in $2s$ rather than in $4$ dimensions.
\par
By performing the change of basis to the block diagonal form of the
matrices we can verify that $C_{ij} - Z_{ij}$ generate the $\so(4)$
subalgebra while $C_{ij} + Z_{ij}$ together with $Q_{mn}$ and
$Y_{im}$ generate the subalgebra $\so(4+2s)$.
\par
The full set of generators is ordered in the following way:
\begin{equation}
  T_\Lambda \, = \, \left\{ \underbrace{\mathcal{H}_i}_{r} \, , \, \underbrace{K^{ij}_- }_{6} \, , \, \underbrace{K^{ij}_+}_{6} \, , \,
  \underbrace{K^i_m}_{8s} \, , \, \underbrace{Z^{ij}}_{6} \, , \, \underbrace{C^{ij}}_{ 6} \, , \, \underbrace{Y^i_{m}}_{8s} \, ,
  \, \underbrace{Q_{mn}}_{s(2s-1)} \right\}
\label{allgenera}
\end{equation}
and satisfy the trace relation:
\begin{eqnarray}
  \mbox{Tr} \,\left ( T_\Lambda \, T_\Sigma \right) & = & g_{\Lambda \Sigma}~, \nonumber\\
g_{\Lambda \Sigma
  } & = & 2 \, \mbox{diag} \left(\underbrace{+,+,\dots, +}_{16+8s},
  \underbrace{-,-,\dots,-}_{12 + 8s + 2s^2 -s} \right)~.
\label{tracciagenera}
\end{eqnarray}
\subsection{Decompositions with respect to  the $\mathbb{H}^\star$ subalgebra and to the Ehlers subalgebra}
As it happens in all $\mathcal{N}=2$ theories there are three decompositions of the $\mathbb{U}_{D=4} = \so(4,4+2s)$ Lie algebra that we have to consider at the same time: that with respect to the maximal compact subalgebra $\mathbb{H} \, = \, \so(4)\oplus \so(4+2s)$, that with respect
to its non-compact counter part $\mathbb{H}^\star \, = \, \so(2,2)\oplus \so(2,2+2s)$ and that with respect to the Ehlers subalgebra times the original $\mathbb{U}_{D=4}$ Lie algebra namely $\slal(2)_E \oplus \slal(2) \oplus \so(2,2+2s)$. The three decompositions have the following form and interpretation:
\begin{eqnarray}
  \so(4,4+2s) &=& \underbrace{\so(4)\oplus \so(4+2s)}_{\mathbb{H}} \oplus \underbrace{(\mathbf{4},\mathbf{4+2s})}_{\mathbb{K}}\label{decompo1} \\
   \so(4,4+2s) &=& \underbrace{\so(2,2)\oplus \so(2,2+2s)}_{\mathbb{H}^\star} \oplus \underbrace{(\mathbf{4},\mathbf{4+2s})}_{\mathbb{K}^\star \, \sim \, \Delta^{\alpha|A}}\label{decompo2}\\
  \so(4,4+2s) &=& \underbrace{\slal(2,\mathbb{R})_E}_{\mbox{Ehlers} }\oplus \underbrace{\slal(2,\mathbb{R}) \oplus \so(2,2+2s)}_{\mathbb{U}_{D=4}} \oplus \underbrace{\left(\mathbf{2},\left(\mathbf{2,2+2s}\right)\right)}_{(\mathbf{2},\mathbf{W})}\label{decompo3}
\end{eqnarray}
where $\mathbb{K}$ and $\mathbb{K}^\star$ denote the complementary orthogonal spaces to the isotropy subalgebras in the two coset cases (riemannian and non riemannian) and encompass all possible Lax operators for the corresponding coset.  On the other hand $ (\mathbf{2},\mathbf{W})$ denote the universal form of the generators associated with vector fields in the dimensional reduction from $D=4$ to $D=3$. By $\mathbf{W}$ we always denote the symplectic representation of the $\mathbb{U}_{D=3}$ Lie algebra which enters the construction of special geometry.
\par
The decomposition (\ref{decompo1}) was discussed in the previous subsection; the remaining two are the goal of the present subsection.
\par
A fundamental universal feature of $\mathcal{N}=2$ models is that the subalgebra $\mathbb{H}^\star$ and $\slal(2)_E \times \mathbb{U}_{D=4}$ are always isomorphic although, inside $\mathbb{U}_{D=4}$ they correspond to distinct algebras singled out by two different procedures.
In the present case this isomorphism is easily seen recalling the well known isomorphisms:
\begin{eqnarray}
  \so(2,2) &\simeq& \slal(2,\mathbb{R}) \oplus \slal(2,\mathbb{R}) \\
  \su(1,1) &=& \slal(2,\mathbb{R})
\end{eqnarray}
Once more there are two universal procedures to perform the two decompositions under consideration and to single out the two distinct but isomorphic Lie algebras $\mathbb{H}^\star$ and $\slal(2)_E \times \mathbb{U}_{D=4}$:
\begin{description}
  \item[a] The $\slal(2)_E \times \mathbb{U}_{D=4}$ subalgebra is found decomposing $\mathbb{U}_{D=4}$ with respect to its highest root $\hat{\alpha}_h$, since the Ehlers subalgebra is universally associated with the Chevalley triple of the highest root. Hence we have just to consider the highest root and the set of all roots orthogonal to it; by definition these latter compose the root system of $\mathbb{U}_{D=3}$. The remaining generators of $\mathbb{U}_{D=4}$ that have a grading both with respect to the Ehlers Cartan $\mathcal{H}_{\hat{\alpha}_h}$ and to the Cartans in $\mathbb{U}_{D=4}$ form the representation $ (\mathbf{2},\mathbf{W})$.
  \item[b] The $\mathbb{H}^\star$ subalgebra is found by introducing a suitable diagonal $\eta_d$ tensor with the appropriate signature and then by defining the subset of Lie algebra elements $\Lambda$ that in addition to the general condition (\ref{Lambdatdefi}) satisfy also the condition
      \begin{equation}\label{lamdaHstar}
        \Lambda^T \eta_d \, + \, \eta_d \Lambda \, = \, 0
      \end{equation}
      An important point to stress is that the choice of $\eta$ is not an independent element of the construction. The subalgebra $\mathbb{H}^\star$
      is uniquely dictated by the Wick rotation (\ref{viccus}) which maps the quaternionic manifold into its lorentzian counterpart corresponding to time-like dimensional reductions.
      \end{description}
\subsubsection{The Ehlers decomposition}
We begin with the Ehlers decomposition.
\par
An   intrinsic property of the $D_\ell$ Lie algebras is that the highest root has the following form in terms of the simple roots:
\begin{equation}\label{altarutta}
    \hat{\alpha}_h \, = \, \hat{\alpha}_1 + 2\hat{\alpha}_2 +2\hat{\alpha}_3 + \dots + 2\hat{\alpha}_{\ell-2} + \hat{\alpha}_{\ell -1} + \hat{\alpha}_\ell
\end{equation}
In the orthonormal basis that we use for the $\hat{\alpha}$ roots this means  that:
\begin{equation}\label{alfona}
    \hat{\alpha}_h \, = \, \epsilon_1 \, + \, \epsilon_2
\end{equation}
Utilizing this information, the Ehlers decomposition becomes very easy and immediate at the Dynkin diagram level. It suffices to remove the simple root $\hat{\alpha}_2$ and substitute it with the highest one $\hat{\alpha}_h$. The two roots $\hat{\alpha}_1 \, = \, \epsilon_1 - \epsilon_2$ and $\hat{\alpha}_h \, = \, \epsilon_1 + \epsilon_2$ are orthogonal among themselves and define a system $A_1 \oplus A_1 \sim \so(2,2)$. They are also orthogonal to the remaining simple roots $\hat{\alpha}_3 \dots \hat{\alpha}_\ell$ which form a $D_{2+s}$ system and therefore are associated, in the real form that we consider, with a subalgebra $\so(2,2+2s)$. In this way we see that we have:
\begin{equation}
\begin{array}{rcccl}
 \hat{\alpha}_h & \Leftrightarrow & A_1 &\Rightarrow & \underbrace{\slal(2)_E}_{\mbox{Ehlers alg.}} \\
 \hat{\alpha}_1  \, , \, \hat{\alpha}_3 \, , \, \dots \, , \, \hat{\alpha}_{4+s}& \Leftrightarrow & A_1 \oplus D_{2+s} &\Rightarrow& \underbrace{\su(1,1) \oplus \so(2,2+2s)}_{\mathbb{U}_{\mathrm{D=4}}}\
 \end{array}
 \end{equation}
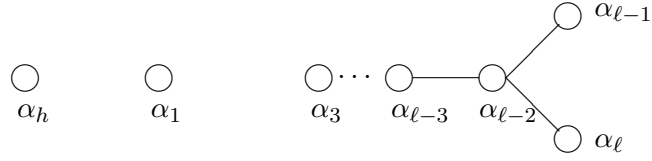
\begin{figure}
\caption{\it Removing the $\alpha_2$ root and adding the highest root $\alpha_h$ we embed the Lie algebra $A_1 \oplus A_1 \oplus D_{\ell -2}$ into the  $D_\ell$ Lie algebra.
\label{Delpiccola}}
\centering
\begin{picture}(60,130)
\put (-70,70){\circle {10}}
\put (-73,55){$\alpha_h$}
\put (-20,70){\circle {10}}
\put (-23,55){$\alpha_1$}
\put (40,70){\circle {10}}
\put (37,55){$\alpha_3$}
\put (47,70){$\dots$}
\put (70,70){\circle {10}}
\put (67,55){$\alpha_{\ell-3}$}
\put (75,70){\line (1,0){25}}
\put (105,70){\circle {10}}
\put (100,55){$\alpha_{\ell-2}$}
\put (110,70){\line (1,1){20}}
\put (110,70){\line (1,-1){20}}
\put (133.2,93.2){\circle {10}}
\put (133.2,46.8){\circle {10}}
\put (143.2,93.2){$\alpha_{\ell-1}$}
\put (143.2,43.8){$\alpha_\ell$}
\end{picture}
\end{figure}
This procedure is graphically illustrated in fig. \ref{Delpiccola}.
\par
Once we have singled out both the Ehlers algebra and the $\mathbb{U}_{\mathrm{D=4}}$ subalgebra inside the Lie algebra $\mathbb{U}_{\mathrm{D=}3} \, = \, \so(4,4+2s)$ the remaining Lie algebra generators span the representation $(2,W)$ and those corresponding to the positive weight of the $\slal(2)_E$ doublet form the generators $\mathcal{W}^M$ of the solvable Lie algebra associated with the dimensional reduction of vector fields. In other words with the above information we are in the position to write the general form of the solvable coset representative advocated in \cite{noig22}, namely;
\begin{equation}\label{genaforma}
    \mathbb{L}(\Phi) \, = \, \exp\left[ - a \, L_+^E \right]\, \exp\left[ \sqrt{2}\, Z^M \,  \mathcal{W}_M \right] \,
    \mathbb{L}_4(\phi) \, \exp\left[ U \, L_0^E \right]
\end{equation}
where $U$ is the warp factor parameterizing  the $D=4$ metric, $a$ is associated with the Taub-NUT charge, $\phi$ are the scalar fields in $D=4$,
$Z^M$ are the $D=3$ scalars produced by the dimensional reduction of the $D=4$ vector fields. We do not dwell on the details of the $D=4$ oxidation thoroughly described in \cite{noig22} neither we use the integration algorithm in order to produce explicit solutions. In the present paper which has a purely algebraic scope we confine ourselves to the above illustration which shows how the relevant basis of generators advocated by the construction of the solvable coset representative is uniquely defined in intrinsic Lie algebra terms and is ready to use. Our main goal is the algebraic classification of nilpotent orbits of Lax operators and on this task we concentrate.
\subsubsection{The $\mathbb{H}^\star$ decomposition}
\label{bumbu}
First of all we begin by defining the basis of generators of the solvable Lie algebra ${Solv} \subset \so(4,4+2s)$. This is extracted from the construction of section \ref{tspara} in the following way. As generators of the solvable Lie algebra we take all the non-compact Cartan generators plus the step operators associated with positive roots that are not orthogonal to the non-compact CSA, namely that have non-vanishing Tits Satake projection onto the $B_4$ system. As order, after the Cartan generators, we take the lexicographic one in the orthonormal basis, listing first the roots of long type (in the projection) and secondly those of short type (also in the projection). So we set
\begin{equation}\label{tsolvi}
    T^{Solv}_A \, = \, \left (\underbrace{\mathcal{H}_i}_{4}\, , \,\underbrace{ E^{\epsilon^i \pm \epsilon^j}}_{12} , \underbrace{E^{\epsilon^i \pm \epsilon^p}}_{8s},\right)
\end{equation}
Defining a generic element of the solvable Lie algebra as:
\begin{equation}\label{genisol}
    Solv \, \ni \, \mathfrak{B} \, = \, \sum_{A=1}^{16+8s} \, \phi^A \, T^{Solv}_A
\end{equation}
in the upper triangular basis we find:
\begin{equation}\label{Bsol}
 \mathfrak{B} \, = \,  \left(
\begin{array}{llllllllll}
 \phi _1 & \phi _5 & \phi _6 & \phi _7 & \phi _{17} & \phi _{18} & \phi _{13} & \phi _{12}
   & \phi _{11} & 0 \\
 0 & \phi _2 & \phi _8 & \phi _9 & \phi _{19} & \phi _{20} & \phi _{15} & \phi _{14} & 0 &
   -\phi _{11} \\
 0 & 0 & \phi _3 & \phi _{10} & \phi _{21} & \phi _{22} & \phi _{16} & 0 & -\phi _{14} &
   -\phi _{12} \\
 0 & 0 & 0 & \phi _4 & \phi _{23} & \phi _{24} & 0 & -\phi _{16} & -\phi _{15} & -\phi
   _{13} \\
 0 & 0 & 0 & 0 & 0 & 0 & -\phi _{23} & -\phi _{21} & -\phi _{19} & -\phi _{17} \\
 0 & 0 & 0 & 0 & 0 & 0 & -\phi _{24} & -\phi _{22} & -\phi _{20} & -\phi _{18} \\
 0 & 0 & 0 & 0 & 0 & 0 & -\phi _4 & -\phi _{10} & -\phi _9 & -\phi _7 \\
 0 & 0 & 0 & 0 & 0 & 0 & 0 & -\phi _3 & -\phi _8 & -\phi _6 \\
 0 & 0 & 0 & 0 & 0 & 0 & 0 & 0 & -\phi _2 & -\phi _5 \\
 0 & 0 & 0 & 0 & 0 & 0 & 0 & 0 & 0 & -\phi _1
\end{array}
\right)
\end{equation}
where we have used the case $s=1$ as a mean of illustration.
\par
In the same upper triangular basis the appropriate $\eta$-tensor which singles out the $\mathbb{H}^\star$-subalgebra defined by the Wick rotation (\ref{viccus}) is the following one:
\begin{equation}\label{etadiag}
  \eta_d \, = \,  \left(
\begin{array}{llll|llllll|llll}
 1 & 0 & 0 & 0 & 0 & 0  & \ldots& 0 &0&0 & 0
   & 0 & 0 \\
 0 & 1 & 0 & 0 & 0 & 0  &\ldots& 0 &0  &0& 0 & 0
   & 0 & 0 \\
 0 & 0 & -1 & 0 & 0 & 0  &\ldots& 0 &0 &0 & 0 & 0
   & 0 & 0 \\
 0 & 0 & 0 & -1 & 0 & 0 &\ldots& 0  & 0&0 & 0 & 0
   & 0 & 0 \\
   \hline
 0 & 0 & 0 & 0 & 1 & 0 &\ldots& 0  & 0&0 & 0 & 0
   & 0 & 0 \\
 0 & 0 & 0 & 0 & 0 & 1 &\ldots& 0 & 0 &0 & 0 & 0
   & 0 & 0 \\
\vdots & \vdots & \vdots & \vdots &0 & 0 & \vdots & \vdots &\vdots&\vdots & \vdots & \vdots
   & \vdots& \vdots \\
   \vdots & \vdots & \vdots & \vdots &\vdots & \vdots & \vdots & \vdots &\vdots&\vdots & \vdots & \vdots
   & \vdots& \vdots \\
    \vdots & \vdots & \vdots & \vdots &\vdots & \vdots & \vdots & 0 &1&0 & \vdots & \vdots
   & \vdots& \vdots \\
 0 & 0 & 0 & 0 & 0 & 0&\ldots& 0&0&1 & 0 & 0
   & 0 & 0 \\
   \hline
 0 & 0 & 0 & 0 & 0 & 0 & \ldots & 0&0&0  & -1 & 0
   & 0 & 0 \\
 0 & 0 & 0 & 0 & 0 & 0 & \ldots& 0 &0&0 & 0 & -1
   & 0 & 0 \\
 0 & 0 & 0 & 0 & 0 & 0 & \ldots & 0 &0&0 & 0 & 0
   & 1 & 0 \\
 0 & 0 & 0 & 0 & 0 & 0 & \ldots & 0 &0&0 & 0 & 0
   & 0 & 1
\end{array}
\right)
\end{equation}
Using this tensor we define the general form of the Lax operator by setting:
\begin{equation}\label{franciscus}
    L(\phi) \, = \, \frac{1}{2\sqrt{2}} \, \left( \mathfrak{B} \, + \, \eta_d \, \mathfrak{B} \, \eta_d \right ) \, \equiv \,
     2 \,\sum_{A=1}^{4} \phi^A\, \mathfrak{K}_A \, + \, \sum_{A=5}^{16+8s} \phi^A\, \mathfrak{K}_A
\end{equation}
which defines the basis of $\mathbb{K}$-generators, namely of the generators spanning the complementary orthogonal subspace in the decomposition
of the $\so(4,4+2s)$ Lie algebra with respect to the $\mathbb{H}^\star$ subalgebra:
\begin{equation}\label{hstarrodecompo}
    \so(4,4+2s) \, = \, \mathbb{H}^\star \oplus \mathbb{K}
\end{equation}
With the above introduced definition, the $\mathfrak{K}_A$ generators are normalized to $\pm 1$ with respect to the trace:
\begin{equation}\label{norma}
    \mbox{Tr} \left (\mathfrak{K}_A \, , \, \mathfrak{K}_B \right ) \, = \, \pm \, \delta_{AB}
\end{equation}
and we always have $8+4s$ plus signs and $8+4s$ minus signs. This means that $8+4s$ generators of the $\mathbb{K}$ basis are compact and just as many are non-compact.
\par
Using again the case $s=1$ as an illustration we find:
\begin{eqnarray}
 L(\phi) &=   &
\left(
\begin{array}{llllllllll}
 2 \phi _1 & \phi _5 & \phi _6 & \phi _7 & \phi _{17} & \phi _{18} & \phi _{13} & \phi
   _{12} & \phi _{11} & 0 \\
 \phi _5 & 2 \phi _2 & \phi _8 & \phi _9 & \phi _{19} & \phi _{20} & \phi _{15} & \phi
   _{14} & 0 & -\phi _{11} \\
 -\phi _6 & -\phi _8 & 2 \phi _3 & \phi _{10} & \phi _{21} & \phi _{22} & \phi _{16} & 0 &
   -\phi _{14} & -\phi _{12} \\
 -\phi _7 & -\phi _9 & \phi _{10} & 2 \phi _4 & \phi _{23} & \phi _{24} & 0 & -\phi _{16} &
   -\phi _{15} & -\phi _{13} \\
 \phi _{17} & \phi _{19} & -\phi _{21} & -\phi _{23} & 0 & 0 & -\phi _{23} & -\phi _{21} &
   -\phi _{19} & -\phi _{17} \\
 \phi _{18} & \phi _{20} & -\phi _{22} & -\phi _{24} & 0 & 0 & -\phi _{24} & -\phi _{22} &
   -\phi _{20} & -\phi _{18} \\
 -\phi _{13} & -\phi _{15} & \phi _{16} & 0 & \phi _{23} & \phi _{24} & -2 \phi _4 & -\phi
   _{10} & -\phi _9 & -\phi _7 \\
 -\phi _{12} & -\phi _{14} & 0 & -\phi _{16} & \phi _{21} & \phi _{22} & -\phi _{10} & -2
   \phi _3 & -\phi _8 & -\phi _6 \\
 \phi _{11} & 0 & \phi _{14} & \phi _{15} & -\phi _{19} & -\phi _{20} & \phi _9 & \phi _8 &
   -2 \phi _2 & -\phi _5 \\
 0 & -\phi _{11} & \phi _{12} & \phi _{13} & -\phi _{17} & -\phi _{18} & \phi _7 & \phi _6
   & -\phi _5 & -2 \phi _1
\end{array}
\right) \nonumber\\
\label{laxop}
\end{eqnarray}
Next we consider the general form of an element of the $\mathbb{H}^\star$ subalgebra. To this effect we introduce the following  basis of $ 12+8s+s(2s^2-1) $ generators:
\begin{eqnarray}
   \mathfrak{H}_ I &=& \ft 12 \, \left ( T_{4+I}^{Solv}-\eta_d \,  T_{4+I}^{Solv}\, \eta_d \right) \quad; \quad  \left ( I\, = \, 1,\dots , 12  +8*s\right )\nonumber\\
  \mathfrak{H}_ {12+8s + mn} &=& Q_{mn} \quad ; \quad \left ( [mn]\, = \, 1 ,\dots , s(2s^2-1\right )\label{Hstargen}
\end{eqnarray}
where $Q_{mn}$ are the previously introduced generators of the Paint Group spanning the  $\so(2s)$ Lie algebra and the pair of antisymmetric indices $mn$ are enumerated in lexicographic order.
\par
Therefore a generic element of the $\mathbb{H}^\star$ Lie algebra:
\begin{equation}\label{circum}
    \mathfrak{W} \, = \, \sum_{I=1}^{12+8*s} \omega^I \, \mathfrak{H}_ I  \, + \, \sum_{i=1}^{s(2s-1)} \rho^i \, \mathfrak{H}_ {12 + 8s +i}
\end{equation}
has the following appearance (using once again the case $s=1$ as an illustration):
\begin{eqnarray}\label{omegone}
 \mathfrak{W} & =  &\left(
\begin{array}{llllllllll}
 0 & \frac{\omega _1}{2} & \frac{\omega _2}{2} & \frac{\omega _3}{2} & \frac{\omega
   _{13}}{2} & \frac{\omega _{14}}{2} & \frac{\omega _9}{2} & \frac{\omega _8}{2} &
   \frac{\omega _7}{2} & 0 \\
 -\frac{\omega _1}{2} & 0 & \frac{\omega _4}{2} & \frac{\omega _5}{2} & \frac{\omega
   _{15}}{2} & \frac{\omega _{16}}{2} & \frac{\omega _{11}}{2} & \frac{\omega _{10}}{2} & 0
   & -\frac{\omega _7}{2} \\
 \frac{\omega _2}{2} & \frac{\omega _4}{2} & 0 & \frac{\omega _6}{2} & \frac{\omega
   _{17}}{2} & \frac{\omega _{18}}{2} & \frac{\omega _{12}}{2} & 0 & -\frac{\omega
   _{10}}{2} & -\frac{\omega _8}{2} \\
 \frac{\omega _3}{2} & \frac{\omega _5}{2} & -\frac{\omega _6}{2} & 0 & \frac{\omega
   _{19}}{2} & \frac{\omega _{20}}{2} & 0 & -\frac{\omega _{12}}{2} & -\frac{\omega
   _{11}}{2} & -\frac{\omega _9}{2} \\
 -\frac{\omega _{13}}{2} & -\frac{\omega _{15}}{2} & \frac{\omega _{17}}{2} & \frac{\omega
   _{19}}{2} & 0 & \frac{\rho _1}{\sqrt{2}} & -\frac{\omega _{19}}{2} & -\frac{\omega
   _{17}}{2} & -\frac{\omega _{15}}{2} & -\frac{\omega _{13}}{2} \\
 -\frac{\omega _{14}}{2} & -\frac{\omega _{16}}{2} & \frac{\omega _{18}}{2} & \frac{\omega
   _{20}}{2} & -\frac{\rho _1}{\sqrt{2}} & 0 & -\frac{\omega _{20}}{2} & -\frac{\omega
   _{18}}{2} & -\frac{\omega _{16}}{2} & -\frac{\omega _{14}}{2} \\
 \frac{\omega _9}{2} & \frac{\omega _{11}}{2} & -\frac{\omega _{12}}{2} & 0 & -\frac{\omega
   _{19}}{2} & -\frac{\omega _{20}}{2} & 0 & -\frac{\omega _6}{2} & -\frac{\omega _5}{2} &
   -\frac{\omega _3}{2} \\
 \frac{\omega _8}{2} & \frac{\omega _{10}}{2} & 0 & \frac{\omega _{12}}{2} & -\frac{\omega
   _{17}}{2} & -\frac{\omega _{18}}{2} & \frac{\omega _6}{2} & 0 & -\frac{\omega _4}{2} &
   -\frac{\omega _2}{2} \\
 -\frac{\omega _7}{2} & 0 & -\frac{\omega _{10}}{2} & -\frac{\omega _{11}}{2} &
   \frac{\omega _{15}}{2} & \frac{\omega _{16}}{2} & -\frac{\omega _5}{2} & -\frac{\omega
   _4}{2} & 0 & -\frac{\omega _1}{2} \\
 0 & \frac{\omega _7}{2} & -\frac{\omega _8}{2} & -\frac{\omega _9}{2} & \frac{\omega
   _{13}}{2} & \frac{\omega _{14}}{2} & -\frac{\omega _3}{2} & -\frac{\omega _2}{2} &
   \frac{\omega _1}{2} & 0
\end{array}
\right) \nonumber\\
\end{eqnarray}
The parameters of $\mathbb{H}^\star$ that belong to the Paint group subalgebra have been denoted with the letter $\rho$ in order to distinguish them from the others.
\par In order to facilitate the identification of the tensor structure of the Lax operator, it is convenient to perform the backward transformation from the upper triangular basis to the standard diagonal basis by means of the inverse of the matrix (\ref{Omega}).
In this basis both the invariant eta tensor of the $\so(4,4+2s)$ subalgebra and that which singles out the $\mathbb{H}^\star$ subalgebra are diagonal. Using once again the $s=1$ case as an illustration we find:
\begin{eqnarray}\label{cirenaica}
    \Omega \, L(\phi) \, \Omega^T &= & \left(
\begin{array}{llllllllll}
 0 & 0 & 0 & 0 & 0 & 0 & \Delta _{1,1} & \Delta _{1,2} & \Delta _{1,3} & \Delta _{1,4} \\
 0 & 0 & 0 & 0 & 0 & 0 & \Delta _{2,1} & \Delta _{2,2} & \Delta _{2,3} & \Delta _{2,4} \\
 0 & 0 & 0 & 0 & 0 & 0 & \Delta _{3,1} & \Delta _{3,2} & \Delta _{3,3} & \Delta _{3,4} \\
 0 & 0 & 0 & 0 & 0 & 0 & \Delta _{4,1} & \Delta _{4,2} & \Delta _{4,3} & \Delta _{4,4} \\
 0 & 0 & 0 & 0 & 0 & 0 & \Delta _{5,1} & \Delta _{5,2} & \Delta _{5,3} & \Delta _{5,4} \\
 0 & 0 & 0 & 0 & 0 & 0 & \Delta _{6,1} & \Delta _{6,2} & \Delta _{6,3} & \Delta _{6,4} \\
 \Delta _{1,1} & \Delta _{2,1} & \Delta _{3,1} & \Delta _{4,1} & -\Delta _{5,1} & -\Delta
   _{6,1} & 0 & 0 & 0 & 0 \\
 \Delta _{1,2} & \Delta _{2,2} & \Delta _{3,2} & \Delta _{4,2} & -\Delta _{5,2} & -\Delta
   _{6,2} & 0 & 0 & 0 & 0 \\
 -\Delta _{1,3} & -\Delta _{2,3} & -\Delta _{3,3} & -\Delta _{4,3} & \Delta _{5,3} & \Delta
   _{6,3} & 0 & 0 & 0 & 0 \\
 -\Delta _{1,4} & -\Delta _{2,4} & -\Delta _{3,4} & -\Delta _{4,4} & \Delta _{5,4} & \Delta
   _{6,4} & 0 & 0 & 0 & 0
\end{array}
\right) \nonumber\\
\end{eqnarray}
where the relation between the components of the tensor $\Delta^{a|I}$ and the fields $\phi$ parameterizing the Lax operator are displayed below:
\begin{equation}\label{fallus}
\begin{array}{lclcl}
 \Delta _{5,2} = \phi _5-\phi _{11} & ; &
 \Delta _{5,3} = \phi _6+\phi _{12} & ; &
 \Delta _{5,4} = \phi _7+\phi _{13} \\
 \Delta _{6,1} = \phi _5+\phi _{11} & ; &
 \Delta _{6,3} = \phi _8+\phi _{14} & ; &
 \Delta _{6,4} = \phi _9+\phi _{15} \\
 \Delta _{7,1} = \phi _{12}-\phi _6 & ; &
 \Delta _{7,2} = \phi _{14}-\phi _8 & ; &
 \Delta _{7,4} = \phi _{10}+\phi _{16} \\
 \Delta _{8,1} = \phi _{13}-\phi _7 & ; &
 \Delta _{8,2} = \phi _{15}-\phi _9 & ; &
 \Delta _{8,3} = \phi _{10}-\phi _{16} \\
 \Delta _{1,1} = \sqrt{2} \phi _{17} & ; &
 \Delta _{1,2} = \sqrt{2} \phi _{21} & ; &
 \Delta _{1,3} = -\sqrt{2} \phi _{25} \\
 \Delta _{1,4} = -\sqrt{2} \phi _{29} & ; &
 \Delta _{2,1} = \sqrt{2} \phi _{18} & ; &
 \Delta _{2,2} = \sqrt{2} \phi _{22} \\
 \Delta _{2,3} = -\sqrt{2} \phi _{26} & ; &
 \Delta _{2,4} = -\sqrt{2} \phi _{30} & ; &
 \Delta _{3,1} = \sqrt{2} \phi _{19} \\
 \Delta _{3,2} = \sqrt{2} \phi _{23} & ; &
 \Delta _{3,3} = -\sqrt{2} \phi _{27}& ; &
 \Delta _{3,4} = -\sqrt{2} \phi _{31} \\
 \Delta _{4,1} = \sqrt{2} \phi _{20} & ; &
 \Delta _{4,2} = \sqrt{2} \phi _{24} & ; &
 \Delta _{4,3} = -\sqrt{2} \phi _{28} \\
 \Delta _{4,4} = -\sqrt{2} \phi _{32} & ; &
 \Delta _{5,1} = 2 \phi _1 & ; &
 \Delta _{6,2} = 2 \phi _2 \\
 \Delta _{7,3} = 2 \phi _3 & ; &
 \Delta _{8,4} = 2 \phi _4 & \null & \null\
\end{array}
\end{equation}
If we perform the same change of basis on the generic element of the Lie subalgebra $\mathbb{H}^\star$ displayed in eq. (\ref{omegone}) we obtain:
\begin{eqnarray}\label{goniata}
    \Omega \, \mathfrak{W} \, \Omega^T & = & \left(
\begin{array}{llllllllll}
 0 & t_1 & t_6 & t_{10} & t_{13} & t_{15} & 0 & 0 & 0 & 0 \\
 -t_1 & 0 & t_2 & t_7 & t_{11} & t_{14} & 0 & 0 & 0 & 0 \\
 -t_6 & -t_2 & 0 & t_3 & t_8 & t_{12} & 0 & 0 & 0 & 0 \\
 -t_{10} & -t_7 & -t_3 & 0 & t_4 & t_9 & 0 & 0 & 0 & 0 \\
 t_{13} & t_{11} & t_8 & t_4 & 0 & t_5 & 0 & 0 & 0 & 0 \\
 t_{15} & t_{14} & t_{12} & t_9 & -t_5 & 0 & 0 & 0 & 0 & 0 \\
 0 & 0 & 0 & 0 & 0 & 0 & 0 & \chi _1 & \chi _4 & \chi _6 \\
 0 & 0 & 0 & 0 & 0 & 0 & -\chi _1 & 0 & \chi _2 & \chi _5 \\
 0 & 0 & 0 & 0 & 0 & 0 & \chi _4 & \chi _2 & 0 & \chi _3 \\
 0 & 0 & 0 & 0 & 0 & 0 & \chi _6 & \chi _5 & -\chi _3 & 0
\end{array}
\right)\nonumber\\
\end{eqnarray}
where the relation between the standard matrix entries $t_i, \chi_i$ and the original parameters $\omega_i,\rho_i$ is displayed below for the case $s=1$ chosen for illustration.
\begin{equation}\label{convertino}
\begin{array}{lclcl}
 \omega _1 = t_3+\chi _1 &;&
 \omega _2 = t_8+\chi _4 &;&
 \omega _3 = t_{12}+\chi _6 \\
 \omega _4 = t_4+\chi _2 &;&
 \omega _5 = t_9+\chi _5 &;&
 \omega _6 = t_5+\chi _3 \\
 \omega _7 = t_3-\chi _1 &;&
 \omega _8 = \chi _4-t_8 &;&
 \omega _9 = \chi _6-t_{12} \\
 \omega _{10} = \chi _2-t_4 &;&
 \omega _{11} = \chi _5-t_9 &;&
 \omega _{12} = \chi _3-t_5 \\
 \omega _{13} = -\sqrt{2} t_6 &;&
 \omega _{14} = -\sqrt{2} t_2 &;&
 \omega _{15} = -\sqrt{2} t_{10} \\
 \omega _{16} = -\sqrt{2} t_7 &;&
 \omega _{17} = \sqrt{2} t_{13} &;&
 \omega _{18} = \sqrt{2} t_{11} \\
 \omega _{19} = \sqrt{2} t_{15} &;&
 \omega _{20} = \sqrt{2} t_{14} &;&
 \rho _{1} = \sqrt{2} t_1\\
\end{array}
\end{equation}
Having established the above vocabulary between the tensor notation and that intrinsic to the Cartan Weyl basis of the Lie algebra, one can  proceed to define a set of tensor classifiers that, hopefully might distinguish  different nilpotent orbits of Lax operators, just as it was the case in the $\mathfrak{g}_{(2,2)}$ model. In section \ref{tensoreanale} we will construct a rich set of such classifiers and we will measure them on the representatives of nilpotent orbits constructed with the Weyl group method. If we confine ourselves to a boolean analysis  (the tensor is zero = 0, the tensor does not vanish = 1) we  will show that the tensor classifiers are not able to separate all the orbits. A finer analysis of the invariants associated to these tensor structures is therefore required. This result suffices to answer in the negative the question whether the tensor methods might be alternative to the standard triple method, which therefore seems anavoidable.
\subsection{Choosing a Cartan subalgebra contained in $\mathbb{H}^\star$ and diagonalization of its adjoint action}
The first step necessary to implement the standard triple method of nilpotent orbit construction  consists of selecting a new CSA inside the $\mathbb{H}^\star$ subalgebra and of diagonalizing its adjoint action both on the subspace $\mathbb{K}$ and on the subalgebra $\mathbb{H}^\star$. Obviously the eigenvalues cannot be anything else but the roots of the abstract Lie algebra and in this way we obtain new step operators $E[\alpha]$ which either belong to $\mathbb{K}$ or to $\mathbb{H}^\star$. There is however a caveat that has to be taken into account. What we want to diagonalize is not the full Cartan subalgebra but only its non compact part. This means that the relevant roots are only the universal ones of the Tits-Satake projection and the corresponding eigenspaces will have a multiplicity related to the paint group. The very nice and deep result is that the pattern of this decomposition is universal and emphasizes the concept of universality classes associated with the Tits-Satake projection. What happens is the following. Each root-space has either dimensionality $1$ or dimensionality $2s$ and which is the case depends only on the root and not on the value of $s$. The combinations corresponding to dimensionality $1$ are universal, while for those roots with multiplicity $2s$ the entire eigenspace transforms in the irreducible vector representation of the paint group $\mathrm{SO}(2s)$.
\par
As new Cartan subalgebra, referring to the standard form (\ref{goniata}) we take the following ones:
\begin{eqnarray}
  \mathcal{C}_1 &=& \frac{\partial}{\partial\chi_4} \mathfrak{W} \\
  \mathcal{C}_2 &=& \frac{\partial}{\partial\chi_5} \mathfrak{W}  \\
  \mathcal{C}_3 &=& \frac{\partial}{\partial t_{8+s}} \mathfrak{W}  \\
  \mathcal{C}_4 &=& \frac{\partial}{\partial t_{9+s}} \mathfrak{W} \label{newCSA}
\end{eqnarray}
where it is understood that in the operator $\mathfrak{W}$ defined in eq.(\ref{omegone}) the inverse of the substitution (\ref{convertino}) has been made. Diagonalizing the adjoint action of this Cartan subalgebra on the $\mathbb{K}$ space we find that the following 10 roots (and their negatives) are represented in the tangent space:
\begin{equation}\label{frantoio}
    \begin{array}{lllll}
 1 & \alpha _2 & \mbox{multiplicity} & = & 1 \\
 2 & \alpha _1+\alpha _2 & \mbox{multiplicity}& = & 1 \\
 3 & \alpha _2+\alpha _3 & \mbox{multiplicity} & = & 1 \\
 4 & \alpha _1+\alpha _2+\alpha _3 & \mbox{multiplicity} & = & 1 \\
 5 & \alpha _2+\alpha _3+\alpha _4 & \mbox{multiplicity} & = & 2s \\
 6 & \alpha _1+\alpha _2+\alpha _3+\alpha _4 & \mbox{multiplicity} & = & 2s \\
 7 & \alpha _2+\alpha _3+2 \alpha _4 & \mbox{multiplicity} & = & 1 \\
 8 & \alpha _1+\alpha _2+\alpha _3+2 \alpha _4 & \mbox{multiplicity} & = & 1 \\
 9 & \alpha _2+2 \alpha _3+2 \alpha _4 & \mbox{multiplicity} & = & 1 \\
 10 & \alpha _1+\alpha _2+2 \alpha _3+2 \alpha _4 & \mbox{multiplicity} & = & 1
\end{array}
\end{equation}
while the remaining six (and their negatives) are represented in the $\mathbb{H}^\star$ subalgebra:
\begin{equation}\label{olla}
    \begin{array}{lllll}
 1 & \alpha _1 & \mbox{multiplicity} & = & 1 \\
 2 & \alpha _3 & \mbox{multiplicity} & = & 1 \\
 3 & \alpha _4 & \mbox{multiplicity} & = & 2s \\
 4 & \alpha _3+\alpha _4 & \mbox{multiplicity} & = & 2s \\
 5 & \alpha _3+2 \alpha _4 & \mbox{multiplicity} & = & 1 \\
 6 & \alpha _1+2 \alpha _2+2 \alpha _3+2 \alpha _4 & \mbox{multiplicity} & = & 1
\end{array}
\end{equation}
The explicit linear combination of $\mathfrak{K}_A$ or $\mathfrak{H}_a$ operators that form the eigenspaces corresponding to the various positive and negative roots are now listed. In this linear combinations we introduce a parameter $\sigma_{a,i}$ for the positive roots and a parameter $\tau_{a,i}$ for the negative ones that takes into account the multiplicity.
\par
For the positive root eigenspaces belonging to $\mathbb{K}$ we find the following result:
\begin{equation}\label{positrone}
\begin{array}{lll}
 E\left[\alpha _2\right] & = &
   \left(\mathfrak{K}_5-\mathfrak{K}_7+\mathfrak{K}_8+\mathfrak{K}_{10}-\mathfrak{K}_{11}-\mathfrak{K}_{13}-\mathfrak{K}_{14}+\mathfrak{K}_{16}\right) \sigma _{1,1} \\
 E\left[\alpha _1+\alpha _2\right] & = &
   \left(-\frac{\mathfrak{K}_1}{\sqrt{2}}-\frac{\mathfrak{K}_3}{\sqrt{2}}+\mathfrak{K}_{12}\right) \sigma _{2,1} \\
 E\left[\alpha _2+\alpha _3\right] & = &
   \left(-\frac{\mathfrak{K}_2}{\sqrt{2}}-\frac{\mathfrak{K}_4}{\sqrt{2}}+\mathfrak{K}_{15}\right) \sigma _{3,1} \\
 E\left[\alpha _1+\alpha _2+\alpha _3\right] & = &
   \left(-\mathfrak{K}_5-\mathfrak{K}_7+\mathfrak{K}_8-\mathfrak{K}_{10}-\mathfrak{K}_{11}+\mathfrak{K}_{13}+\mathfrak{K}_{14}+\mathfrak{K}_{16}\right) \sigma _{4,1} \\
 E\left[\alpha _2+\alpha _3+\alpha _4\right] & = &
   \left(\mathfrak{K}_{20}+\mathfrak{K}_{24}\right) \sigma _{5,1}+\left(\mathfrak{K}_{19}+\mathfrak{K}_{23}\right) \sigma _{5,2}
   \\
 E\left[\alpha _1+\alpha _2+\alpha _3+\alpha _4\right] & = &
   \left(\mathfrak{K}_{18}+\mathfrak{K}_{22}\right) \sigma _{6,1}+\left(\mathfrak{K}_{17}+\mathfrak{K}_{21}\right) \sigma _{6,2}
   \\
 E\left[\alpha _2+\alpha _3+2 \alpha _4\right] & = &
   \left(-\frac{\mathfrak{K}_2}{\sqrt{2}}+\frac{\mathfrak{K}_4}{\sqrt{2}}+\mathfrak{K}_9\right) \sigma _{7,1} \\
 E\left[\alpha _1+\alpha _2+\alpha _3+2 \alpha _4\right] & = &
   \left(\mathfrak{K}_5-\mathfrak{K}_7-\mathfrak{K}_8-\mathfrak{K}_{10}+\mathfrak{K}_{11}+\mathfrak{K}_{13}-\mathfrak{K}_{14}+\mathfrak{K}_{16}\right) \sigma _{8,1} \\
 E\left[\alpha _2+2 \alpha _3+2 \alpha _4\right] & = &
   \left(-\mathfrak{K}_5+\mathfrak{K}_7+\mathfrak{K}_8+\mathfrak{K}_{10}+\mathfrak{K}_{11}+\mathfrak{K}_{13}-\mathfrak{K}_{14}+\mathfrak{K}_{16}\right) \sigma _{9,1} \\
 E\left[\alpha _1+\alpha _2+2 \alpha _3+2 \alpha _4\right] & = &
   \left(-\frac{\mathfrak{K}_1}{\sqrt{2}}+\frac{\mathfrak{K}_3}{\sqrt{2}}+\mathfrak{K}_6\right) \sigma _{10,1}
\end{array}
\end{equation}
while for the negative ones belonging to the same space we get:
\begin{equation}\label{negatrone}
    \begin{array}{lll}
 E\left[-\alpha _2\right] & = &
   \left(\mathfrak{K}_5+\mathfrak{K}_7-\mathfrak{K}_8+\mathfrak{K}_{10}-\mathfrak{K}_{11}+\mathfrak{K}_{13}+\mathfrak{K}_{14}+\mathfrak{K}_{16}\right) \tau _{1,1} \\
 E\left[-\alpha _1-\alpha _2\right] & = &
   \left(\frac{\mathfrak{K}_1}{\sqrt{2}}+\frac{\mathfrak{K}_3}{\sqrt{2}}+\mathfrak{K}_{12}\right) \tau _{2,1} \\
 E\left[-\alpha _2-\alpha _3\right] & = &
   \left(\frac{\mathfrak{K}_2}{\sqrt{2}}+\frac{\mathfrak{K}_4}{\sqrt{2}}+\mathfrak{K}_{15}\right) \tau _{3,1} \\
 E\left[-\alpha _1-\alpha _2-\alpha _3\right] & = &
   -\left(\mathfrak{K}_5-\mathfrak{K}_7+\mathfrak{K}_8+\mathfrak{K}_{10}+\mathfrak{K}_{11}+\mathfrak{K}_{13}+\mathfrak{K}_{14}-\mathfrak{K}_{16}\right) \tau _{4,1} \\
 E\left[-\alpha _2-\alpha _3-\alpha _4\right] & = &
   \left(\mathfrak{K}_{24}-\mathfrak{K}_{20}\right) \tau _{5,1}+\left(\mathfrak{K}_{23}-\mathfrak{K}_{19}\right) \tau _{5,2} \\
 E\left[-\alpha _1-\alpha _2-\alpha _3-\alpha _4\right] & = &
   \left(\mathfrak{K}_{22}-\mathfrak{K}_{18}\right) \tau _{6,1}+\left(\mathfrak{K}_{21}-\mathfrak{K}_{17}\right) \tau _{6,2} \\
 E\left[-\alpha _2-\alpha _3-2 \alpha _4\right] & = &
   \left(\frac{\mathfrak{K}_2}{\sqrt{2}}-\frac{\mathfrak{K}_4}{\sqrt{2}}+\mathfrak{K}_9\right) \tau _{7,1} \\
 E\left[-\alpha _1-\alpha _2-\alpha _3-2 \alpha _4\right] & = &
   \left(\mathfrak{K}_5+\mathfrak{K}_7+\mathfrak{K}_8-\mathfrak{K}_{10}+\mathfrak{K}_{11}-\mathfrak{K}_{13}+\mathfrak{K}_{14}+\mathfrak{K}_{16}\right) \tau _{8,1} \\
 E\left[-\alpha _2-2 \alpha _3-2 \alpha _4\right] & = &
   \left(-\mathfrak{K}_5-\mathfrak{K}_7-\mathfrak{K}_8+\mathfrak{K}_{10}+\mathfrak{K}_{11}-\mathfrak{K}_{13}+\mathfrak{K}_{14}+\mathfrak{K}_{16}\right) \tau _{9,1} \\
 E\left[-\alpha _1-\alpha _2-2 \alpha _3-2 \alpha _4\right] & = &
   \left(\frac{\mathfrak{K}_1}{\sqrt{2}}-\frac{\mathfrak{K}_3}{\sqrt{2}}+\mathfrak{K}_6\right) \tau _{10,1}
\end{array}
\end{equation}
For the root eigenspaces belonging to the $\mathbb{H}^\star$ subalgebra we find instead the following results. For the positive roots we have:
\begin{equation}\label{hpositon}
    \begin{array}{lll}
 E\left[\alpha _1\right] & = &
   \left(\mathfrak{H}_1+\mathfrak{H}_3+\mathfrak{H}_4+\mathfrak{H}_6-\mathfrak{H}_7+\mathfrak{H}_9+\mathfrak{H}_{10}+\mathfrak{H}_{12}\right) \mu _{1,1} \\
 E\left[\alpha _3\right] & = &
   \left(-\mathfrak{H}_1-\mathfrak{H}_3-\mathfrak{H}_4-\mathfrak{H}_6-\mathfrak{H}_7+\mathfrak{H}_9+\mathfrak{H}_{10}+\mathfrak{H}_{12}\right) \mu _{2,1} \\
 E\left[\alpha _4\right] & = & \left(\mathfrak{H}_{16}+\mathfrak{H}_{20}\right) \mu
   _{3,1}+\left(\mathfrak{H}_{15}+\mathfrak{H}_{19}\right) \mu _{3,2} \\
 E\left[\alpha _3+\alpha _4\right] & = & \left(\mathfrak{H}_{14}+\mathfrak{H}_{18}\right) \mu
   _{4,1}+\left(\mathfrak{H}_{13}+\mathfrak{H}_{17}\right) \mu _{4,2} \\
 E\left[\alpha _3+2 \alpha _4\right] & = &
   \left(\mathfrak{H}_1-\mathfrak{H}_3+\mathfrak{H}_4-\mathfrak{H}_6+\mathfrak{H}_7+\mathfrak{H}_9-\mathfrak{H}_{10}+\mathfrak{H}_{12}\right) \mu _{5,1} \\
 E\left[\alpha _1+2 \alpha _2+2 \alpha _3+2 \alpha _4\right] & = &
   \left(-\mathfrak{H}_1+\mathfrak{H}_3-\mathfrak{H}_4+\mathfrak{H}_6+\mathfrak{H}_7+\mathfrak{H}_9-\mathfrak{H}_{10}+\mathfrak{H}_{12}\right) \mu _{6,1}
\end{array}
\end{equation}
while for the negative ones we get:
\begin{equation}\label{hnegaton}
    \begin{array}{lll}
 E\left[-\alpha _1\right] & = &
   \left(\mathfrak{H}_1-\mathfrak{H}_3-\mathfrak{H}_4+\mathfrak{H}_6-\mathfrak{H}_7-\mathfrak{H}_9-\mathfrak{H}_{10}+\mathfrak{H}_{12}\right) \lambda _{1,1} \\
 E\left[-\alpha _3\right] & = &
   \left(-\mathfrak{H}_1+\mathfrak{H}_3+\mathfrak{H}_4-\mathfrak{H}_6-\mathfrak{H}_7-\mathfrak{H}_9-\mathfrak{H}_{10}+\mathfrak{H}_{12}\right) \lambda _{2,1} \\
 E\left[-\alpha _4\right] & = & \left(\mathfrak{H}_{20}-\mathfrak{H}_{16}\right) \lambda
   _{3,1}+\left(\mathfrak{H}_{19}-\mathfrak{H}_{15}\right) \lambda _{3,2} \\
 E\left[-\alpha _3-\alpha _4\right] & = & \left(\mathfrak{H}_{18}-\mathfrak{H}_{14}\right)
   \lambda _{4,1}+\left(\mathfrak{H}_{17}-\mathfrak{H}_{13}\right) \lambda _{4,2} \\
 E\left[-\alpha _3-2 \alpha _4\right] & = &
   \left(\mathfrak{H}_1+\mathfrak{H}_3-\mathfrak{H}_4-\mathfrak{H}_6+\mathfrak{H}_7-\mathfrak{H}_9+\mathfrak{H}_{10}+\mathfrak{H}_{12}\right) \lambda _{5,1} \\
 E\left[-\alpha _1-2 \alpha _2-2 \alpha _3-2 \alpha _4\right] & = &
   \left(-\mathfrak{H}_1-\mathfrak{H}_3+\mathfrak{H}_4+\mathfrak{H}_6+\mathfrak{H}_7-\mathfrak{H}_9+\mathfrak{H}_{10}+\mathfrak{H}_{12}\right) \lambda _{6,1}
\end{array}
\end{equation}
In the above formulae we used as an illustration the case $s=1$. Yet, as we already stressed, the linear combinations are universal for the roots with multiplicity $1$ while they are simply prolonged with more terms for the roots with multiplicity $2s$.
\section{Nilpotent orbits for the coset manifolds $\mathcal{QM}^\star_{(4,4+2s)}$ }
 In order to implement the  algorithm described in section \ref{praticone} to the case under consideration, the first step to be fulfilled is the determination of the Weyl group for the $B_4$ root system. We have:
 \begin{equation}\label{weylusb4}
    \mathcal{W} \, = \,   S_4 \,  \ltimes \,\mathbb{Z}_2^{\phantom{2}4} \quad \Rightarrow \quad  |\mathcal{W}| \, = \,384
 \end{equation}
where $S_4$ denotes the permutation group of four objects. The semidirect structure of this Weyl group is best described by spelling out its action on a four component euclidian vector:
\begin{equation}\label{describo}
   \left \{x_1,\, x_2, \, x_3, \, x_4\right \}
\end{equation}
For each of the $24$ permutations of the symmetric subgroup
\begin{equation}\label{fornicino}
\forall \mathfrak{p} \, \in \, S_4 \quad \Leftrightarrow \quad\mathfrak{p}
\, = \,\left ( \begin{array}{cccc}
1 & 2 & 3 & 4 \\
\downarrow & \downarrow & \downarrow & \downarrow \\
P(1) & P(2) & P(3) & P(4)
\end{array}\right)
\end{equation}
the  action on the euclidian vector is the corresponding permutation of its entries:
\begin{equation}
\mathfrak{p} \, \left \{x_1,\, x_2, \, x_3, \, x_4\right \} \, = \, \left \{x_{P(1)},\, x_{P(2)}, \, x_{P(3)}, \, x_{P(4)}\right \}
\label{fornicino2}
\end{equation}
The four $\mathbb{Z}_2$ subgroups, act instead as flips of sign of the four entries of the euclidian vector:
\begin{equation}\label{ciangiasegno}
    \mathbb{Z}_2^{\phantom{2}4}\, \quad : \quad \, \left \{x_1,\, x_2, \, x_3, \, x_4\right \} \, \Rightarrow \, \left \{\pm x_1,\, \pm x_2, \, \pm x_3 \, \pm x_4\right \}
\end{equation}
Considering the root system composed of the 32 euclidian vectors (\ref{fronte}), we easily verify that it is invariant under the action of the above defined group which is indeed generated by the reflections along all the roots.
\par
The next step is the determination of the subgroup $\mathcal{W}_H \subset \mathcal{W} $ which respects the splitting of the $32$ root system into the order $20$ subset composed by the $\mathbb{K}$-type roots (\ref{frantoio}) plus their negatives and the order $12$ subset composed by
the $\mathbb{H}^\star$-type roots (\ref{olla}) plus their negatives. The answer is very simple. By looking at the explicit components of the vectors belonging to the two sets one easily realizes that the searched for subgroup is:
\begin{equation}\label{weylsubbo}
    \mathcal{W}_H \, = \, \left [  S_2 \otimes \, S_2 \right]\, \ltimes \,\mathbb{Z}_2^{\phantom{2}4}
\end{equation}
The action of the $\mathbb{Z}_2^{\phantom{2}4}$ factor on the vector $\left \{x_1,\, x_2, \, x_3, \, x_4\right \}$ is obviously the same as in eq.(\ref{fornicino2}), while reduction to the subgroup $ S_2 \otimes \, S_2 \subset S_4$ means that we confine ourselves to the following four permutations:
\begin{equation}\label{fradicio}
    \begin{array}{cc}
       \left \{x_1,\, x_2, \, x_3, \, x_4\right \} & \left \{x_2,\, x_1, \, x_3, \, x_4\right \} \\
       \left \{x_1,\, x_2, \, x_4, \, x_3\right \} & \left \{x_2,\, x_1, \, x_4, \, x_3\right \}
     \end{array}
\end{equation}
\par
Since $384/64 = 6$ we expect that the Weyl group splits into $6$ lateral classes $g \cdot \mathcal{W}_H$. An easy way of choosing a standard representative for each lateral class is provided by mentioning its action on the vector $\left \{x_1,\, x_2, \, x_3, \, x_4\right \}$. Then the six classes can be described as follows:
\begin{equation}\label{Whclasses}
\begin{array}{lcllll}
g_1\cdot \mathcal{W}_H&\simeq& \left\{ x_1, \right. & x_2, & x_3, &\left. x_4 \right \}\\
g_2 \cdot \mathcal{W}_H& \simeq&\left\{ x_1,\right. & x_3, & x_2, &\left. x_4 \right \}\\
g_3 \cdot \mathcal{W}_H& \simeq& \left\{x_1 ,\right.& x_4, & x_2, &\left. x_3 \right \}\\
g_4 \cdot \mathcal{W}_H& \simeq&\left\{ x_2, \right.& x_3, & x_1, &\left. x_4 \right \}\\
g_5 \cdot \mathcal{W}_H& \simeq&\left\{ x_2, \right.& x_4, & x_1, &\left. x_3 \right \}\\
g_6\cdot \mathcal{W}_H& \simeq& \left\{x_3, \right.& x_4, & x_1, &\left. x_2\right \}
\end{array}
\end{equation}
\par
As for the generalized Weyl group $\mathcal{GW}$ it contains $3072$ elements. The normal subgroup $\mathcal{HW}$ which, by definition, stabilizes each element $h$ of the Cartan subalgebra, is of order $8$ and has the structure $\mathcal{HW} \sim \mathbb{Z}_2 \times \mathbb{Z}_2 \times \mathbb{Z}_2$. It is obviously impossible to present here all the elements of  $\mathcal{GW}$, but they are easily constructed by means of a computer programme.
\par
The third step in our construction is the determination of the possible branching laws of the fundamental representation of the Tits Satake
subalgebra $\mathbb{U}_{\mathrm{TS}}\, \equiv \, \mathbb{G}_R \, = \, \so(4,5)$ into irreducible representations of $\slal(2)$. As we already observed, this problem is equivalent to the problem of finding the partitions of an integer into integers.
\par
The dimension of the fundamental representations of $\so(4,5)$ is obviously $9$. Hence the possible embeddings  $\slal(2)\hookrightarrow\so(4,5)$ are associated with the partitions of $9$ into integers. These latter are thirty and precisely the following ones:
\begin{eqnarray}\label{filinini}
\mathfrak{P}[9] & = &\left [\{9\}, \{8, 1\}, \{7, 2\}, \{7, 1^2\}, \{6, 3\}, \{6, 2, 1\}, \{6, 1^3\}, \{5, 4\}, \{5,
    3, 1\}, \right.\nonumber\\
&&\left.\{5, 2^2\}, \{5, 2, 1^2\}, \{5, 1^4\}, \{4^2, 1\}, \{4, 3, 2\}, \{4,
     3, 1^2\}, \right.\nonumber\\
&& \left.\{4, 2^2, 1\}, \{4, 2, 1^3\}, \{4, 1^4\}, \{3^3\}, \{3^2, 2, 1\}, \{3^2, 1^3\}, \{3, 2^3\}, \{3, 2^2, 1^2\},\right.\nonumber\\
&&\left. \{3, 2,1^4\}, \{3, 1^6\}, \{2^4, 1\}, \{2^3,  1^3,\}, \{2^2,  1^5\}, \{2, 1^7\}, \{1^9\}\right]
\end{eqnarray}
The main simplifying information that we take from mathematical books is that for the algebras $\so(2p+1)$ we have to consider only those partitions where \textit{each even addend appears an even number of times}.
\par
Such a restriction deletes nineteen of the thirty partitions. Furthermore the partitions made only of $1$.s is to be excluded
 because it means no $\slal(2)$ embedding and therefore no standard triple. This leaves with the following twelve partitions:
\begin{eqnarray}\label{filinucci}
\widehat{\mathfrak{P}}[9] & = &\left [\{9\}, \{7, 1^2\},  \{5,
    3, 1\}, \{5, 2^2\}, \{5, 1^4\}, \{4^2, 1\},    \{3^3\}, \right.\nonumber\\
     &&\left.\{3^2, 1^3\},  \{3, 2^2, 1^2\}, \{3, 1^6\}, \{2^4, 1\},  \{2^2,  1^5\}\right]
\end{eqnarray}
Given the set of partitions for each of them we know the possible eigenvalues of the $h$-element of the standard triple which we put into the Cartan subalgebra. Under the action of the Weyl group, each choice of the eigenvalues generates a Weyl orbit of such $h$-operators which contains $384$ elements.
\par
By $a$-label we just denote the integer valued four vector:
\begin{equation}\label{franceschio}
    a-\mbox{label} \ \, \equiv \, \{\alpha_1(h),\alpha_2(h),\alpha_3(h),\alpha_4(h)\} \, = \,\{a_1,a_2,a_3,a_4\}
\end{equation}
where $\alpha_i$ are the simple roots.
\par
Since $|\mathcal{W}_H|=64$ every Weyl orbit splits into six $\mathcal{W}_H$-suborbits of $64$ elements each, corresponding to the six equivalence classes of the coset $\mathcal{W}/\mathcal{W}_H$ (see eq.(\ref{Whclasses})). For some choices of the eigenvalues the number of distinct elements in the various $\mathcal{W}_H$-suborbits can be less than $64$ if a subgroup of $\mathcal{W}_H$ leaves that particular four vector  $\{a_1,a_2,a_3,a_4\}$ invariant.
\par
Having grouped in this way the $a$-labels of the  central elements $h$ for the candidate standard triples $\{h,x,y\}$, by means of a computer programme we have verified in which $\mathcal{W}_H$-suborbits, the $x$-element can be constructed inside $\mathbb{K}$, by solving the two equations (\ref{pagnato1},\ref{pagnato2}).
\paragraph{The computer implemented algorithm}
The logical structure of our algorithm is the following. Given an $a$-label the computer considers the corresponding $h_a$ central element of the candidate triple and verifies whether equation (\ref{pagnato1}) can be solved in $\mathbb{K}$, calculating also the degeneracy of such a the solution. In other words the computer determines the eigenspace of $\mbox{adj}_\mathbb{K}(h_a)$, corresponding to eigenvalue $\lambda =1$:
\begin{eqnarray}\label{eigenx}
    M_1(h_a) & \equiv & \mbox{Eigenspace}\left[\lambda=1,\mbox{adj}_\mathbb{K}(h_a)\right] \nonumber\\
    dg[h_a]&\equiv& \mbox{dim} \, M_1(h_a)
\end{eqnarray}
The result for $dg[h_a]$ depends only on the chosen partition $\{j_1 , \dots , j_n\}$ and on the $\mathcal{W}_H$ class of the considered $a$-label. For all representatives inside the same $\mathcal{W}_H$ class the degeneracy of the eigenspace is the same. In case $dg[h_a]\, = \,0$ the entire $\mathcal{W}_H$ class of $a$-labels is discarded and the computer goes to the next. If $dg[h_a]\, > \,0$ the computer chooses a standard representative inside the
considered $\mathcal{W}_H$ class (which one is irrelevant) and goes to equation (\ref{pagnato2}). This latter is a set of $16$ quadratic equations for $dg[h_a]$ unknowns, the number $16$ being the dimension of $\so(2,2)\oplus\so(2,3)$, \textit{i.e.}  of $\mathbb{H}^\star_{\mathrm{TS}}$. It is clear that depending on the case there may be no solutions or several. In case there are no solutions the entire $\mathcal{W}_H$ class of $a$-labels is discarded and the computer goes to the next. In case solutions exist their multiplicity varies very much from case to case. Their set has to be organized in distinct orbits.
Similarly to the case of $\mathfrak{g}_{(2,2)}$ this is done by means of the $\beta$ labels, defined in eq.s(\ref{betlab1},\ref{betlab2}). As we know the set of available $\beta$-labels for each $h$ is equal to the number of available $\gamma$-labels appearing in the partition to which the considered $h$ belongs. Hence we can calculate a priori the set of $\gamma$-labels in each partition $\{j_1 , \dots , j_n\}$. Since there are six lateral classes of $\mathcal{W}/\mathcal{W}_H$ the maximal number of $\gamma\beta$-labels which can appear for each $\{j_1 , \dots , j_n\}$ is six but it might be less since for specific partitions several classes can coincide as we already noted. Furthermore we restrict our attention only to those $\gamma$-labels for which a solution for $x$ can be found. This means that the set of $\gamma\beta$-labels varies in length from partition to partition. In the next paragraph we discuss them and this discussion provides the means to emphasize the Tits Satake universality mechanism.
\paragraph{$\gamma$ -labels and the Tits Satake universality at work.}
In our case the dimension of the relevant subalgebras $\mathbb{H}$ or $\mathbb{H}^\star$ (which are equal)  is given by:
\begin{equation}
    \mbox {dim} \left(\begin{array}{c}\mathbb{H}\\
   \mathbb{H}^\star\\
   \end{array}\right)  \, = \underbrace{\, 12 \,}_{\# \, \mbox{of long roots } }
    \, + \,\underbrace{\, 8 \, s \,}_{ \# \, \mbox{of short roots with mult. $2s$}} \, + \, \underbrace{s(2 s -1) }_{\mbox{dim paint group}}
    \label{cucuruza}
\end{equation}
For the compact subalgebra the counting (\ref{cucuruza}) is easily understood. The compact generators are obtained in the form:
  $  E^{\alpha} - E^{-\alpha}$ for all available roots, long or short (counted with their multiplicity). In addition one has to add the generators of the paint group. Since $\mathbb{H}$ or $\mathbb{H}^\star$ are different real forms of the same complex Lie algebra the same counting applies also to $\mathbb{H}^\star$.
\par
We can now take a generic element of the Cartan subalgebra chosen in $\mathbb{H}^\star$:
\begin{equation}\label{genericus}
    h \, \in \, \mathrm{CSA} \, \subset \, \mathbb{H}^\star
\end{equation}
and calculate the spectrum of its adjoint action on $\mathbb{H}^\star$, which, according to eq.(\ref{gamlab}) is the definition of $\gamma$-labels.
For the entire class of Lie Algebras $\so(4,4+2s)$ we obtain the following universal result:
\begin{equation}
    \mbox{Spectrum}\left[\mbox{adj}_{\mathbb{H}^\star}(h) \right] \, = \, \left \{ \begin{array}{l}
                                                                                    \left[0 \right]_{s (2 s-1)+4} \\
                                                                                     \left[(\pm \alpha_1(h)\right ]_{1} \\
                                                                                    \left[\pm \alpha_3(h)\right ])_{1} \\
                                                                                    \left[\pm \alpha_4(h)\right ]_{2 s} \\
                                                                                     \left[\pm \left(\alpha_3(h)\, + \, \alpha_4(h)\right)\right ]_{2 s} \\
                                                                                    \left[\pm \left(\alpha_3(h)\, + \, 2\alpha_4(h)\right)\right ]_{1}\\
                                                                                     \left[\pm \left(\alpha_1(h)\, + \, 2\alpha_2(h)\, + \, 2\alpha_3(h)\, + \, 2\alpha_4(h)
                                                                                    \right) \right ]_{1}\\ \\
                                                                                  \end{array}
    \right\}\label{univgamma}
\end{equation}
where $\alpha_i(h)$ denotes the value of the $i$-th simple root of the Tits Satake subalgebra $\so(4,5)$ on the chosen $h$. The subscript in the symbol of the eigenvalues denotes their multiplicity and one easily verifies the sum rule:
\begin{equation}\label{summinorullo}
    s (2 s-1)+4 + 2\times 1 + 2\times 1 + 2\times 2s + 2\times 2s + 2\times 1 +2\times 1 \,=\, 12 +8s + s (2 s-1)
\end{equation}
Inspection of the result (\ref{univgamma}) reveals its rational. Apart from $0$ the non vanishing eigenvalues correspond to the subset of $6$ roots of the Tits-Satake system $B_4$ that appear in the $\mathbb{H}^\star$ subalgebra when we diagonalize the Cartan Subalgebra chosen  $\mathbb{H}^\star$ (compare with eq.(\ref{olla}) ). The degeneracy of the non-vanishing eigenvalues is just the multiplicity of the roots. The structure of the spectrum shows that the rank of the adjoint matrix $\mbox{adj}_{\mathbb{H}^\star}(h)$ is always at most $8+8s$, the paint group taking no part in the deal.
\par
In table \ref{gammatabba} we have listed  the gamma labels found for the $12$ partitions. Note that in order to obtain integer rather then half-integer eigenvalues we have listed the $\gamma$-labels of $2 h_a$ rather than $h_a$. Note also that we have listed only those $\gamma$-label for which a solution for $x$ in $\mathbb{K}$ can be found. The presented table clarifies that the concept of $\gamma$-labels coincides with that of lateral classes of the Weyl group $\mathcal{W}$ with respect to the stability subgroup $\mathcal{W}_H$.
\begin{table}
\begin{center}
{\scriptsize
$$
\begin{array}{||l|l|l|l||}
\hline
\hline
N & \alpha -\mbox{label} & \gamma\beta -\mbox{labels} &  \mathcal{W}_H -\mbox{classes}\\
\hline
\hline
1 & \mbox{[j=4]} &\gamma\beta_1 = \left\{0_{s (2 s-1)+4},\pm 8_1,\pm 4_1,\pm
   4_{2 s},\pm 8_{2 s},\pm 12_1,\pm
   4_1\right\} & \left(\times,\times,\times,\times,\gamma_1,\times\right)\\
 \hline
2 &\mbox{[j=3]$\times $2[j=0]} & \begin{array}{lll}
                                  \gamma\beta_1&=&\left\{0_{s (2 s-1)+4},\pm 8_1,\pm 4_1,\pm
   0_{2 s},\pm 4_{2 s},\pm 4_1,\pm
   4_1\right\} \\
                                  \gamma\beta_2&=&\left\{0_{s (2 s-1)+4},\pm 4_1,\pm 4_1,\pm
   2_{2 s},\pm 6_{2 s},\pm 8_1,\pm
   4_1\right\}
                                 \end{array}
& \left(\times,\times,\gamma_1,\times,\gamma_2,\times\right)\\
 \hline
3 & \mbox{[j=2]$\times $2[j=1/2]} & \gamma\beta_1 =\left\{0_{s (2 s-1)+4},\pm 3_1,\pm 3_1,\pm
   1_{2 s},\pm 4_{2 s},\pm 5_1,\pm
   1_1\right\}& \left(\times,\times,\times,\gamma_1,\gamma_1,\times\right)\\
   \hline
4 & \mbox{2[j=3/2]$\times $[j=0])} &\gamma\beta_1 =  \left\{0_{s (2 s-1)+4},\pm 4_1,\pm 2_1,\pm
   1_{2 s},\pm 3_{2 s},\pm 4_1,\pm
   2_1\right\} & \left(\times,\gamma_1,\gamma_1,\gamma_1,\gamma_1,\times\right)\\
   \hline
5 & \mbox{3[j=1]} & \gamma\beta_1 = \left\{0_{s (2 s-1)+4},\pm 2_1,\pm 0_1,\pm
   2_{2 s},\pm 2_{2 s},\pm 4_1,\pm
   2_1\right\}& \left(\times,\times,\gamma_1,\times,\gamma_1,\times\right)\\
 \hline
6 & \mbox{[j=1]$\times $2[j=1/2]$\times $2[j=0]} & \begin{array}{lll}
                                                    \gamma\beta_1   & = & \left\{0_{s (2 s-1)+4},\pm 3_1,\pm 1_1,\pm
   0_{2 s},\pm 1_{2 s},\pm 1_1,\pm
   1_1\right\}\\
                                                    \gamma\beta_2  & = & \left\{0_{s (2 s-1)+4},\pm 1_1,\pm 1_1,\pm
   1_{2 s},\pm 2_{2 s},\pm 3_1,\pm
   1_1\right\}
                                                  \end{array}&\left(\gamma_1,\gamma_1,\times,\times,\gamma_2,\gamma_2\right)\\
   \hline
7 & \mbox{2[j=1]$\times $3[j=0]} & \begin{array}{lll}
                                    \gamma\beta_1& = & \left\{0_{s (2 s-1)+4},\pm 4_1,\pm 0_1,\pm
   0_{2 s},\pm 0_{2 s},\pm 0_1,\pm
   0_1\right\} \\
                                    \gamma\beta_2 & =&\left\{0_{s (2 s-1)+4},\pm 2_1,\pm 2_1,\pm
   0_{2 s},\pm 2_{2 s},\pm 2_1,\pm
   2_1\right\} \\
                                    \gamma\beta_3 & = & \left\{0_{s (2 s-1)+4},\pm 0_1,\pm 0_1,\pm
   2_{2 s},\pm 2_{2 s},\pm 4_1,\pm
   0_1\right\}
                                    \end{array}& \left(\gamma_1,\gamma_2,\gamma_2,\gamma_2,\gamma_2,\gamma_3\right)\\
   \hline
8 &\mbox{4[j=1/2]$\times $[j=0]} & \gamma\beta_1=\left\{0_{s (2 s-1)+4},\pm 2_1,\pm 0_1,\pm
   1_{2 s},\pm 1_{2 s},\pm 2_1,\pm
   0_1\right\}
  & \left(\gamma_1,\gamma_1,\gamma_1,\gamma_1,\gamma_1,\gamma_1\right)\\
  \hline
9 & \mbox{[j=1]$\times $6[j=0]} & \begin{array}{lll}
                                   \gamma\beta_1 & = & \left\{0_{s (2 s-1)+4},\pm 2_1,\pm 0_1,\pm
   0_{2 s},\pm 0_{2 s},\pm 0_1,\pm
   2_1\right\} \\
                                   \gamma\beta_2 & = & \left\{0_{s (2 s-1)+4},\pm 0_1,\pm 2_1,\pm
   0_{2 s},\pm 2_{2 s},\pm 2_1,\pm
   0_1\right\}
                                 \end{array}& \left(\gamma_1,\gamma_1,\gamma_1,\gamma_2,\gamma_2,\gamma_2\right)\\
   \hline
10 &\mbox{2[j=1/2]$\times $5[j=0]} &\gamma\beta_1 = \left\{0_{s (2 s-1)+4},\pm 1_1,\pm 1_1,\pm
   0_{2 s},\pm 1_{2 s},\pm 1_1,\pm
   1_1\right\}& \left(\times,\gamma_1,\gamma_1,\gamma_1,\gamma_1,\times\right)\\
 \hline
11 &\mbox{[j=2]$\times $[j=1]$\times $[j=0]} &\begin{array}{lll}
                                           \gamma\beta_1 & = & \left\{0_{s (2 s-1)+4},\pm 4_1,\pm 0_1,\pm
   2_{2 s},\pm 2_{2 s},\pm 4_1,\pm
   4_1\right\} \\
                                           \gamma\beta_2 & = &\left\{0_{s (2 s-1)+4},\pm 4_1,\pm 4_1,\pm
   0_{2 s},\pm 4_{2 s},\pm 4_1,\pm
   0_1\right\}\\
                                           \gamma\beta_3 & = &\left\{0_{s (2 s-1)+4},\pm 2_1,\pm 2_1,\pm
   2_{2 s},\pm 4_{2 s},\pm 6_1,\pm
   2_1\right\}\\
                                           \end{array} &\left(\times,\times,\gamma_1,\gamma_2,\gamma_3,\gamma_3\right)\\
   \hline
12 &\mbox{[j=2]$\times $4[j=0]} & \begin{array}{lll}
                                           \gamma\beta_1 & = & \left\{0_{s (2 s-1)+4},\pm 4_1,\pm 2_1,\pm
   0_{2 s},\pm 2_{2 s},\pm 2_1,\pm
   4_1\right\} \\
                                           \gamma\beta_2 & = & \left\{0_{s (2 s-1)+4},\pm 2_1,\pm 4_1,\pm
   0_{2 s},\pm 4_{2 s},\pm 4_1,\pm
   2_1\right\}
                                           \end{array}& \left(\times,\gamma_1,\gamma_1,\gamma_2,\gamma_2,\times\right)\\
   \hline
   \hline
\end{array}
$$
}
\caption{The set of $\gamma\beta$-labels available for each partition. In each partition we list only those $\gamma$-labels for which an $x$ completing the triple $\{h,x,x^T\}$ can be found in the subspace $\mathbb{K}$. The last column of the table lists the $\mathcal{W}_H$ lateral classes and specifies in which of them the listed $\gamma$-label are located. Repetition of the same $\gamma$-label  in more than one lateral class means that for the chosen partition those lateral classes coincide. The symbol $\times$ means that in that class no $x$ can be found in $\mathbb{K}$ that completes the triple.\label{gammatabba}}
\end{center}
\end{table}
Another important consequence of this analysis concerns the mechanism of  Tits-Satake universality classes. Since the $\beta$-labels coincide with the $\gamma$-ones it follows that they also have the universal structure displayed in eq.(\ref{univgamma}). This means that the adjoint matrix $\mbox{adj}_{\mathbb{H}}(x-y)$ has always rank less or equal to $8+8s$ and that the paint group plays no role. Said differently we always have:
\begin{equation}\label{carciofino}
 \emptyset \, =\,    \left\{\left(x\,-\,y\right) \right \}\bigcap \so(2s) \subset \mathbb{H}
\end{equation}
Clearly this is not the choice taken by mathematicians when they classify nilpotent orbits for the real form of the algebra $\so(4,4+2s)$. Starting from $\beta$-labels one would introduce many more possibilities where the compact generator $(x-y)$ has also legs on the paint group. The corresponding orbits, when found, depend on the specific value of $s$, yet most of them are irrelevant because their nilpotent operator $x$ is not entirely contained in $\mathbb{K}$. The procedure we have adopted deletes from the start all these irrelevant orbits and shows that the relevant ones are universal, depending only on the Tits Satake universality class of the considered coset manifold.
\subsection{The table of $37$ universal nilpotent orbits}
The complete result of our computed aided classification of nilpotent orbits  yields a final list of $37$ orbits that are reported  in table \ref{tablizzona}.
\begin{table}
\begin{center}
{\scriptsize
$$
\begin{array}{||l||c|l|l|c||}
\hline
\hline
N & d_n &\alpha -\mbox{label} & \gamma\beta -\mbox{labels} & \mbox{Orbits} \\
\hline
\hline
1 & 9 &\mbox{[j=4]} &\gamma\beta_1 =  \left\{\pm 0_{p_s},\pm 4_{2 s+2},\pm
   8_{2 s+1},\pm 12_1\right\}& \mathcal{O}_1^1\\
 \hline
2 &7  &\mbox{[j=3]$\times $2[j=0]} & \begin{array}{l}
                                  \gamma\beta_1=\left\{\pm 0_{2 s+p_s},\pm 4_{2 s+3},\pm
   8_1\right\}\\
                                  \gamma\beta_2=\left\{\pm 0_{p_s},\pm 2_{2 s},\pm
   4_3,\pm 6_{2 s},\pm 8_1\right\}
                                 \end{array}
&\begin{array}{l|cc}
  \null & \beta_1 & \beta_2 \\
  \hline
   \gamma_1 & \mathcal{O}^2_{1,1} & \mathcal{O}^2_{1,2} \\
   \gamma_2 & \mathcal{O}^2_{2,1} &\mathcal{O}^2_{2,2}
 \end{array}\\
 \hline
3 & 5 &\mbox{[j=2]$\times $2[j=1/2]} & \gamma\beta_1 = \left\{\pm 0_{p_s},\pm 1_{2 s+1},\pm
   3_2,\pm 4_{2 s},\pm 5_1\right\} &\mathcal{O}_1^3 \\
   \hline
4 & 4 &\mbox{2[j=3/2]$\times $[j=0])} &\gamma\beta_1 =  \left\{\pm 0_{p_s},\pm 1_{2 s},\pm
   2_2,\pm 3_{2 s},\pm 4_2\right\}&\mathcal{O}_1^4\\
   \hline
5 & 3 &\mbox{3[j=1]} & \gamma\beta_1 =\left\{\pm 0_{p_s+1},\pm 2_{4 s+2},\pm
   4_1\right\}&\mathcal{O}_1^5\\
 \hline
6 & 3 &\mbox{[j=1]$\times $2[j=1/2]$\times $2[j=0]} & \begin{array}{l}
                                                    \gamma\beta_1   = \left\{\pm 0_{2 s+p_s},\pm 1_{2 s+3},\pm
   3_1\right\}\\
                                                    \gamma\beta_2  = \left\{\pm 0_{p_s},\pm 1_{2 s+3},\pm
   2_{2 s},\pm 3_1\right\}
                                                  \end{array}&\begin{array}{l|cc}
  \null & \beta_1 & \beta_2 \\
  \hline
   \gamma_1 & \mathcal{O}^6_{1,1} & \mathcal{O}^6_{1,2} \\
   \gamma_2 & \mathcal{O}^6_{2,1} &\mathcal{O}^6_{2,2}
 \end{array}\\
   \hline
7 & 3 &\mbox{2[j=1]$\times $3[j=0]} & \begin{array}{l}
                                    \gamma\beta_1 = \left\{\pm 0_{4 s+p_s+3},\pm 4_1\right\} \\
                                    \gamma\beta_2 = \left\{\pm 0_{2 s+p_s},\pm 2_{2
   s+4}\right\} \\
                                    \gamma\beta_3 =  \left\{\pm 0_{p_s+3},\pm 2_{4 s},\pm
   4_1\right\}
                                  \end{array}&\begin{array}{l|ccc}
                                                 & \beta_1 & \beta_2& \beta_3 \\
                                                \hline
                                                \gamma_1 &\mathcal{ O}^7_{1,1}& \mathcal{ O}^7_{1,2} &\mathcal{ O}^7_{1,3} \\
                                                \gamma_2 & \mathcal{ O}^7_{2,1}& \mathcal{ O}^7_{2,2} & \mathcal{ O}^7_{2,3}\\
                                                \gamma_3 & \mathcal{ O}^7_{3,1} & \mathcal{ O}^7_{3,2} & \mathcal{ O}^7_{3,3}
                                              \end{array}
                                  \\
   \hline
8 &2 &\mbox{4[j=1/2]$\times $[j=0]} & \gamma\beta_1=\left\{\pm 0_{p_s+2},\pm 1_{4 s},\pm
   2_2\right\}
  &\mathcal{O}_1^8 \\
  \hline
9 & 3 &\mbox{[j=1]$\times $6[j=0]} & \begin{array}{l}
                                   \gamma\beta_1 =  \left\{\pm 0_{4 s+p_s+2},\pm 2_2\right\} \\
                                   \gamma\beta_2 = \left\{\pm 0_{2 s+p_s+2},\pm 2_{2
   s+2}\right\}
                                 \end{array}&\begin{array}{l|cc}
  \null & \beta_1 & \beta_2 \\
  \hline
   \gamma_1 & \mathcal{O}^6_{1,1} & \mathcal{O}^6_{1,2} \\
   \gamma_2 & \mathcal{O}^6_{2,1} &\mathcal{O}^6_{2,2}
 \end{array}\\
   \hline
10 &2  &\mbox{2[j=1/2]$\times $5[j=0]} &\gamma\beta_1 = \left\{\pm 0_{2 s+p_s},\pm 1_{2
   s+4}\right\} &\mathcal{O}_1^{10} \\
 \hline
11 &5  &\mbox{[j=2]$\times $[j=1]$\times $[j=0]} &\begin{array}{l}
                                                \gamma\beta_1 = \left\{\pm 0_{p_s+1},\pm 2_{4 s},\pm
   4_3\right\}\\
                                                \gamma\beta_2 = \left\{\pm 0_{2 s+p_s+1},\pm 4_{2
   s+3}\right\} \\
                                                \gamma\beta_3 = \left\{\pm 0_{p_s},\pm 2_{2 s+3},\pm
   4_{2 s},\pm 6_1\right\}
                                              \end{array}&\begin{array}{l|ccc}
                                                 & \beta_1 & \beta_2& \beta_3 \\
                                                \hline
                                                \gamma_1 &\mathcal{ O}^{11}_{1,1}& \mathcal{ O}^{11}_{1,2} &\times \\
                                                \gamma_2 & \mathcal{ O}^{11}_{2,1}& \times & \mathcal{ O}^{11}_{2,3}\\
                                                \gamma_3 & \times & \mathcal{ O}^{11}_{3,2} & \mathcal{ O}^{11}_{3,3}
                                              \end{array}\\
   \hline
12 &5  &\mbox{[j=2]$\times $4[j=0]} &\begin{array}{l}
                                           \gamma\beta_1 = \left\{\pm 0_{2 s+p_s},\pm 2_{2 s+2},\pm
   4_2\right\} \\
                                           \gamma\beta_2 = \left\{\pm 0_{2 s+p_s},\pm 2_2,\pm 4_{2
   s+2}\right\}
                                         \end{array}&\begin{array}{l|cc}
  \null & \beta_1 & \beta_2 \\
  \hline
   \gamma_1 & \mathcal{O}^{12}_{1,1} & \mathcal{O}^{12}_{1,2} \\
   \gamma_2 & \mathcal{O}^{12}_{2,1} &\mathcal{O}^{12}_{2,2}
 \end{array}\\
   \hline
   \hline
\end{array}
$$
}
\caption{The $37$ nilpotent orbits for the manifolds $\frac{\mathrm{SO(4,4+2s)}}{\mathrm{SO(2,2)}\times \mathrm{SO(2,2+2s)}}$ which all belong to the same Tits Satake universality class $\frac{\mathrm{SO(4,5)}}{\mathrm{SO(2,2)}\times \mathrm{SO(2,5)}}$.  The classification depends only on the universality class and it is presented according to $\beta$ and $\gamma$-labels. In the above table $p_s$ is a short hand for the following number
$p_s = s(2s -1)+4$. \label{tablizzona}}
\end{center}
\end{table}
In  this table we have mentioned the explicit form of the $\gamma\beta$-labels in a shortened notation with respect to the notation of table \ref{gammatabba} by grouping together the identical eigenvalues that come from different roots. The assignment of each $\gamma$-label to $\mathcal{W}/\mathcal{W}_H$ lateral classes (cosets)  is no longer mentioned since it was already displayed in table \ref{gammatabba}.
\par
There are three important observations that emerge by inspection of this table.
\begin{itemize}
  \item The first is that the degree of nilpotency of the $x$ operators is just the dimension of the highest spin representation contained in the partition, as it was   anticipated in eq.(\ref{banfo}).
  \item The second observation concerns the results for the partition $11$. There we find three $\gamma$-labels, yet when we assign $h$ to one of them we do not find three solutions for $x$ corresponding to the three available $\beta$-labels. For each $\gamma$ there are, in $\mathbb{K}$ only two $\beta$.s. This counter example is a warning that every time one has to check explicitly which of the available $\beta$-labels are actually populated for each choice of gamma.
      \item In a similar way to the $\mathfrak{g}_{(2,2)}$ case the partition that contains the BPS and non BPS regular black holes is the partition $2\times [j=1] \times 3 \times [j=0]$. The BPS black holes come from a certain $\gamma$-label while the non BPS ones come from the other $\gamma$-labels. This can be easily shown considering the generating (or seed) geodesic for regular extremal black holes constructed in  \cite{Bergshoeff:2008be}. Using this universal construction of the seed geodesic
          it is possible to show that the regular solutions correspond to the diagonal entries of the $\gamma\beta$-table for which $\gamma$-label$=$ $\beta$-label. Solutions within the off-diagonal orbits are characterized by the warp factor $e^{-U}$ vanishing at finite $\tau$, thus signalling a singularity. We shall illustrate this result in a forthcoming work.
\end{itemize}
The complete list of the $37$ nilpotent operators, which is the main result of our paper, is given in appendix \ref{puppo}.
\subsection{About orbit stability subalgebras}
Since our conclusion is that $\mathrm{H}^\star$-orbits of nilpotent operators are classified by the triplet of  labels $\alpha$, $\gamma$ and $\beta$ we find it convenient to utilize the same to classify the corresponding  subalgebra$/$subgroup of $ \mathbb{H}^\star\,/\,\mathrm{H}^\star$ introducing the  notation $\mathfrak{S}^\alpha_{\gamma,\beta} \, / \,{\mathcal{S}}^\alpha_{\gamma,\beta} $. Given the standard representative $\mathcal{O}^\alpha_{\gamma,\beta}$ of the considered nilpotent orbit, its stability subalgebra is defined as follows:
\begin{equation}\label{defisubbo}
    \mathfrak{h} \, \in \, \mathfrak{S}^\alpha_{\gamma,\beta} \, \subset \, \mathbb{H}^\star \quad \Leftrightarrow \quad
    \left [ \mathcal{O}^\alpha_{\gamma,\beta} \, , \, \mathfrak{h}\right ] \, = \, 0
\end{equation}
For any other representative of the same orbit $\mathcal{O}^{\alpha\prime}_{\gamma,\beta} \, = \, h \, \mathcal{O}^\alpha_{\gamma,\beta} \, h^{-1}$, where $h\in \mathrm{H}^\star$ is a group element of the denominator group, the stability subalgebra is  conjugate to that of the standard representative and therefore isomorphic to it:
\begin{equation}\label{falara}
    \mathfrak{S}^{\alpha\prime}_{\gamma,\beta} \, = \, h \,  \mathfrak{S}^\alpha_{\gamma,\beta}\, h^{-1}
\end{equation}
It follows that each orbit of nilpotent $\mathbb{K}$-operators is isomorphic to the coset manifold:
\begin{equation}\label{orbitonen}
    \mathcal{M}^\alpha_{\gamma,\beta}\, = \, \frac{\mathrm{H}^\star}{{\mathcal{S}}^\alpha_{\gamma,\beta} }
\end{equation}
and the dimension of the orbit $\mathcal{O}_{\left[j_1,\dots,j_n\right] }$ is just the dimension of that coset:
\begin{equation}\label{dimensionen}
    \mbox{dim} \,\mathcal{O}^\alpha_{\gamma,\beta} \, =\,\mbox{dim} \,\mathcal{M}^\alpha_{\gamma,\beta}
\end{equation}
For this reason it is of particular relevance to calculate the stability subalgebras of the various orbits and study their abstract structure. For the first largest orbits $\mathcal{O}^\alpha_{\gamma,\beta} $ this task is fairly simple, why for the smaller ones it becomes increasingly complicated and requires some attention. We have already obtained some partial results but the presentation of the complete result is postponed to a forthcoming publication \cite{fortocomo}.
\par
What we would like to anticipate here is the challenging implications of the above simple observations. The classification of the $37$ nilpotent orbits amounts to the classifications of a list of $37$ families of coset manifolds:
\begin{equation}\label{listacosetta}
    \mathcal{M}^\alpha_{\gamma,\beta}\, = \, \frac{\mathrm{SO(2,2)} \times \mathrm{SO(2,2+2s)}}{{\mathcal{S}}^\alpha_{\gamma,\beta} }
\end{equation}
which generically turn up to be neither symmetric nor reducible. Yet as a result of our construction each of them constitutes a dynamical system which, through the embedding in the father system, we are able to integrate. The consequences of this might be far reaching and open new perspectives on integrability per se, not only in relation with the classification of black-holes.
\section{Tensor classifiers  for the $\mathcal{QM}^\star_{(4,4+2s)}$  nilpotent orbits}
\label{tensoreanale}
In the case of the $\mathfrak{g}_{(2,2)}$ model we were able to separate all the classified orbits by means of tensor classifiers, whose signatures provide an equivalent way of classification. It is a natural question whether the same is true also in the more complicated case of the $\mathcal{QM}^\star_{(4,4+2s)}$ spaces. Tensor classifiers very similar to those of the $\mathfrak{g}_{(2,2)}$ case can be constructed also here, as we anticipated in \cite{noig22}, yet, as we will see, their pattern is not able to separate all the $37$ orbits. By means of them we achieve only a partial classification. Let us see that in some detail.
\subsection{Structure of the Tensor classifiers }
In section \ref{bumbu} we arranged the Lax operator into a double tensor $\Delta^{i|I}$ where the index $i$ takes four values and spans the fundamental defining vector representation of $\mathrm{SO(2,2)}$, while the index $I$ spans the fundamental vector representation of $\mathrm{SO(2,2+2s)}$. In order to define the tensor classifiers we have to split the vector representation of $\mathrm{SO(2,2)}$ as the tensor product of two fundamental representation of $\mathrm{SL(2)_1}$ and $\mathrm{SL(2)_2}$. This is done in the following way.
Consider a generic element of the $\so(2,2)$ Lie algebra of the form it appears in the block diagonal decomposition of $\mathbb{H}^\star$  as given in eq.(\ref{goniata}), namely:
\begin{equation}
\label{cucaracia}
    \mathfrak{a} \, = \, \left(
\begin{array}{llll}
 0 & \chi _1 & \chi _4 & \chi _6 \\
 -\chi _1 & 0 & \chi _2 & \chi _5 \\
 \chi _4 & \chi _2 & 0 & \chi _3 \\
 \chi _6 & \chi _5 & -\chi _3 & 0
\end{array}
\right)
\end{equation}
its splitting in the two standard $\slal(2)$ Lie algebras is performed by setting:
\begin{equation}\label{decompi}
    \mathfrak{a} \, = \, \sum_{i=1}^2 \left( \gamma_{i,0} L_0^i + \gamma_{i,1} L_+^i+ \gamma_{i,-1} L_{-}^i\right)
\end{equation}
where the standard generators of the two $\slal(2)$ Lie algebra are normalized as follows:
\begin{equation}\label{algella}
    \left [ L_0^i \, , \, L_{\pm }^i \right ] \, = \, \pm L_{\pm }^i \quad ; \quad \left [ L_{+}^i \, , \, L_{- }^i \right ] \, = \, 2 L_0^i   \quad ; \quad \left [ L_a^1 \, , \, L_b^2 \right ] \, = \, 0
\end{equation}
and the relation between the two set of parameters is the following one:
\begin{equation}\label{chigamma}
\begin{array}{lcl}
 \chi _1\to \frac{1}{2} \left(\gamma _{1,-1}-\gamma _{1,1}+\gamma _{2,-1}-\gamma
   _{2,1}\right) &;&
 \chi _2\to \frac{1}{2} \left(\gamma _{1,-1}+\gamma _{1,1}+\gamma _{2,-1}+\gamma
   _{2,1}\right) \\
 \chi _3\to \frac{1}{2} \left(-\gamma _{1,-1}+\gamma _{1,1}+\gamma _{2,-1}-\gamma
   _{2,1}\right) &;&
 \chi _4\to \frac{1}{2} \left(-\gamma _{1,0}-\gamma _{2,0}\right) \\
 \chi _5\to \frac{1}{2} \left(\gamma _{2,0}-\gamma _{1,0}\right) &;&
 \chi _6\to \frac{1}{2} \left(-\gamma _{1,-1}-\gamma _{1,1}+\gamma _{2,-1}+\gamma
   _{2,1}\right)
\end{array}
\end{equation}
Correspondingly if $\Lambda^{\alpha,\dot{\beta}} $ is an object transforming in the tensor product of the fundamental representation of $\slal(2)_1$ and of the fundamental representation of $\slal(2)_2$ the components of the $\so(2,2)$ vector are as follows:
\begin{equation}\label{duecod}
    v_{\so(2,2)} \, = \, \left(
\begin{array}{l}
 \Lambda ^{1,1} \\
 \Lambda ^{1,2} \\
 \Lambda ^{2,1} \\
 \Lambda ^{2,2}
\end{array}
\right)
\end{equation}
This means that the tensor $\Delta^{i|I}$ representing the Lax operator can be reinterpreted as a three index object $\Delta^{\alpha,\dot{\beta}|I}$ according to the following rule:
\begin{equation}\label{duecodicini}
    \left(
\begin{array}{l}
 \Delta ^{1|I} \\
 \Delta ^{2|I} \\
 \Delta ^{3|I} \\
 \Delta ^{4|I}
\end{array}
\right)\, = \, \left(
\begin{array}{l}
 \Delta ^{1,1|I} \\
 \Delta ^{1,2|I} \\
 \Delta ^{2,1|I} \\
 \Delta ^{2,2|I}
\end{array}
\right) \, = \, \Delta^{\alpha,\dot{\beta}|I}
\end{equation}
where the index $I$ spans the fundamental representation of $\so(2,2+2s)$.
\par
Equipped with these conversion vocabulary, according to the scheme developed in \cite{noig22}, we can  define the following tensor classifiers:
\paragraph{The quadratic hamiltonian}
\begin{equation}\label{hamillo}
    \mathfrak{H}_{quadr} \, = \, \Delta^{\alpha,\dot{\beta}|I} \, \Delta^{\gamma,\dot{\delta}|J} \, \epsilon_{\alpha\gamma} \,\epsilon_{\dot{\beta}\dot{\delta}} \, \eta_{IJ}
\end{equation}
where:
\begin{equation}\label{etalunga}
  \eta \, = \,   \left(
\begin{array}{llllll}
 1 & 0 & 0 & 0 & 0 & 0 \\
 0 & 1 & 0 & 0 & 0 & 0 \\
 \vdots & \ddots & \ddots & \ddots & \vdots & \vdots \\
 0 & 0 & 0 & 1 & 0 & 0 \\
 0 & 0 & 0 & 0 & -1 & 0 \\
 0 & 0 & 0 & 0 & 0 & -1
\end{array}
\right)
\end{equation}
is the diagonal $\so(2,2+2s)$ invariant metric and:
\begin{equation}\label{epsili}
    \epsilon^{\alpha\gamma} \, = \, \epsilon^{\dot{\beta}\dot{\delta}} \, = \, \left( \begin{array}{cc}
                                                                                        0 & 1 \\
                                                                                        -1 & 0
                                                                                      \end{array}
    \right)
\end{equation}
are the standard invariant tensors of the $\slal(2)$ Lie algebras.
\paragraph{The irreducible quadratic $\mathcal{T}$-tensors}
In this case we can define three irreducible quadratic $\mathcal{T}$-tensors.
\begin{eqnarray}\label{confucio}
    \mathcal{T}^{\dot{\beta}\dot{\delta}|IJ}_{1} & = &  \Delta^{\alpha,\dot{\beta}|I} \, \Delta^{\gamma,\dot{\delta}|J} \, \epsilon_{\alpha\gamma} \,- \, \ft 12 \,\ft{1}{4+2s}\, \epsilon^{\dot{\beta}\dot{\delta}} \, \eta^{IJ} \, \mathfrak{H}_{quadr}\nonumber\\
    \mathcal{T}^{\alpha\gamma|IJ}_{2} & = &  \Delta^{\alpha,\dot{\beta}|I} \, \Delta^{\gamma,\dot{\delta}|J} \, \epsilon_{\dot{\beta}\dot{\delta}} \,- \, \ft 12 \,\ft{1}{4+2s}\, \epsilon^{\alpha\gamma} \, \eta^{IJ} \, \mathfrak{H}_{quadr}\nonumber\\
    \mathcal{T}^{IJ}_{3} & = &  \Delta^{\alpha,\dot{\beta}|I} \, \Delta^{\gamma,\dot{\delta}|J} \,\epsilon_{\alpha\gamma}  \,\epsilon_{\dot{\beta}\dot{\delta}} \,- \, \ft{1}{4+2s}\,  \eta^{IJ} \, \mathfrak{H}_{quadr}\nonumber\\
\end{eqnarray}
The vanishing of the second of these  operators, according to the results of \cite{noig22}, is the necessary and sufficient condition for a Lax operator to define a BPS black-hole solution:
\paragraph{The quadratic $W$ tensors}
Utilizing the projection operators from the tensor product of two spinor representation of $\slal(2)$ to the vector representation of
$\so(1,2)$ introduced in \cite{noig22}
\begin{equation}\label{paffuto}
    \Pi^x_{\alpha\beta} \, = \, \left\{\left(
\begin{array}{ll}
 \frac{1}{2} & 0 \\
 0 & 0
\end{array}
\right)\, , \, \left(
\begin{array}{ll}
 0 & \frac{1}{4} \\
 \frac{1}{4} & 0
\end{array}
\right) \, , \, \left(
\begin{array}{ll}
 0 & 0 \\
 0 & \frac{1}{2}
\end{array}
\right)\right\}
\end{equation}
we construct the quadratic symmetric tensor $W$:
\begin{equation}\label{Wcumulator}
    W^{x|\dot{\beta},I,\dot{\delta},J} \, = \, \Pi^x_{\alpha\gamma} \, \Delta^{\alpha,\dot{\beta}|I} \, \Delta^{\gamma,\dot{\delta}|J}
\end{equation}
In addition from $W$ we construct the following derived tensors:
\begin{eqnarray}
  W^{x|y||IJ} &=& \Pi^y_{\dot{\beta}\dot{\gamma} } \,  W^{x|\dot{\beta},I,\dot{\delta},J} \\
  W^{x|y} &=& W^{x|y||IJ} \, \eta_{IJ} \label{fricandella}
\end{eqnarray}
\paragraph{The quartic $\mathbb{T}$ tensor}
The quartic $\mathbb{T}$-tensor is now defined as follows:
\begin{equation}\label{tfattus}
    \mathbb{T}^{xy} \, = \, W^{x|\dot{\alpha},I,\dot{\gamma},J} \, W^{y|\dot{\beta},K,\dot{\delta},L}\, \epsilon_{\dot{\alpha}\dot{\beta}} \,\epsilon_{\dot{\gamma}\dot{\delta}} \, \eta_{IJ}\, \, \eta_{KL}
\end{equation}
\paragraph{The quartic invariant}
A quartic invariant with respect to $\mathbb{H}^\star$ subalgebra can now be constructed by setting:
\begin{equation}\label{quarticI}
    \mathcal{I}_4 \, = \, \mathbb{T}^{xy}  \, \eta_{xy}
\end{equation}
where the $\so(1,2)$ invariant metric in the chosen basis is the following one:
\begin{equation}\label{friggi}
    \eta_{xy} \, = \, \left(
\begin{array}{lll}
 0 & 0 & 1 \\
 0 & -2 & 0 \\
 1 & 0 & 0
\end{array}
\right)
\end{equation}
\paragraph{The quartic $\mathfrak{T}$-tensors}
Following the procedure of \cite{noig22} we introduce the following two $\mathfrak{T}$-tensors, the first being a representation of $\sl(2)_2$ and a singlet of $\so(2,2+2s)$, the second viceversa.
\begin{equation}\label{Tgoth1}
    \mathfrak{T}^{\dot{\alpha},\dot{\beta},\dot{\gamma},\dot{\delta}}_{[1]} \, = \, W^{x|\dot{\alpha},I,\dot{\beta},J} \,W^{y|\dot{\gamma},K,\dot{\delta},L} \, \eta_{xy} \, \eta_{IJ} \, \eta_{KL}
\end{equation}
\begin{equation}\label{Tgoth2}
    \mathfrak{T}^{IJKL}_{[2]} \, = \, W^{x|\dot{\alpha},I,\dot{\beta},J} \,W^{y|\dot{\gamma},K,\dot{\delta},L} \, \eta_{xy} \, \epsilon_{\dot{\alpha}\dot{\gamma}}\,\epsilon_{\dot{\beta}\dot{\delta}}
\end{equation}
\subsection{Evaluation of the tensor classifiers on the representatives of $\mathcal{QM}^\star_{(4,4+2s)}$ nilpotent orbits}
Having defined the tensor classifiers we can evaluate them on the representatives of the $37$ nilpotent orbits. Introducing a dichotomic indicator scheme, namely assigning  $1$ to a tensor classifier that does not vanish  and assigning $0$ to a tensor that vanish we obtain a finite collection of patterns according to which we can group the orbits. The result is displayed in table \ref{tensoranal}.
\begin{table}
\begin{center}
$$
\begin{array}{||l|l|l|l|l|l|l||l||}
\hline
\hline
 \mathcal{T}_1 & \mathcal{T}_2 & W^{x|y||IJ} &W^{x|y|} &
   \mathbb{T} & \mathfrak{T}_{[1]} & \mathfrak{T}_{[2]} &
   \mbox{group of orbits} \\
   \hline
   \hline
 1 & 1 & 1 & 1 & 1 &  1 & 1 &
   \left\{O_{1,1},O_{1,1}^2,O_{1,2}^2,O_{2,1}^2,O_{2,2}^2,O_{1,1}^{1
   1},O_{1,2}^{11},O_{1,1}^{12}\right\} \\
   \hline
 1 & 1 & 1 & 1 & 0 & 1 & 1 &
   \left\{O_{1,1}^4,O_{2,1}^{11},O_{2,3}^{11}\right\} \\
   \hline
 1 & 1 & 1 & 1 & 0 & 0 & 1 &
   \left\{O_{1,1}^3,O_{1,1}^5,O_{2,1}^7,O_{2,2}^7,O_{2,3}^7,O_{3,2}^
   {11},O_{3,3}^{11}\right\} \\
   \hline
 1 & 1 & 1 & 0 & 0 & 0 & 1 &
   \left\{O_{2,1}^6,O_{2,2}^6,O_{3,1}^7,O_{3,2}^7,O_{3,3}^7,O_{2,1}^
   9,O_{2,2}^9\right\} \\
   \hline
 1 & 0 & 1 & 1 & 0  & 1 & 0 &
   \left\{O_{1,1}^7,O_{1,2}^7,O_{1,3}^7\right\} \\
   \hline
 1 & 0 & 1 & 1 & 0  & 0 & 0 & \left\{O_{1,1}^6,O_{1,2}^6\right\}
   \\
   \hline
 1 & 0 & 1 & 0 & 0  & 0 & 0 & \left\{O_{1,1}^8\right\} \\
 \hline
 0 & 0 & 1 & 1 & 0  & 0 & 0 & \left\{O_{1,1}^9,O_{1,2}^9\right\}
   \\
   \hline
 0 & 0 & 1 & 0 & 0 & 0 & 0 & \left\{O_{1,1}^{10}\right\}\\
 \hline
 \hline
\end{array}
$$
\end{center}
\caption{Grouping of the $37$ nilpotent orbits according to their tensor classifier patterns. The number $1$ means that the corresponding tensor does not vanish identically on that orbit, while the number $0$ means that the corresponding tensor is just zero for all members of the orbit. \label{tensoranal}}
\end{table}
It is a matter of fact that the tensor classifiers are not able, at least at the level of this coarse analysis, to separate all the orbits. Yet the discovered grouping is certainly meaningful and
deserves further investigation which we postpone to the
future publication where we plan to
analyse the stability subgroups \cite{fortocomo}.
\section{Conclusions}
\label{concludo}
In this paper we have constructed the list of  $\mathrm{H}^\star$ nilpotent orbits  in $\mathbb{K}$  for the series of pseudo-quaternionic manifolds (\ref{p2qmanstarro}). These latter  are the $c^\star$-map image  of the special geometry series (\ref{skgseries}); in other words, by time-like reduction,  they emerge from supergravity models where the vector multiplets  are coupled according to such homogeneous symmetric special geometries  which occur  in many instances of superstring compactifications and in particular correspond to the large radius limit of  several Calabi Yau moduli spaces.
\par
As we emphasized in the introduction and at all levels in the course of our construction, the most important result revealed by our analysis is the universal character of the  $\mathbb{K}$-based nilpotent orbits. Their pattern  is a common feature of all manifolds belonging to the same Tits Satake universality class. Indeed it depends only on the structure of the Tits Satake subalgebra: $\mathbb{U}_{\mathrm{TS}} \subset \mathbb{U}$.
\par
The method we employed to work out our classification and explicitly construct the representatives of the nilpotent orbits is based on the Weyl group of $\mathbb{U}_{\mathrm{TS}}$. We showed that the splitting:
\begin{equation}\label{splittusone}
    \mathbb{U} \, = \, \mathbb{H}^\star \, \oplus \, \mathbb{K}
\end{equation}
defines a proper subgroup $\mathcal{W}_H \subset \mathcal{W}$ of the Weyl group and that the coset $\frac{\mathcal{W}}{\mathcal{W}_H}$ is at the root of the concept of $\gamma$-labels. The so called $\alpha$-labels are nothing else but the entire Weyl orbit  of  possible spectra of the angular momentum third component (the central element $h$ of a standard triple $\{h,x,y\}$), which is fixed once a branching rule of the fundamental representation of  $\mathbb{U}_{\mathrm{TS}}$ with respect to the $\slal(2)$ subalgebra $\{h,x,y\}$  is given.  Hence $\alpha$-labels enumerate the different available branching rules for the embedding:
 \begin{equation}\label{frescofino}
    \slal(2) \, \hookrightarrow \, \mathbb{U}_{\mathrm{TS}}
 \end{equation}
 The Weyl   orbit of $h$-spectra corresponding to a given branching rule, ($\alpha$-label)  splits in $m$-suborbits with respect to the sub-Weyl group $\mathcal{W}_H$, where we named  $m$ the number of lateral classes in the coset  $\frac{\mathcal{W}}{\mathcal{W}_H}$. These suborbits correspond to the possible $\gamma$-labels. The set  of available $\beta$-labels,  that, by definition, are the possible spectra of the compact operator $x-y$, coincides with the set of $\gamma$-labels. The various nilpotent orbits are thus classified by the triplet of labels $\alpha,\gamma,\beta$.
 \par
 Since the Tits Satake universality classes of symmetric special geometries are just five, it suffices to derive the list of nilpotent orbits for the five universal manifolds appearing in the third column of table \ref{skTSstarro} and, when the corresponding class contains more than one element determine the regular embedding of the universal orbit within the ambient algebra. This is precisely what we did in this paper for the fourth class of table \ref{skTSstarro}. Not only we determined the list of nilpotent orbits for the universal manifold $\mathrm{\SO(4,5)/\SO(2,3)\times \SO(2,2)}$ but we also showed how they are generically embedded in the infinite series of manifolds $\mathrm{\SO(4+2s,4)/\SO(2,2+2s)\times \SO(2,2)}$ filling one half of the class for $p=2s$, even.  Obviously we  also have the odd case, $p=2s+1$, yet it is quite evident that by means of simple modifications the embedding of the universal orbits can be extended also to such manifolds.
 \par
 The case of the second class ($\mathfrak{g}_{(2,2)}$) is done and requires no further  study of embeddings since it is a one-element class.
 \par
 A very interesting Lie algebra problem is provided by the fifth and last of the Tits-Satake universality classes of table \ref{skTSstarro} that contains three additional elements besides the universal manifold $\frac{\mathrm{F_{(4,4)}}}{\mathrm{Sp(6)\times SU(1,1)}}$. The list of nilpotent orbits for the latter was recently derived by one of us in a different collaboration \cite{marioetalF4}. It is now a challenging problem to embed these universal classes in the three remaining members of the class.
 \paragraph{Stability subalgebras and the orbit coset manifolds} As we already stressed in the text, the classification of orbits amounts also to a classification of very special subalgebras of $\mathbb{H}^\star$ which are the stability subalgebras $\mathfrak{S}^\alpha_{\beta\gamma}$, leading to a series of coset manifolds, equivalent to the nilpotent orbit manifolds
 \begin{equation}\label{cosettini}
    \frac{\mathrm{H}^\star}{\mathcal{S}^\alpha_{\beta\gamma}}
 \end{equation}
 whose structure and properties are intriguing. In particular, in view of the integrability of the ambient manifold, this rich class of special cosets might provide new unexpected examples of integrable models.  In a forthcoming publication,  as already announced, we plan to work out the list of these cosets
  \begin{equation}\label{the4classa}
     \frac{\mathrm{SO(2+2s,2)\times SO(2,2)}}{\mathcal{S}^\alpha_{\beta\gamma}}
  \end{equation}
  for the fourth Tits Satake universality class and we already possess some partial results. Obviously the same classification has to be worked out for the other universality classes. Challenging, as usual, is the fifth class where the list of nilpotent orbits $\mathcal{O}^\alpha_{\gamma\beta}$ for $\frac{\mathrm{F_{(4,4)}}}{\mathrm{Sp(6)\times SU(1,1)}}$ singles out an equal number of coset manifolds of the form:
 \begin{equation}\label{cassettonicomodoni}
    \frac{\mathrm{Sp(6)\times SU(1,1)}}{\mathcal{S}^\alpha_{\beta\gamma}}\quad ; \quad
     \frac{\mathrm{SU(3,3)\times SU(1,1)}}{\bar{{\mathcal{S}}}^\alpha_{\beta\gamma}}\quad ; \quad
    \frac{\mathrm{SO^\star(12)\times SU(1,1)}}{\tilde{\mathcal{S}}^\alpha_{\beta\gamma}}\quad ; \quad
      \frac{\mathrm{E_{(7,-25)}\times SU(1,1)}}{\hat{{\mathcal{S}}}^\alpha_{\beta\gamma}}
 \end{equation}
 \paragraph{Branching rules and regular black hole solutions} As it is known regular extremal black-holes are associated with Lax operators of nilpotency degree $3$. Our analysis has revealed that the nilpotency degree is just $d_g \, = \, 2 \, j_{max} +1$ where $j_{max}$ is the highest spin contained in the decomposition of the fundamental representation of $\mathbb{U}_{\mathrm{TS}}$ with respect to the embedded $\slal(2)$ subalgebra of the standard triple $\{h,x,y\}$. Hence regular extremal black-holes correspond to $j_{max} \, = \, 1$. In general there are several branching rules satisfying this condition, yet comparison of the results for $\mathfrak{g}_{(2,2)}$ and for $\so(4,5)$ seems to suggest that both BPS and non BPS regular solutions come from a universal $\alpha$-label:
 $$2\times [j=1]\times p\times [j=0]$$
 Whether this conjecture is correct or not has to be verified by comparison with the results for the remaining universality classes. In case it is true, this fact deserves careful consideration and requires an explanation.
 \par
 It appears that nilpotent orbits corresponding to higher degree of nilpotency and hence with $j_{max} >1$ might play a role in the construction of multi-center solutions \cite{Bossard:2011kz}. It is challenging to understand the relation between the spin content of the $\alpha$-label and physical properties of the corresponding hole, in particular the entropy.
 \paragraph{Fake-superpotential, fixed points and nilpotent orbits}
 A further direction of future investigation stimulated by our results concerns the relation between the classification of nilpotent orbits and the parallel classification of attraction points of the geodesic potential in special K\"ahler geometry. A bridge between the two approaches to  black holes might be provided by the classification of the stability subalgebras $\mathcal{S}^\alpha_{\beta\gamma} \subset \mathrm{H}^\star$.  Indeed $\mathrm{H}^\star \sim \su(1,1) \times \mathrm{U}_{\mathrm{D=4}}$ and $\mathrm{U_{D=4}}$ is the symplectic group acting on the quantized charges $(p,q)$ which define the geodesic potential. It is quite possible that in each nilpotent orbit $\mathcal{S}^\alpha_{\beta\gamma}$ or its subgroup at vanishing Taub-Nut charge might be a symmetry of the geodesic potential as well. Also this point is in our agenda for the next coming publication \cite{fortocomo}.
 \par
 \vskip 2cm
{\bf Acknowledgments}
The work of A.S. was partially supported by the RFBR Grants No.
11-02-01335-?, 11-02-12232-$\mathrm{ofi\_m}$-2011, 09-02-00725-a,
09-02-91349-NNIO\_a; DFG grant No 436 RUS/113/669,
the Heisenberg-Landau, INFN-BLTP/JINR and CERN-BLTP/JINR
Programs.
\newpage
\appendix
\section{The list of the representatives for the $37$ nilpotent orbits of $\mathcal{QM}^\star_{(4,4+2s)}$}
\label{puppo}
In this appendix we display the explicit form of one representative for each of the classified $37$ nilpotent orbits. They are given as $10 \times 10$ matrices since by setting $s=1$ we chose the lowest lying member of the Tits-Satake universality class. For all the other members of the same universality class we could write a similar list of $37$ matrices whose $\alpha$,$\gamma$ and $\beta$ labels are the same.
{\scriptsize
\begin{equation}\label{nil1}
   \mathcal{O}_{1,1}\, = \,\left(
\begin{array}{llllllllll}
 -\sqrt{2} &
   -\frac{\sqrt{\frac{7}{2}}}{2} &
   \sqrt{2} &
   -\frac{\sqrt{\frac{7}{2}}}{2} & 0 & 0
   & \frac{\sqrt{\frac{7}{2}}}{2} & 0 &
   -\frac{\sqrt{\frac{7}{2}}}{2} & 0 \\
 -\frac{\sqrt{\frac{7}{2}}}{2} &
   -\frac{3}{\sqrt{2}} &
   -\frac{\sqrt{\frac{7}{2}}}{2} & 0 & 0
   & -\sqrt{5} & -\frac{3}{\sqrt{2}} &
   -\frac{\sqrt{\frac{7}{2}}}{2} & 0 &
   \frac{\sqrt{\frac{7}{2}}}{2} \\
 -\sqrt{2} &
   \frac{\sqrt{\frac{7}{2}}}{2} &
   \sqrt{2} &
   \frac{\sqrt{\frac{7}{2}}}{2} & 0 & 0
   & -\frac{\sqrt{\frac{7}{2}}}{2} & 0 &
   \frac{\sqrt{\frac{7}{2}}}{2} & 0 \\
 \frac{\sqrt{\frac{7}{2}}}{2} & 0 &
   \frac{\sqrt{\frac{7}{2}}}{2} &
   -\frac{3}{\sqrt{2}} & 0 & -\sqrt{5} &
   0 & \frac{\sqrt{\frac{7}{2}}}{2} &
   \frac{3}{\sqrt{2}} &
   -\frac{\sqrt{\frac{7}{2}}}{2} \\
 0 & 0 & 0 & 0 & 0 & 0 & 0 & 0 & 0 & 0
   \\
 0 & -\sqrt{5} & 0 & \sqrt{5} & 0 & 0 &
   \sqrt{5} & 0 & \sqrt{5} & 0 \\
 -\frac{\sqrt{\frac{7}{2}}}{2} &
   \frac{3}{\sqrt{2}} &
   -\frac{\sqrt{\frac{7}{2}}}{2} & 0 & 0
   & -\sqrt{5} & \frac{3}{\sqrt{2}} &
   -\frac{\sqrt{\frac{7}{2}}}{2} & 0 &
   \frac{\sqrt{\frac{7}{2}}}{2} \\
 0 & \frac{\sqrt{\frac{7}{2}}}{2} & 0 &
   \frac{\sqrt{\frac{7}{2}}}{2} & 0 & 0
   & -\frac{\sqrt{\frac{7}{2}}}{2} &
   -\sqrt{2} &
   \frac{\sqrt{\frac{7}{2}}}{2} &
   -\sqrt{2} \\
 -\frac{\sqrt{\frac{7}{2}}}{2} & 0 &
   -\frac{\sqrt{\frac{7}{2}}}{2} &
   -\frac{3}{\sqrt{2}} & 0 & \sqrt{5} &
   0 & -\frac{\sqrt{\frac{7}{2}}}{2} &
   \frac{3}{\sqrt{2}} &
   \frac{\sqrt{\frac{7}{2}}}{2} \\
 0 & \frac{\sqrt{\frac{7}{2}}}{2} & 0 &
   \frac{\sqrt{\frac{7}{2}}}{2} & 0 & 0
   & -\frac{\sqrt{\frac{7}{2}}}{2} &
   \sqrt{2} &
   \frac{\sqrt{\frac{7}{2}}}{2} &
   \sqrt{2}
\end{array}
\right)
\end{equation}
\begin{equation}\label{nil2}
    \mathcal{O}_{1,1}^2\, = \,\left(
\begin{array}{llllllllll}
 -\sqrt{\frac{3}{2}} &
   -\frac{\sqrt{\frac{5}{2}}}{2} &
   -\sqrt{\frac{3}{2}} &
   -\frac{\sqrt{\frac{5}{2}}}{2} & 0 & 0
   & -\frac{\sqrt{\frac{5}{2}}}{2} & 0 &
   \frac{\sqrt{\frac{5}{2}}}{2} & 0 \\
 -\frac{\sqrt{\frac{5}{2}}}{2} & 0 &
   \frac{\sqrt{\frac{5}{2}}}{2} &
   -\sqrt{\frac{3}{2}} & 0 & 0 &
   \sqrt{\frac{3}{2}} &
   -\frac{\sqrt{\frac{5}{2}}}{2} & 0 &
   -\frac{\sqrt{\frac{5}{2}}}{2} \\
 \sqrt{\frac{3}{2}} &
   -\frac{\sqrt{\frac{5}{2}}}{2} &
   \sqrt{\frac{3}{2}} &
   -\frac{\sqrt{\frac{5}{2}}}{2} & 0 & 0
   & -\frac{\sqrt{\frac{5}{2}}}{2} & 0 &
   \frac{\sqrt{\frac{5}{2}}}{2} & 0 \\
 \frac{\sqrt{\frac{5}{2}}}{2} &
   \sqrt{\frac{3}{2}} &
   -\frac{\sqrt{\frac{5}{2}}}{2} &
   -\sqrt{6} & 0 & 0 & 0 &
   \frac{\sqrt{\frac{5}{2}}}{2} &
   -\sqrt{\frac{3}{2}} &
   \frac{\sqrt{\frac{5}{2}}}{2} \\
 0 & 0 & 0 & 0 & 0 & 0 & 0 & 0 & 0 & 0
   \\
 0 & 0 & 0 & 0 & 0 & 0 & 0 & 0 & 0 & 0
   \\
 \frac{\sqrt{\frac{5}{2}}}{2} &
   -\sqrt{\frac{3}{2}} &
   -\frac{\sqrt{\frac{5}{2}}}{2} & 0 & 0
   & 0 & \sqrt{6} &
   \frac{\sqrt{\frac{5}{2}}}{2} &
   \sqrt{\frac{3}{2}} &
   \frac{\sqrt{\frac{5}{2}}}{2} \\
 0 & \frac{\sqrt{\frac{5}{2}}}{2} & 0 &
   \frac{\sqrt{\frac{5}{2}}}{2} & 0 & 0
   & \frac{\sqrt{\frac{5}{2}}}{2} &
   -\sqrt{\frac{3}{2}} &
   -\frac{\sqrt{\frac{5}{2}}}{2} &
   \sqrt{\frac{3}{2}} \\
 \frac{\sqrt{\frac{5}{2}}}{2} & 0 &
   -\frac{\sqrt{\frac{5}{2}}}{2} &
   \sqrt{\frac{3}{2}} & 0 & 0 &
   -\sqrt{\frac{3}{2}} &
   \frac{\sqrt{\frac{5}{2}}}{2} & 0 &
   \frac{\sqrt{\frac{5}{2}}}{2} \\
 0 & -\frac{\sqrt{\frac{5}{2}}}{2} & 0 &
   -\frac{\sqrt{\frac{5}{2}}}{2} & 0 & 0
   & -\frac{\sqrt{\frac{5}{2}}}{2} &
   -\sqrt{\frac{3}{2}} &
   \frac{\sqrt{\frac{5}{2}}}{2} &
   \sqrt{\frac{3}{2}}
\end{array}
\right)
\end{equation}
\begin{equation}\label{nil3}
    \mathcal{O}_{1,2}^2\, = \,\left(
\begin{array}{llllllllll}
 -\sqrt{\frac{3}{2}} & -\frac{\sqrt{\frac{5}{2}}}{2} & -\sqrt{\frac{3}{2}} & -\frac{\sqrt{\frac{5}{2}}}{2} & 0 & 0 & -\frac{\sqrt{\frac{5}{2}}}{2} & 0 &
   \frac{\sqrt{\frac{5}{2}}}{2} & 0 \\
 -\frac{\sqrt{\frac{5}{2}}}{2} & -\sqrt{6} & \frac{\sqrt{\frac{5}{2}}}{2} & \sqrt{\frac{3}{2}} & 0 & 0 & \sqrt{\frac{3}{2}} & -\frac{\sqrt{\frac{5}{2}}}{2} & 0 &
   -\frac{\sqrt{\frac{5}{2}}}{2} \\
 \sqrt{\frac{3}{2}} & -\frac{\sqrt{\frac{5}{2}}}{2} & \sqrt{\frac{3}{2}} & -\frac{\sqrt{\frac{5}{2}}}{2} & 0 & 0 & -\frac{\sqrt{\frac{5}{2}}}{2} & 0 &
   \frac{\sqrt{\frac{5}{2}}}{2} & 0 \\
 \frac{\sqrt{\frac{5}{2}}}{2} & -\sqrt{\frac{3}{2}} & -\frac{\sqrt{\frac{5}{2}}}{2} & 0 & 0 & 0 & 0 & \frac{\sqrt{\frac{5}{2}}}{2} & -\sqrt{\frac{3}{2}} &
   \frac{\sqrt{\frac{5}{2}}}{2} \\
 0 & 0 & 0 & 0 & 0 & 0 & 0 & 0 & 0 & 0 \\
 0 & 0 & 0 & 0 & 0 & 0 & 0 & 0 & 0 & 0 \\
 \frac{\sqrt{\frac{5}{2}}}{2} & -\sqrt{\frac{3}{2}} & -\frac{\sqrt{\frac{5}{2}}}{2} & 0 & 0 & 0 & 0 & \frac{\sqrt{\frac{5}{2}}}{2} & -\sqrt{\frac{3}{2}} &
   \frac{\sqrt{\frac{5}{2}}}{2} \\
 0 & \frac{\sqrt{\frac{5}{2}}}{2} & 0 & \frac{\sqrt{\frac{5}{2}}}{2} & 0 & 0 & \frac{\sqrt{\frac{5}{2}}}{2} & -\sqrt{\frac{3}{2}} & -\frac{\sqrt{\frac{5}{2}}}{2} &
   \sqrt{\frac{3}{2}} \\
 \frac{\sqrt{\frac{5}{2}}}{2} & 0 & -\frac{\sqrt{\frac{5}{2}}}{2} & \sqrt{\frac{3}{2}} & 0 & 0 & \sqrt{\frac{3}{2}} & \frac{\sqrt{\frac{5}{2}}}{2} & \sqrt{6} &
   \frac{\sqrt{\frac{5}{2}}}{2} \\
 0 & -\frac{\sqrt{\frac{5}{2}}}{2} & 0 & -\frac{\sqrt{\frac{5}{2}}}{2} & 0 & 0 & -\frac{\sqrt{\frac{5}{2}}}{2} & -\sqrt{\frac{3}{2}} & \frac{\sqrt{\frac{5}{2}}}{2} &
   \sqrt{\frac{3}{2}}
\end{array}
\right)
\end{equation}
\begin{equation}\label{nil4}
    \mathcal{O}_{2,1}^2\, = \,\left(
\begin{array}{llllllllll}
 -\sqrt{\frac{3}{2}} & -\frac{\sqrt{\frac{5}{2}}}{2} & \sqrt{\frac{3}{2}} & -\frac{\sqrt{\frac{5}{2}}}{2} & 0 & 0 & \frac{\sqrt{\frac{5}{2}}}{2} & 0 &
   -\frac{\sqrt{\frac{5}{2}}}{2} & 0 \\
 -\frac{\sqrt{\frac{5}{2}}}{2} & -\sqrt{6} & -\frac{\sqrt{\frac{5}{2}}}{2} & \sqrt{\frac{3}{2}} & 0 & 0 & -\sqrt{\frac{3}{2}} & -\frac{\sqrt{\frac{5}{2}}}{2} & 0 &
   \frac{\sqrt{\frac{5}{2}}}{2} \\
 -\sqrt{\frac{3}{2}} & \frac{\sqrt{\frac{5}{2}}}{2} & \sqrt{\frac{3}{2}} & \frac{\sqrt{\frac{5}{2}}}{2} & 0 & 0 & -\frac{\sqrt{\frac{5}{2}}}{2} & 0 &
   \frac{\sqrt{\frac{5}{2}}}{2} & 0 \\
 \frac{\sqrt{\frac{5}{2}}}{2} & -\sqrt{\frac{3}{2}} & \frac{\sqrt{\frac{5}{2}}}{2} & 0 & 0 & 0 & 0 & \frac{\sqrt{\frac{5}{2}}}{2} & \sqrt{\frac{3}{2}} &
   -\frac{\sqrt{\frac{5}{2}}}{2} \\
 0 & 0 & 0 & 0 & 0 & 0 & 0 & 0 & 0 & 0 \\
 0 & 0 & 0 & 0 & 0 & 0 & 0 & 0 & 0 & 0 \\
 -\frac{\sqrt{\frac{5}{2}}}{2} & \sqrt{\frac{3}{2}} & -\frac{\sqrt{\frac{5}{2}}}{2} & 0 & 0 & 0 & 0 & -\frac{\sqrt{\frac{5}{2}}}{2} & -\sqrt{\frac{3}{2}} &
   \frac{\sqrt{\frac{5}{2}}}{2} \\
 0 & \frac{\sqrt{\frac{5}{2}}}{2} & 0 & \frac{\sqrt{\frac{5}{2}}}{2} & 0 & 0 & -\frac{\sqrt{\frac{5}{2}}}{2} & -\sqrt{\frac{3}{2}} & \frac{\sqrt{\frac{5}{2}}}{2} &
   -\sqrt{\frac{3}{2}} \\
 -\frac{\sqrt{\frac{5}{2}}}{2} & 0 & -\frac{\sqrt{\frac{5}{2}}}{2} & -\sqrt{\frac{3}{2}} & 0 & 0 & \sqrt{\frac{3}{2}} & -\frac{\sqrt{\frac{5}{2}}}{2} & \sqrt{6} &
   \frac{\sqrt{\frac{5}{2}}}{2} \\
 0 & \frac{\sqrt{\frac{5}{2}}}{2} & 0 & \frac{\sqrt{\frac{5}{2}}}{2} & 0 & 0 & -\frac{\sqrt{\frac{5}{2}}}{2} & \sqrt{\frac{3}{2}} & \frac{\sqrt{\frac{5}{2}}}{2} &
   \sqrt{\frac{3}{2}}
\end{array}
\right)
\end{equation}
\begin{equation}\label{nil5}
   \mathcal{O}_{2,2}^2\, = \,\left(
\begin{array}{llllllllll}
 -\sqrt{\frac{3}{2}} & -\frac{\sqrt{\frac{5}{2}}}{2} & \sqrt{\frac{3}{2}} & -\frac{\sqrt{\frac{5}{2}}}{2} & 0 & 0 & \frac{\sqrt{\frac{5}{2}}}{2} & 0 &
   -\frac{\sqrt{\frac{5}{2}}}{2} & 0 \\
 -\frac{\sqrt{\frac{5}{2}}}{2} & 0 & -\frac{\sqrt{\frac{5}{2}}}{2} & -\sqrt{\frac{3}{2}} & 0 & 0 & -\sqrt{\frac{3}{2}} & -\frac{\sqrt{\frac{5}{2}}}{2} & 0 &
   \frac{\sqrt{\frac{5}{2}}}{2} \\
 -\sqrt{\frac{3}{2}} & \frac{\sqrt{\frac{5}{2}}}{2} & \sqrt{\frac{3}{2}} & \frac{\sqrt{\frac{5}{2}}}{2} & 0 & 0 & -\frac{\sqrt{\frac{5}{2}}}{2} & 0 &
   \frac{\sqrt{\frac{5}{2}}}{2} & 0 \\
 \frac{\sqrt{\frac{5}{2}}}{2} & \sqrt{\frac{3}{2}} & \frac{\sqrt{\frac{5}{2}}}{2} & -\sqrt{6} & 0 & 0 & 0 & \frac{\sqrt{\frac{5}{2}}}{2} & \sqrt{\frac{3}{2}} &
   -\frac{\sqrt{\frac{5}{2}}}{2} \\
 0 & 0 & 0 & 0 & 0 & 0 & 0 & 0 & 0 & 0 \\
 0 & 0 & 0 & 0 & 0 & 0 & 0 & 0 & 0 & 0 \\
 -\frac{\sqrt{\frac{5}{2}}}{2} & \sqrt{\frac{3}{2}} & -\frac{\sqrt{\frac{5}{2}}}{2} & 0 & 0 & 0 & \sqrt{6} & -\frac{\sqrt{\frac{5}{2}}}{2} & \sqrt{\frac{3}{2}} &
   \frac{\sqrt{\frac{5}{2}}}{2} \\
 0 & \frac{\sqrt{\frac{5}{2}}}{2} & 0 & \frac{\sqrt{\frac{5}{2}}}{2} & 0 & 0 & -\frac{\sqrt{\frac{5}{2}}}{2} & -\sqrt{\frac{3}{2}} & \frac{\sqrt{\frac{5}{2}}}{2} &
   -\sqrt{\frac{3}{2}} \\
 -\frac{\sqrt{\frac{5}{2}}}{2} & 0 & -\frac{\sqrt{\frac{5}{2}}}{2} & -\sqrt{\frac{3}{2}} & 0 & 0 & -\sqrt{\frac{3}{2}} & -\frac{\sqrt{\frac{5}{2}}}{2} & 0 &
   \frac{\sqrt{\frac{5}{2}}}{2} \\
 0 & \frac{\sqrt{\frac{5}{2}}}{2} & 0 & \frac{\sqrt{\frac{5}{2}}}{2} & 0 & 0 & -\frac{\sqrt{\frac{5}{2}}}{2} & \sqrt{\frac{3}{2}} & \frac{\sqrt{\frac{5}{2}}}{2} &
   \sqrt{\frac{3}{2}}
\end{array}
\right)
\end{equation}
\begin{equation}\label{nil6}
    \mathcal{O}_{1,1}^3\, = \,\left(
\begin{array}{llllllllll}
 -1 & 0 & 1 & 0 & 0 & -\sqrt{\frac{3}{2}} & 0 & 0 & 0 & 0 \\
 0 & -\frac{1}{2} & 0 & \frac{1}{2} & 0 & 0 & 0 & 0 & 0 & 0 \\
 -1 & 0 & 1 & 0 & 0 & \sqrt{\frac{3}{2}} & 0 & 0 & 0 & 0 \\
 0 & -\frac{1}{2} & 0 & \frac{1}{2} & 0 & 0 & 0 & 0 & 0 & 0 \\
 0 & 0 & 0 & 0 & 0 & 0 & 0 & 0 & 0 & 0 \\
 -\sqrt{\frac{3}{2}} & 0 & -\sqrt{\frac{3}{2}} & 0 & 0 & 0 & 0 & -\sqrt{\frac{3}{2}} & 0 & \sqrt{\frac{3}{2}} \\
 0 & 0 & 0 & 0 & 0 & 0 & -\frac{1}{2} & 0 & -\frac{1}{2} & 0 \\
 0 & 0 & 0 & 0 & 0 & \sqrt{\frac{3}{2}} & 0 & -1 & 0 & -1 \\
 0 & 0 & 0 & 0 & 0 & 0 & \frac{1}{2} & 0 & \frac{1}{2} & 0 \\
 0 & 0 & 0 & 0 & 0 & \sqrt{\frac{3}{2}} & 0 & 1 & 0 & 1
\end{array}
\right)
\end{equation}
\begin{equation}\label{nil7}
    \mathcal{O}_{1,1}^4\, = \,\left(
\begin{array}{llllllllll}
 0 & 0 & 0 & 0 & 0 & 0 & -\frac{\sqrt{3}}{2} & 0 & \frac{\sqrt{3}}{2} & 0 \\
 0 & -1 & \frac{\sqrt{3}}{2} & 1 & 0 & 0 & 0 & 0 & 0 & -\frac{\sqrt{3}}{2} \\
 0 & -\frac{\sqrt{3}}{2} & 0 & -\frac{\sqrt{3}}{2} & 0 & 0 & 0 & 0 & 0 & 0 \\
 0 & -1 & -\frac{\sqrt{3}}{2} & 1 & 0 & 0 & 0 & 0 & 0 & \frac{\sqrt{3}}{2} \\
 0 & 0 & 0 & 0 & 0 & 0 & 0 & 0 & 0 & 0 \\
 0 & 0 & 0 & 0 & 0 & 0 & 0 & 0 & 0 & 0 \\
 \frac{\sqrt{3}}{2} & 0 & 0 & 0 & 0 & 0 & -1 & \frac{\sqrt{3}}{2} & -1 & 0 \\
 0 & 0 & 0 & 0 & 0 & 0 & \frac{\sqrt{3}}{2} & 0 & -\frac{\sqrt{3}}{2} & 0 \\
 \frac{\sqrt{3}}{2} & 0 & 0 & 0 & 0 & 0 & 1 & \frac{\sqrt{3}}{2} & 1 & 0 \\
 0 & -\frac{\sqrt{3}}{2} & 0 & -\frac{\sqrt{3}}{2} & 0 & 0 & 0 & 0 & 0 & 0
\end{array}
\right)
\end{equation}
\begin{equation}\label{nil8}
    \mathcal{O}_{1,1}^5\, = \,\left(
\begin{array}{llllllllll}
 0 & -\frac{1}{2 \sqrt{2}} & 0 & -\frac{1}{2 \sqrt{2}} & 0 & -\frac{1}{\sqrt{2}} & -\frac{1}{2 \sqrt{2}} & 0 & \frac{1}{2 \sqrt{2}} & 0 \\
 -\frac{1}{2 \sqrt{2}} & -\frac{1}{\sqrt{2}} & \frac{1}{2 \sqrt{2}} & \frac{1}{\sqrt{2}} & 0 & 0 & 0 & -\frac{1}{2 \sqrt{2}} & 0 & -\frac{1}{2 \sqrt{2}} \\
 0 & -\frac{1}{2 \sqrt{2}} & 0 & -\frac{1}{2 \sqrt{2}} & 0 & \frac{1}{\sqrt{2}} & -\frac{1}{2 \sqrt{2}} & 0 & \frac{1}{2 \sqrt{2}} & 0 \\
 \frac{1}{2 \sqrt{2}} & -\frac{1}{\sqrt{2}} & -\frac{1}{2 \sqrt{2}} & \frac{1}{\sqrt{2}} & 0 & 0 & 0 & \frac{1}{2 \sqrt{2}} & 0 & \frac{1}{2 \sqrt{2}} \\
 0 & 0 & 0 & 0 & 0 & 0 & 0 & 0 & 0 & 0 \\
 -\frac{1}{\sqrt{2}} & 0 & -\frac{1}{\sqrt{2}} & 0 & 0 & 0 & 0 & -\frac{1}{\sqrt{2}} & 0 & \frac{1}{\sqrt{2}} \\
 \frac{1}{2 \sqrt{2}} & 0 & -\frac{1}{2 \sqrt{2}} & 0 & 0 & 0 & -\frac{1}{\sqrt{2}} & \frac{1}{2 \sqrt{2}} & -\frac{1}{\sqrt{2}} & \frac{1}{2 \sqrt{2}} \\
 0 & \frac{1}{2 \sqrt{2}} & 0 & \frac{1}{2 \sqrt{2}} & 0 & \frac{1}{\sqrt{2}} & \frac{1}{2 \sqrt{2}} & 0 & -\frac{1}{2 \sqrt{2}} & 0 \\
 \frac{1}{2 \sqrt{2}} & 0 & -\frac{1}{2 \sqrt{2}} & 0 & 0 & 0 & \frac{1}{\sqrt{2}} & \frac{1}{2 \sqrt{2}} & \frac{1}{\sqrt{2}} & \frac{1}{2 \sqrt{2}} \\
 0 & -\frac{1}{2 \sqrt{2}} & 0 & -\frac{1}{2 \sqrt{2}} & 0 & \frac{1}{\sqrt{2}} & -\frac{1}{2 \sqrt{2}} & 0 & \frac{1}{2 \sqrt{2}} & 0
\end{array}
\right)
\end{equation}
\begin{equation}\label{nil9}
   \mathcal{O}_{1,1}^6\, = \,\left(
\begin{array}{llllllllll}
 0 & -\frac{1}{4} & 0 & -\frac{1}{4} & 0 & 0 & \frac{3}{4} & 0 & \frac{1}{4} & 0 \\
 -\frac{1}{4} & 0 & \frac{1}{4} & 0 & 0 & 0 & 0 & -\frac{1}{4} & 0 & -\frac{1}{4} \\
 0 & -\frac{1}{4} & 0 & \frac{3}{4} & 0 & 0 & -\frac{1}{4} & 0 & \frac{1}{4} & 0 \\
 \frac{1}{4} & 0 & \frac{3}{4} & 0 & 0 & 0 & 0 & \frac{1}{4} & 0 & -\frac{3}{4} \\
 0 & 0 & 0 & 0 & 0 & 0 & 0 & 0 & 0 & 0 \\
 0 & 0 & 0 & 0 & 0 & 0 & 0 & 0 & 0 & 0 \\
 -\frac{3}{4} & 0 & -\frac{1}{4} & 0 & 0 & 0 & 0 & -\frac{3}{4} & 0 & \frac{1}{4} \\
 0 & \frac{1}{4} & 0 & \frac{1}{4} & 0 & 0 & -\frac{3}{4} & 0 & -\frac{1}{4} & 0 \\
 \frac{1}{4} & 0 & -\frac{1}{4} & 0 & 0 & 0 & 0 & \frac{1}{4} & 0 & \frac{1}{4} \\
 0 & -\frac{1}{4} & 0 & \frac{3}{4} & 0 & 0 & -\frac{1}{4} & 0 & \frac{1}{4} & 0
\end{array}
\right)
\end{equation}
\begin{equation}\label{nil10}
    \mathcal{O}_{1,2}^6\, = \,\left(
\begin{array}{llllllllll}
 0 & -\frac{3}{4} & 0 & \frac{1}{4} & 0 & 0 & \frac{1}{4} & 0 & -\frac{1}{4} & 0 \\
 -\frac{3}{4} & 0 & -\frac{1}{4} & 0 & 0 & 0 & 0 & -\frac{3}{4} & 0 & \frac{1}{4} \\
 0 & \frac{1}{4} & 0 & \frac{1}{4} & 0 & 0 & \frac{1}{4} & 0 & \frac{3}{4} & 0 \\
 -\frac{1}{4} & 0 & \frac{1}{4} & 0 & 0 & 0 & 0 & -\frac{1}{4} & 0 & -\frac{1}{4} \\
 0 & 0 & 0 & 0 & 0 & 0 & 0 & 0 & 0 & 0 \\
 0 & 0 & 0 & 0 & 0 & 0 & 0 & 0 & 0 & 0 \\
 -\frac{1}{4} & 0 & \frac{1}{4} & 0 & 0 & 0 & 0 & -\frac{1}{4} & 0 & -\frac{1}{4} \\
 0 & \frac{3}{4} & 0 & -\frac{1}{4} & 0 & 0 & -\frac{1}{4} & 0 & \frac{1}{4} & 0 \\
 -\frac{1}{4} & 0 & -\frac{3}{4} & 0 & 0 & 0 & 0 & -\frac{1}{4} & 0 & \frac{3}{4} \\
 0 & \frac{1}{4} & 0 & \frac{1}{4} & 0 & 0 & \frac{1}{4} & 0 & \frac{3}{4} & 0
\end{array}
\right)
\end{equation}
\begin{equation}\label{nil11}
   \mathcal{O}_{2,1}^6\, = \,\left(
\begin{array}{llllllllll}
 0 & -\frac{3}{4} & 0 & \frac{1}{4} & 0 & 0 & -\frac{1}{4} & 0 & \frac{1}{4} & 0 \\
 -\frac{3}{4} & 0 & \frac{1}{4} & 0 & 0 & 0 & 0 & -\frac{3}{4} & 0 & -\frac{1}{4} \\
 0 & -\frac{1}{4} & 0 & -\frac{1}{4} & 0 & 0 & \frac{1}{4} & 0 & \frac{3}{4} & 0 \\
 -\frac{1}{4} & 0 & -\frac{1}{4} & 0 & 0 & 0 & 0 & -\frac{1}{4} & 0 & \frac{1}{4} \\
 0 & 0 & 0 & 0 & 0 & 0 & 0 & 0 & 0 & 0 \\
 0 & 0 & 0 & 0 & 0 & 0 & 0 & 0 & 0 & 0 \\
 \frac{1}{4} & 0 & \frac{1}{4} & 0 & 0 & 0 & 0 & \frac{1}{4} & 0 & -\frac{1}{4} \\
 0 & \frac{3}{4} & 0 & -\frac{1}{4} & 0 & 0 & \frac{1}{4} & 0 & -\frac{1}{4} & 0 \\
 \frac{1}{4} & 0 & -\frac{3}{4} & 0 & 0 & 0 & 0 & \frac{1}{4} & 0 & \frac{3}{4} \\
 0 & -\frac{1}{4} & 0 & -\frac{1}{4} & 0 & 0 & \frac{1}{4} & 0 & \frac{3}{4} & 0
\end{array}
\right)
\end{equation}
\begin{equation}\label{nil12}
    \mathcal{O}_{2,2}^6\, = \,\left(
\begin{array}{llllllllll}
 0 & -\frac{1}{4} & 0 & -\frac{1}{4} & 0 & 0 & -\frac{3}{4} & 0 & -\frac{1}{4} & 0 \\
 -\frac{1}{4} & 0 & -\frac{1}{4} & 0 & 0 & 0 & 0 & -\frac{1}{4} & 0 & \frac{1}{4} \\
 0 & \frac{1}{4} & 0 & -\frac{3}{4} & 0 & 0 & -\frac{1}{4} & 0 & \frac{1}{4} & 0 \\
 \frac{1}{4} & 0 & -\frac{3}{4} & 0 & 0 & 0 & 0 & \frac{1}{4} & 0 & \frac{3}{4} \\
 0 & 0 & 0 & 0 & 0 & 0 & 0 & 0 & 0 & 0 \\
 0 & 0 & 0 & 0 & 0 & 0 & 0 & 0 & 0 & 0 \\
 \frac{3}{4} & 0 & -\frac{1}{4} & 0 & 0 & 0 & 0 & \frac{3}{4} & 0 & \frac{1}{4} \\
 0 & \frac{1}{4} & 0 & \frac{1}{4} & 0 & 0 & \frac{3}{4} & 0 & \frac{1}{4} & 0 \\
 -\frac{1}{4} & 0 & -\frac{1}{4} & 0 & 0 & 0 & 0 & -\frac{1}{4} & 0 & \frac{1}{4} \\
 0 & \frac{1}{4} & 0 & -\frac{3}{4} & 0 & 0 & -\frac{1}{4} & 0 & \frac{1}{4} & 0
\end{array}
\right)\
\end{equation}
\begin{equation}\label{nil13}
    \mathcal{O}_{1,1}^7\, = \,\left(
\begin{array}{llllllllll}
 0 & 0 & 0 & -\frac{1}{2} & 0 & 0 & \frac{1}{2} & 0 & 0 & 0 \\
 0 & 0 & -\frac{1}{2} & 0 & 0 & 0 & 0 & \frac{1}{2} & 0 & 0 \\
 0 & \frac{1}{2} & 0 & 0 & 0 & 0 & -1 & 0 & -\frac{1}{2} & 0 \\
 \frac{1}{2} & 0 & 0 & 0 & 0 & 0 & 0 & 1 & 0 & -\frac{1}{2} \\
 0 & 0 & 0 & 0 & 0 & 0 & 0 & 0 & 0 & 0 \\
 0 & 0 & 0 & 0 & 0 & 0 & 0 & 0 & 0 & 0 \\
 -\frac{1}{2} & 0 & -1 & 0 & 0 & 0 & 0 & 0 & 0 & \frac{1}{2} \\
 0 & -\frac{1}{2} & 0 & 1 & 0 & 0 & 0 & 0 & \frac{1}{2} & 0 \\
 0 & 0 & \frac{1}{2} & 0 & 0 & 0 & 0 & -\frac{1}{2} & 0 & 0 \\
 0 & 0 & 0 & \frac{1}{2} & 0 & 0 & -\frac{1}{2} & 0 & 0 & 0
\end{array}
\right)
\end{equation}
\begin{equation}\label{nil14}
   \mathcal{O}_{1,2}^7\, = \,\left(
\begin{array}{llllllllll}
 0 & -\frac{1}{2} & 0 & 0 & 0 & 0 & 0 & 0 & -\frac{1}{2} & 0 \\
 -\frac{1}{2} & 0 & -1 & 0 & 0 & 0 & 0 & 0 & 0 & \frac{1}{2} \\
 0 & 1 & 0 & -\frac{1}{2} & 0 & 0 & -\frac{1}{2} & 0 & 0 & 0 \\
 0 & 0 & -\frac{1}{2} & 0 & 0 & 0 & 0 & \frac{1}{2} & 0 & 0 \\
 0 & 0 & 0 & 0 & 0 & 0 & 0 & 0 & 0 & 0 \\
 0 & 0 & 0 & 0 & 0 & 0 & 0 & 0 & 0 & 0 \\
 0 & 0 & -\frac{1}{2} & 0 & 0 & 0 & 0 & \frac{1}{2} & 0 & 0 \\
 0 & 0 & 0 & \frac{1}{2} & 0 & 0 & \frac{1}{2} & 0 & 1 & 0 \\
 -\frac{1}{2} & 0 & 0 & 0 & 0 & 0 & 0 & -1 & 0 & \frac{1}{2} \\
 0 & \frac{1}{2} & 0 & 0 & 0 & 0 & 0 & 0 & \frac{1}{2} & 0
\end{array}
\right)
\end{equation}
\begin{equation}\label{nil15}
    \mathcal{O}_{1,3}^7\, = \,\left(
\begin{array}{llllllllll}
 0 & -1 & 0 & \frac{1}{2} & 0 & 0 & \frac{1}{2} & 0 & 0 & 0 \\
 -1 & 0 & -\frac{1}{2} & 0 & 0 & 0 & 0 & -\frac{1}{2} & 0 & 0 \\
 0 & \frac{1}{2} & 0 & 0 & 0 & 0 & 0 & 0 & \frac{1}{2} & 0 \\
 -\frac{1}{2} & 0 & 0 & 0 & 0 & 0 & 0 & 0 & 0 & -\frac{1}{2} \\
 0 & 0 & 0 & 0 & 0 & 0 & 0 & 0 & 0 & 0 \\
 0 & 0 & 0 & 0 & 0 & 0 & 0 & 0 & 0 & 0 \\
 -\frac{1}{2} & 0 & 0 & 0 & 0 & 0 & 0 & 0 & 0 & -\frac{1}{2} \\
 0 & \frac{1}{2} & 0 & 0 & 0 & 0 & 0 & 0 & \frac{1}{2} & 0 \\
 0 & 0 & -\frac{1}{2} & 0 & 0 & 0 & 0 & -\frac{1}{2} & 0 & 1 \\
 0 & 0 & 0 & \frac{1}{2} & 0 & 0 & \frac{1}{2} & 0 & 1 & 0
\end{array}
\right)
\end{equation}
\begin{equation}\label{nil16}
    \mathcal{O}_{2,1}^7\, = \,\left(
\begin{array}{llllllllll}
 0 & -\frac{1}{2} & 0 & -\frac{1}{2} & 0 & 0 & \frac{1}{2} & 0 & \frac{1}{2} & 0 \\
 -\frac{1}{2} & 0 & \frac{1}{2} & 0 & 0 & 0 & 0 & -\frac{1}{2} & 0 & -\frac{1}{2} \\
 0 & -\frac{1}{2} & 0 & \frac{1}{2} & 0 & 0 & -\frac{1}{2} & 0 & \frac{1}{2} & 0 \\
 \frac{1}{2} & 0 & \frac{1}{2} & 0 & 0 & 0 & 0 & \frac{1}{2} & 0 & -\frac{1}{2} \\
 0 & 0 & 0 & 0 & 0 & 0 & 0 & 0 & 0 & 0 \\
 0 & 0 & 0 & 0 & 0 & 0 & 0 & 0 & 0 & 0 \\
 -\frac{1}{2} & 0 & -\frac{1}{2} & 0 & 0 & 0 & 0 & -\frac{1}{2} & 0 & \frac{1}{2} \\
 0 & \frac{1}{2} & 0 & \frac{1}{2} & 0 & 0 & -\frac{1}{2} & 0 & -\frac{1}{2} & 0 \\
 \frac{1}{2} & 0 & -\frac{1}{2} & 0 & 0 & 0 & 0 & \frac{1}{2} & 0 & \frac{1}{2} \\
 0 & -\frac{1}{2} & 0 & \frac{1}{2} & 0 & 0 & -\frac{1}{2} & 0 & \frac{1}{2} & 0
\end{array}
\right)
\end{equation}
\begin{equation}\label{nil17}
    \mathcal{O}_{2,2}^7\, = \,\left(
\begin{array}{llllllllll}
 0 & -1 & 0 & 0 & 0 & 0 & 0 & 0 & 0 & 0 \\
 -1 & 0 & 0 & 0 & 0 & 0 & 0 & -1 & 0 & 0 \\
 0 & 0 & 0 & 0 & 0 & 0 & 0 & 0 & 1 & 0 \\
 0 & 0 & 0 & 0 & 0 & 0 & 0 & 0 & 0 & 0 \\
 0 & 0 & 0 & 0 & 0 & 0 & 0 & 0 & 0 & 0 \\
 0 & 0 & 0 & 0 & 0 & 0 & 0 & 0 & 0 & 0 \\
 0 & 0 & 0 & 0 & 0 & 0 & 0 & 0 & 0 & 0 \\
 0 & 1 & 0 & 0 & 0 & 0 & 0 & 0 & 0 & 0 \\
 0 & 0 & -1 & 0 & 0 & 0 & 0 & 0 & 0 & 1 \\
 0 & 0 & 0 & 0 & 0 & 0 & 0 & 0 & 1 & 0
\end{array}
\right)\
\end{equation}
\begin{equation}\label{nil18}
    \mathcal{O}_{2,3}^7\, = \,\left(
\begin{array}{llllllllll}
 0 & -\frac{1}{2} & 0 & -\frac{1}{2} & 0 & 0 & -\frac{1}{2} & 0 & -\frac{1}{2} & 0 \\
 -\frac{1}{2} & 0 & -\frac{1}{2} & 0 & 0 & 0 & 0 & -\frac{1}{2} & 0 & \frac{1}{2} \\
 0 & \frac{1}{2} & 0 & -\frac{1}{2} & 0 & 0 & -\frac{1}{2} & 0 & \frac{1}{2} & 0 \\
 \frac{1}{2} & 0 & -\frac{1}{2} & 0 & 0 & 0 & 0 & \frac{1}{2} & 0 & \frac{1}{2} \\
 0 & 0 & 0 & 0 & 0 & 0 & 0 & 0 & 0 & 0 \\
 0 & 0 & 0 & 0 & 0 & 0 & 0 & 0 & 0 & 0 \\
 \frac{1}{2} & 0 & -\frac{1}{2} & 0 & 0 & 0 & 0 & \frac{1}{2} & 0 & \frac{1}{2} \\
 0 & \frac{1}{2} & 0 & \frac{1}{2} & 0 & 0 & \frac{1}{2} & 0 & \frac{1}{2} & 0 \\
 -\frac{1}{2} & 0 & -\frac{1}{2} & 0 & 0 & 0 & 0 & -\frac{1}{2} & 0 & \frac{1}{2} \\
 0 & \frac{1}{2} & 0 & -\frac{1}{2} & 0 & 0 & -\frac{1}{2} & 0 & \frac{1}{2} & 0
\end{array}
\right)
\end{equation}
\begin{equation}\label{nil19}
   \mathcal{O}_{3,1}^7\, = \,\left(
\begin{array}{llllllllll}
 0 & -1 & 0 & \frac{1}{2} & 0 & 0 & -\frac{1}{2} & 0 & 0 & 0 \\
 -1 & 0 & \frac{1}{2} & 0 & 0 & 0 & 0 & -\frac{1}{2} & 0 & 0 \\
 0 & -\frac{1}{2} & 0 & 0 & 0 & 0 & 0 & 0 & \frac{1}{2} & 0 \\
 -\frac{1}{2} & 0 & 0 & 0 & 0 & 0 & 0 & 0 & 0 & \frac{1}{2} \\
 0 & 0 & 0 & 0 & 0 & 0 & 0 & 0 & 0 & 0 \\
 0 & 0 & 0 & 0 & 0 & 0 & 0 & 0 & 0 & 0 \\
 \frac{1}{2} & 0 & 0 & 0 & 0 & 0 & 0 & 0 & 0 & -\frac{1}{2} \\
 0 & \frac{1}{2} & 0 & 0 & 0 & 0 & 0 & 0 & -\frac{1}{2} & 0 \\
 0 & 0 & -\frac{1}{2} & 0 & 0 & 0 & 0 & \frac{1}{2} & 0 & 1 \\
 0 & 0 & 0 & -\frac{1}{2} & 0 & 0 & \frac{1}{2} & 0 & 1 & 0
\end{array}
\right)
\end{equation}
\begin{equation}\label{nil20}
    \mathcal{O}_{3,2}^7\, = \,\left(
\begin{array}{llllllllll}
 0 & -\frac{1}{2} & 0 & 0 & 0 & 0 & -1 & 0 & -\frac{1}{2} & 0 \\
 -\frac{1}{2} & 0 & 0 & 0 & 0 & 0 & 0 & 0 & 0 & \frac{1}{2} \\
 0 & 0 & 0 & -\frac{1}{2} & 0 & 0 & -\frac{1}{2} & 0 & 0 & 0 \\
 0 & 0 & -\frac{1}{2} & 0 & 0 & 0 & 0 & \frac{1}{2} & 0 & 1 \\
 0 & 0 & 0 & 0 & 0 & 0 & 0 & 0 & 0 & 0 \\
 0 & 0 & 0 & 0 & 0 & 0 & 0 & 0 & 0 & 0 \\
 1 & 0 & -\frac{1}{2} & 0 & 0 & 0 & 0 & \frac{1}{2} & 0 & 0 \\
 0 & 0 & 0 & \frac{1}{2} & 0 & 0 & \frac{1}{2} & 0 & 0 & 0 \\
 -\frac{1}{2} & 0 & 0 & 0 & 0 & 0 & 0 & 0 & 0 & \frac{1}{2} \\
 0 & \frac{1}{2} & 0 & -1 & 0 & 0 & 0 & 0 & \frac{1}{2} & 0
\end{array}
\right)
\end{equation}
\begin{equation}\label{nil21}
    \mathcal{O}_{3,3}^7\, = \,\left(
\begin{array}{llllllllll}
 0 & 0 & 0 & -\frac{1}{2} & 0 & 0 & -\frac{1}{2} & 0 & 0 & 0 \\
 0 & 0 & -\frac{1}{2} & 0 & 0 & 0 & 0 & -\frac{1}{2} & 0 & 0 \\
 0 & \frac{1}{2} & 0 & -1 & 0 & 0 & 0 & 0 & \frac{1}{2} & 0 \\
 \frac{1}{2} & 0 & -1 & 0 & 0 & 0 & 0 & 0 & 0 & \frac{1}{2} \\
 0 & 0 & 0 & 0 & 0 & 0 & 0 & 0 & 0 & 0 \\
 0 & 0 & 0 & 0 & 0 & 0 & 0 & 0 & 0 & 0 \\
 \frac{1}{2} & 0 & 0 & 0 & 0 & 0 & 0 & 1 & 0 & \frac{1}{2} \\
 0 & \frac{1}{2} & 0 & 0 & 0 & 0 & 1 & 0 & \frac{1}{2} & 0 \\
 0 & 0 & -\frac{1}{2} & 0 & 0 & 0 & 0 & -\frac{1}{2} & 0 & 0 \\
 0 & 0 & 0 & -\frac{1}{2} & 0 & 0 & -\frac{1}{2} & 0 & 0 & 0
\end{array}
\right)
\end{equation}
\begin{equation}\label{nil22}
    \mathcal{O}_{1,1}^8\, = \,\left(
\begin{array}{llllllllll}
 -\frac{1}{2} & 0 & 0 & 0 & 0 & 0 & 0 & -\frac{1}{2} & 0 & 0 \\
 0 & -\frac{1}{2} & 0 & \frac{1}{2} & 0 & 0 & 0 & 0 & 0 & 0 \\
 0 & 0 & -\frac{1}{2} & 0 & 0 & 0 & 0 & 0 & 0 & \frac{1}{2} \\
 0 & -\frac{1}{2} & 0 & \frac{1}{2} & 0 & 0 & 0 & 0 & 0 & 0 \\
 0 & 0 & 0 & 0 & 0 & 0 & 0 & 0 & 0 & 0 \\
 0 & 0 & 0 & 0 & 0 & 0 & 0 & 0 & 0 & 0 \\
 0 & 0 & 0 & 0 & 0 & 0 & -\frac{1}{2} & 0 & -\frac{1}{2} & 0 \\
 \frac{1}{2} & 0 & 0 & 0 & 0 & 0 & 0 & \frac{1}{2} & 0 & 0 \\
 0 & 0 & 0 & 0 & 0 & 0 & \frac{1}{2} & 0 & \frac{1}{2} & 0 \\
 0 & 0 & -\frac{1}{2} & 0 & 0 & 0 & 0 & 0 & 0 & \frac{1}{2}
\end{array}
\right)
\end{equation}
\begin{equation}\label{nil23}
   \mathcal{O}_{1,1}^9\, = \,\left(
\begin{array}{llllllllll}
 0 & 0 & 0 & -\frac{1}{2} & 0 & 0 & \frac{1}{2} & 0 & 0 & 0 \\
 0 & 0 & 0 & 0 & 0 & 0 & 0 & 0 & 0 & 0 \\
 0 & 0 & 0 & \frac{1}{2} & 0 & 0 & -\frac{1}{2} & 0 & 0 & 0 \\
 \frac{1}{2} & 0 & \frac{1}{2} & 0 & 0 & 0 & 0 & \frac{1}{2} & 0 & -\frac{1}{2} \\
 0 & 0 & 0 & 0 & 0 & 0 & 0 & 0 & 0 & 0 \\
 0 & 0 & 0 & 0 & 0 & 0 & 0 & 0 & 0 & 0 \\
 -\frac{1}{2} & 0 & -\frac{1}{2} & 0 & 0 & 0 & 0 & -\frac{1}{2} & 0 & \frac{1}{2} \\
 0 & 0 & 0 & \frac{1}{2} & 0 & 0 & -\frac{1}{2} & 0 & 0 & 0 \\
 0 & 0 & 0 & 0 & 0 & 0 & 0 & 0 & 0 & 0 \\
 0 & 0 & 0 & \frac{1}{2} & 0 & 0 & -\frac{1}{2} & 0 & 0 & 0
\end{array}
\right)
\end{equation}
\begin{equation}\label{nil24}
    \mathcal{O}_{1,2}^9\, = \,\left(
\begin{array}{llllllllll}
 0 & -\frac{1}{2} & 0 & 0 & 0 & 0 & 0 & 0 & -\frac{1}{2} & 0 \\
 -\frac{1}{2} & 0 & -\frac{1}{2} & 0 & 0 & 0 & 0 & -\frac{1}{2} & 0 & \frac{1}{2} \\
 0 & \frac{1}{2} & 0 & 0 & 0 & 0 & 0 & 0 & \frac{1}{2} & 0 \\
 0 & 0 & 0 & 0 & 0 & 0 & 0 & 0 & 0 & 0 \\
 0 & 0 & 0 & 0 & 0 & 0 & 0 & 0 & 0 & 0 \\
 0 & 0 & 0 & 0 & 0 & 0 & 0 & 0 & 0 & 0 \\
 0 & 0 & 0 & 0 & 0 & 0 & 0 & 0 & 0 & 0 \\
 0 & \frac{1}{2} & 0 & 0 & 0 & 0 & 0 & 0 & \frac{1}{2} & 0 \\
 -\frac{1}{2} & 0 & -\frac{1}{2} & 0 & 0 & 0 & 0 & -\frac{1}{2} & 0 & \frac{1}{2} \\
 0 & \frac{1}{2} & 0 & 0 & 0 & 0 & 0 & 0 & \frac{1}{2} & 0
\end{array}
\right)
\end{equation}
\begin{equation}\label{nil25}
    \mathcal{O}_{2,1}^9\, = \,\left(
\begin{array}{llllllllll}
 0 & -\frac{1}{2} & 0 & 0 & 0 & 0 & 0 & 0 & \frac{1}{2} & 0 \\
 -\frac{1}{2} & 0 & \frac{1}{2} & 0 & 0 & 0 & 0 & -\frac{1}{2} & 0 & -\frac{1}{2} \\
 0 & -\frac{1}{2} & 0 & 0 & 0 & 0 & 0 & 0 & \frac{1}{2} & 0 \\
 0 & 0 & 0 & 0 & 0 & 0 & 0 & 0 & 0 & 0 \\
 0 & 0 & 0 & 0 & 0 & 0 & 0 & 0 & 0 & 0 \\
 0 & 0 & 0 & 0 & 0 & 0 & 0 & 0 & 0 & 0 \\
 0 & 0 & 0 & 0 & 0 & 0 & 0 & 0 & 0 & 0 \\
 0 & \frac{1}{2} & 0 & 0 & 0 & 0 & 0 & 0 & -\frac{1}{2} & 0 \\
 \frac{1}{2} & 0 & -\frac{1}{2} & 0 & 0 & 0 & 0 & \frac{1}{2} & 0 & \frac{1}{2} \\
 0 & -\frac{1}{2} & 0 & 0 & 0 & 0 & 0 & 0 & \frac{1}{2} & 0
\end{array}
\right)\
\end{equation}
\begin{equation}\label{nil26}
    \mathcal{O}_{2,2}^9\, = \,\left(
\begin{array}{llllllllll}
 0 & 0 & 0 & -\frac{1}{2} & 0 & 0 & -\frac{1}{2} & 0 & 0 & 0 \\
 0 & 0 & 0 & 0 & 0 & 0 & 0 & 0 & 0 & 0 \\
 0 & 0 & 0 & -\frac{1}{2} & 0 & 0 & -\frac{1}{2} & 0 & 0 & 0 \\
 \frac{1}{2} & 0 & -\frac{1}{2} & 0 & 0 & 0 & 0 & \frac{1}{2} & 0 & \frac{1}{2} \\
 0 & 0 & 0 & 0 & 0 & 0 & 0 & 0 & 0 & 0 \\
 0 & 0 & 0 & 0 & 0 & 0 & 0 & 0 & 0 & 0 \\
 \frac{1}{2} & 0 & -\frac{1}{2} & 0 & 0 & 0 & 0 & \frac{1}{2} & 0 & \frac{1}{2} \\
 0 & 0 & 0 & \frac{1}{2} & 0 & 0 & \frac{1}{2} & 0 & 0 & 0 \\
 0 & 0 & 0 & 0 & 0 & 0 & 0 & 0 & 0 & 0 \\
 0 & 0 & 0 & -\frac{1}{2} & 0 & 0 & -\frac{1}{2} & 0 & 0 & 0
\end{array}
\right)
\end{equation}
\begin{equation}\label{nil27}
   \mathcal{O}_{1,1}^{10}\, = \,\left(
\begin{array}{llllllllll}
 -\frac{1}{2} & 0 & 0 & 0 & 0 & 0 & 0 & -\frac{1}{2} & 0 & 0 \\
 0 & 0 & 0 & 0 & 0 & 0 & 0 & 0 & 0 & 0 \\
 0 & 0 & -\frac{1}{2} & 0 & 0 & 0 & 0 & 0 & 0 & \frac{1}{2} \\
 0 & 0 & 0 & 0 & 0 & 0 & 0 & 0 & 0 & 0 \\
 0 & 0 & 0 & 0 & 0 & 0 & 0 & 0 & 0 & 0 \\
 0 & 0 & 0 & 0 & 0 & 0 & 0 & 0 & 0 & 0 \\
 0 & 0 & 0 & 0 & 0 & 0 & 0 & 0 & 0 & 0 \\
 \frac{1}{2} & 0 & 0 & 0 & 0 & 0 & 0 & \frac{1}{2} & 0 & 0 \\
 0 & 0 & 0 & 0 & 0 & 0 & 0 & 0 & 0 & 0 \\
 0 & 0 & -\frac{1}{2} & 0 & 0 & 0 & 0 & 0 & 0 & \frac{1}{2}
\end{array}
\right)
\end{equation}
\begin{equation}\label{nil28}
   \mathcal{O}_{1,1}^{11}\, = \,\left(
\begin{array}{llllllllll}
 -1 & -\frac{\sqrt{3}}{2} & -1 & 0 & 0 & 0 & 0 & 0 & \frac{\sqrt{3}}{2} & 0 \\
 -\frac{\sqrt{3}}{2} & 0 & \frac{\sqrt{3}}{2} & -\frac{1}{2} & 0 & 0 & -\frac{1}{2} & -\frac{\sqrt{3}}{2} & 0 & -\frac{\sqrt{3}}{2} \\
 1 & -\frac{\sqrt{3}}{2} & 1 & 0 & 0 & 0 & 0 & 0 & \frac{\sqrt{3}}{2} & 0 \\
 0 & \frac{1}{2} & 0 & -1 & 0 & 0 & 0 & 0 & \frac{1}{2} & 0 \\
 0 & 0 & 0 & 0 & 0 & 0 & 0 & 0 & 0 & 0 \\
 0 & 0 & 0 & 0 & 0 & 0 & 0 & 0 & 0 & 0 \\
 0 & \frac{1}{2} & 0 & 0 & 0 & 0 & 1 & 0 & \frac{1}{2} & 0 \\
 0 & \frac{\sqrt{3}}{2} & 0 & 0 & 0 & 0 & 0 & -1 & -\frac{\sqrt{3}}{2} & 1 \\
 \frac{\sqrt{3}}{2} & 0 & -\frac{\sqrt{3}}{2} & -\frac{1}{2} & 0 & 0 & -\frac{1}{2} & \frac{\sqrt{3}}{2} & 0 & \frac{\sqrt{3}}{2} \\
 0 & -\frac{\sqrt{3}}{2} & 0 & 0 & 0 & 0 & 0 & -1 & \frac{\sqrt{3}}{2} & 1
\end{array}
\right)
\end{equation}
\begin{equation}\label{nil29}
    \mathcal{O}_{1,2}^{11}\, = \,\left(
\begin{array}{llllllllll}
 -1 & 0 & -1 & -\frac{\sqrt{3}}{2} & 0 & 0 & -\frac{\sqrt{3}}{2} & 0 & 0 & 0 \\
 0 & -1 & 0 & \frac{1}{2} & 0 & 0 & -\frac{1}{2} & 0 & 0 & 0 \\
 1 & 0 & 1 & -\frac{\sqrt{3}}{2} & 0 & 0 & -\frac{\sqrt{3}}{2} & 0 & 0 & 0 \\
 \frac{\sqrt{3}}{2} & -\frac{1}{2} & -\frac{\sqrt{3}}{2} & 0 & 0 & 0 & 0 & \frac{\sqrt{3}}{2} & \frac{1}{2} & \frac{\sqrt{3}}{2} \\
 0 & 0 & 0 & 0 & 0 & 0 & 0 & 0 & 0 & 0 \\
 0 & 0 & 0 & 0 & 0 & 0 & 0 & 0 & 0 & 0 \\
 \frac{\sqrt{3}}{2} & \frac{1}{2} & -\frac{\sqrt{3}}{2} & 0 & 0 & 0 & 0 & \frac{\sqrt{3}}{2} & -\frac{1}{2} & \frac{\sqrt{3}}{2} \\
 0 & 0 & 0 & \frac{\sqrt{3}}{2} & 0 & 0 & \frac{\sqrt{3}}{2} & -1 & 0 & 1 \\
 0 & 0 & 0 & -\frac{1}{2} & 0 & 0 & \frac{1}{2} & 0 & 1 & 0 \\
 0 & 0 & 0 & -\frac{\sqrt{3}}{2} & 0 & 0 & -\frac{\sqrt{3}}{2} & -1 & 0 & 1
\end{array}
\right)
\end{equation}
\begin{equation}\label{nil30}
    \mathcal{O}_{2,1}^{11}\, = \,\left(
\begin{array}{llllllllll}
 0 & -\frac{1}{2} & 0 & -\frac{1}{2} & 0 & \frac{1}{\sqrt{2}} & -\frac{1}{2} & 0 & \frac{1}{2} & 0 \\
 -\frac{1}{2} & 0 & \frac{1}{2} & -\frac{\sqrt{3}}{2} & 0 & 0 & \frac{\sqrt{3}}{2} & -\frac{1}{2} & 0 & -\frac{1}{2} \\
 0 & -\frac{1}{2} & 0 & -\frac{1}{2} & 0 & -\frac{1}{\sqrt{2}} & -\frac{1}{2} & 0 & \frac{1}{2} & 0 \\
 \frac{1}{2} & \frac{\sqrt{3}}{2} & -\frac{1}{2} & -\sqrt{3} & 0 & 0 & 0 & \frac{1}{2} & -\frac{\sqrt{3}}{2} & \frac{1}{2} \\
 0 & 0 & 0 & 0 & 0 & 0 & 0 & 0 & 0 & 0 \\
 \frac{1}{\sqrt{2}} & 0 & \frac{1}{\sqrt{2}} & 0 & 0 & 0 & 0 & \frac{1}{\sqrt{2}} & 0 & -\frac{1}{\sqrt{2}} \\
 \frac{1}{2} & -\frac{\sqrt{3}}{2} & -\frac{1}{2} & 0 & 0 & 0 & \sqrt{3} & \frac{1}{2} & \frac{\sqrt{3}}{2} & \frac{1}{2} \\
 0 & \frac{1}{2} & 0 & \frac{1}{2} & 0 & -\frac{1}{\sqrt{2}} & \frac{1}{2} & 0 & -\frac{1}{2} & 0 \\
 \frac{1}{2} & 0 & -\frac{1}{2} & \frac{\sqrt{3}}{2} & 0 & 0 & -\frac{\sqrt{3}}{2} & \frac{1}{2} & 0 & \frac{1}{2} \\
 0 & -\frac{1}{2} & 0 & -\frac{1}{2} & 0 & -\frac{1}{\sqrt{2}} & -\frac{1}{2} & 0 & \frac{1}{2} & 0
\end{array}
\right)
\end{equation}
\begin{equation}\label{nil31}
    \mathcal{O}_{2,3}^{11}\, = \,\left(
\begin{array}{llllllllll}
 0 & -\frac{1}{2} & 0 & -\frac{1}{2} & 0 & \frac{1}{\sqrt{2}} & -\frac{1}{2} & 0 & \frac{1}{2} & 0 \\
 -\frac{1}{2} & \sqrt{3} & \frac{1}{2} & -\frac{\sqrt{3}}{2} & 0 & 0 & -\frac{\sqrt{3}}{2} & -\frac{1}{2} & 0 & -\frac{1}{2} \\
 0 & -\frac{1}{2} & 0 & -\frac{1}{2} & 0 & -\frac{1}{\sqrt{2}} & -\frac{1}{2} & 0 & \frac{1}{2} & 0 \\
 \frac{1}{2} & \frac{\sqrt{3}}{2} & -\frac{1}{2} & 0 & 0 & 0 & 0 & \frac{1}{2} & \frac{\sqrt{3}}{2} & \frac{1}{2} \\
 0 & 0 & 0 & 0 & 0 & 0 & 0 & 0 & 0 & 0 \\
 \frac{1}{\sqrt{2}} & 0 & \frac{1}{\sqrt{2}} & 0 & 0 & 0 & 0 & \frac{1}{\sqrt{2}} & 0 & -\frac{1}{\sqrt{2}} \\
 \frac{1}{2} & \frac{\sqrt{3}}{2} & -\frac{1}{2} & 0 & 0 & 0 & 0 & \frac{1}{2} & \frac{\sqrt{3}}{2} & \frac{1}{2} \\
 0 & \frac{1}{2} & 0 & \frac{1}{2} & 0 & -\frac{1}{\sqrt{2}} & \frac{1}{2} & 0 & -\frac{1}{2} & 0 \\
 \frac{1}{2} & 0 & -\frac{1}{2} & -\frac{\sqrt{3}}{2} & 0 & 0 & -\frac{\sqrt{3}}{2} & \frac{1}{2} & -\sqrt{3} & \frac{1}{2} \\
 0 & -\frac{1}{2} & 0 & -\frac{1}{2} & 0 & -\frac{1}{\sqrt{2}} & -\frac{1}{2} & 0 & \frac{1}{2} & 0
\end{array}
\right)
\end{equation}
\begin{equation}\label{nil32}
    \mathcal{O}_{3,2}^{11}\, = \,\left(
\begin{array}{llllllllll}
 -1 & 0 & 1 & 0 & 0 & -\sqrt{\frac{3}{2}} & 0 & 0 & 0 & 0 \\
 0 & -1 & 0 & \frac{1}{2} & 0 & 0 & -\frac{1}{2} & 0 & 0 & 0 \\
 -1 & 0 & 1 & 0 & 0 & \sqrt{\frac{3}{2}} & 0 & 0 & 0 & 0 \\
 0 & -\frac{1}{2} & 0 & 0 & 0 & 0 & 0 & 0 & \frac{1}{2} & 0 \\
 0 & 0 & 0 & 0 & 0 & 0 & 0 & 0 & 0 & 0 \\
 -\sqrt{\frac{3}{2}} & 0 & -\sqrt{\frac{3}{2}} & 0 & 0 & 0 & 0 & -\sqrt{\frac{3}{2}} & 0 & \sqrt{\frac{3}{2}} \\
 0 & \frac{1}{2} & 0 & 0 & 0 & 0 & 0 & 0 & -\frac{1}{2} & 0 \\
 0 & 0 & 0 & 0 & 0 & \sqrt{\frac{3}{2}} & 0 & -1 & 0 & -1 \\
 0 & 0 & 0 & -\frac{1}{2} & 0 & 0 & \frac{1}{2} & 0 & 1 & 0 \\
 0 & 0 & 0 & 0 & 0 & \sqrt{\frac{3}{2}} & 0 & 1 & 0 & 1
\end{array}
\right)
\end{equation}
\begin{equation}\label{nil33}
    \mathcal{O}_{3,3}^{11}\, = \,\left(
\begin{array}{llllllllll}
 -1 & 0 & 1 & 0 & 0 & -\sqrt{\frac{3}{2}} & 0 & 0 & 0 & 0 \\
 0 & 0 & 0 & -\frac{1}{2} & 0 & 0 & -\frac{1}{2} & 0 & 0 & 0 \\
 -1 & 0 & 1 & 0 & 0 & \sqrt{\frac{3}{2}} & 0 & 0 & 0 & 0 \\
 0 & \frac{1}{2} & 0 & -1 & 0 & 0 & 0 & 0 & \frac{1}{2} & 0 \\
 0 & 0 & 0 & 0 & 0 & 0 & 0 & 0 & 0 & 0 \\
 -\sqrt{\frac{3}{2}} & 0 & -\sqrt{\frac{3}{2}} & 0 & 0 & 0 & 0 & -\sqrt{\frac{3}{2}} & 0 & \sqrt{\frac{3}{2}} \\
 0 & \frac{1}{2} & 0 & 0 & 0 & 0 & 1 & 0 & \frac{1}{2} & 0 \\
 0 & 0 & 0 & 0 & 0 & \sqrt{\frac{3}{2}} & 0 & -1 & 0 & -1 \\
 0 & 0 & 0 & -\frac{1}{2} & 0 & 0 & -\frac{1}{2} & 0 & 0 & 0 \\
 0 & 0 & 0 & 0 & 0 & \sqrt{\frac{3}{2}} & 0 & 1 & 0 & 1
\end{array}
\right)
\end{equation}
\begin{equation}\label{nil34}
    \mathcal{O}_{1,1}^{12}\, = \,\left(
\begin{array}{llllllllll}
 -1 & -\frac{\sqrt{3}}{2} & -1 & 0 & 0 & 0 & 0 & 0 & \frac{\sqrt{3}}{2} & 0 \\
 -\frac{\sqrt{3}}{2} & 0 & \frac{\sqrt{3}}{2} & 0 & 0 & 0 & 0 & -\frac{\sqrt{3}}{2} & 0 & -\frac{\sqrt{3}}{2} \\
 1 & -\frac{\sqrt{3}}{2} & 1 & 0 & 0 & 0 & 0 & 0 & \frac{\sqrt{3}}{2} & 0 \\
 0 & 0 & 0 & 0 & 0 & 0 & 0 & 0 & 0 & 0 \\
 0 & 0 & 0 & 0 & 0 & 0 & 0 & 0 & 0 & 0 \\
 0 & 0 & 0 & 0 & 0 & 0 & 0 & 0 & 0 & 0 \\
 0 & 0 & 0 & 0 & 0 & 0 & 0 & 0 & 0 & 0 \\
 0 & \frac{\sqrt{3}}{2} & 0 & 0 & 0 & 0 & 0 & -1 & -\frac{\sqrt{3}}{2} & 1 \\
 \frac{\sqrt{3}}{2} & 0 & -\frac{\sqrt{3}}{2} & 0 & 0 & 0 & 0 & \frac{\sqrt{3}}{2} & 0 & \frac{\sqrt{3}}{2} \\
 0 & -\frac{\sqrt{3}}{2} & 0 & 0 & 0 & 0 & 0 & -1 & \frac{\sqrt{3}}{2} & 1
\end{array}
\right)
\end{equation}
\begin{equation}\label{nil35}
    \mathcal{O}_{1,2}^{12}\, = \,\left(
\begin{array}{llllllllll}
 -1 & 0 & -1 & -\frac{\sqrt{3}}{2} & 0 & 0 & -\frac{\sqrt{3}}{2} & 0 & 0 & 0 \\
 0 & 0 & 0 & 0 & 0 & 0 & 0 & 0 & 0 & 0 \\
 1 & 0 & 1 & -\frac{\sqrt{3}}{2} & 0 & 0 & -\frac{\sqrt{3}}{2} & 0 & 0 & 0 \\
 \frac{\sqrt{3}}{2} & 0 & -\frac{\sqrt{3}}{2} & 0 & 0 & 0 & 0 & \frac{\sqrt{3}}{2} & 0 & \frac{\sqrt{3}}{2} \\
 0 & 0 & 0 & 0 & 0 & 0 & 0 & 0 & 0 & 0 \\
 0 & 0 & 0 & 0 & 0 & 0 & 0 & 0 & 0 & 0 \\
 \frac{\sqrt{3}}{2} & 0 & -\frac{\sqrt{3}}{2} & 0 & 0 & 0 & 0 & \frac{\sqrt{3}}{2} & 0 & \frac{\sqrt{3}}{2} \\
 0 & 0 & 0 & \frac{\sqrt{3}}{2} & 0 & 0 & \frac{\sqrt{3}}{2} & -1 & 0 & 1 \\
 0 & 0 & 0 & 0 & 0 & 0 & 0 & 0 & 0 & 0 \\
 0 & 0 & 0 & -\frac{\sqrt{3}}{2} & 0 & 0 & -\frac{\sqrt{3}}{2} & -1 & 0 & 1
\end{array}
\right)
\end{equation}
\begin{equation}\label{nil36}
    \mathcal{O}_{2,1}^{12}\, = \,\left(
\begin{array}{llllllllll}
 -1 & 0 & 1 & -\frac{\sqrt{3}}{2} & 0 & 0 & \frac{\sqrt{3}}{2} & 0 & 0 & 0 \\
 0 & 0 & 0 & 0 & 0 & 0 & 0 & 0 & 0 & 0 \\
 -1 & 0 & 1 & \frac{\sqrt{3}}{2} & 0 & 0 & -\frac{\sqrt{3}}{2} & 0 & 0 & 0 \\
 \frac{\sqrt{3}}{2} & 0 & \frac{\sqrt{3}}{2} & 0 & 0 & 0 & 0 & \frac{\sqrt{3}}{2} & 0 & -\frac{\sqrt{3}}{2} \\
 0 & 0 & 0 & 0 & 0 & 0 & 0 & 0 & 0 & 0 \\
 0 & 0 & 0 & 0 & 0 & 0 & 0 & 0 & 0 & 0 \\
 -\frac{\sqrt{3}}{2} & 0 & -\frac{\sqrt{3}}{2} & 0 & 0 & 0 & 0 & -\frac{\sqrt{3}}{2} & 0 & \frac{\sqrt{3}}{2} \\
 0 & 0 & 0 & \frac{\sqrt{3}}{2} & 0 & 0 & -\frac{\sqrt{3}}{2} & -1 & 0 & -1 \\
 0 & 0 & 0 & 0 & 0 & 0 & 0 & 0 & 0 & 0 \\
 0 & 0 & 0 & \frac{\sqrt{3}}{2} & 0 & 0 & -\frac{\sqrt{3}}{2} & 1 & 0 & 1
\end{array}
\right)
\end{equation}
\begin{equation}\label{nil37}
    \mathcal{O}_{2,2}^{12}\, = \,\left(
\begin{array}{llllllllll}
 -1 & 0 & 1 & 0 & 0 & -\sqrt{\frac{3}{2}} & 0 & 0 & 0 & 0 \\
 0 & 0 & 0 & 0 & 0 & 0 & 0 & 0 & 0 & 0 \\
 -1 & 0 & 1 & 0 & 0 & \sqrt{\frac{3}{2}} & 0 & 0 & 0 & 0 \\
 0 & 0 & 0 & 0 & 0 & 0 & 0 & 0 & 0 & 0 \\
 0 & 0 & 0 & 0 & 0 & 0 & 0 & 0 & 0 & 0 \\
 -\sqrt{\frac{3}{2}} & 0 & -\sqrt{\frac{3}{2}} & 0 & 0 & 0 & 0 & -\sqrt{\frac{3}{2}} & 0 & \sqrt{\frac{3}{2}} \\
 0 & 0 & 0 & 0 & 0 & 0 & 0 & 0 & 0 & 0 \\
 0 & 0 & 0 & 0 & 0 & \sqrt{\frac{3}{2}} & 0 & -1 & 0 & -1 \\
 0 & 0 & 0 & 0 & 0 & 0 & 0 & 0 & 0 & 0 \\
 0 & 0 & 0 & 0 & 0 & \sqrt{\frac{3}{2}} & 0 & 1 & 0 & 1
\end{array}
\right)
\end{equation}
}

\end{document}